\documentclass[12pt,BCOR2mm,final,headsepline,fleqn]{scrbook} 
\usepackage[automark]{scrpage2}
 \pdfoutput=1
\usepackage{graphicx,color,multirow,amsmath,setspace,natbib,longtable}
\usepackage[naustrian,british]{babel}
\usepackage{listings, upgreek}
\usepackage[ansinew]{inputenc} 
\usepackage[T1]{fontenc} 
\usepackage{lmodern}

\usepackage{pgf,tikz}
\usetikzlibrary{arrows}
\usetikzlibrary{mindmap,trees}

\usepackage[pdftex]{hyperref}

\hypersetup{
  linktocpage=true,
  colorlinks=false, 
  urlcolor=blue,
  linkcolor=blue,
  menucolor=blue,
  pdftitle={Exoplanets and Active Stars},
  pdfauthor={J. Weingrill},
  pdfsubject={Doctoral Thesis},
  pdfcreator=pdfTeX,
  pdfproducer={J. Weingrill},
  pdfkeywords={exoplanets} {stars} {stellar activity}
  }

\addtokomafont{caption}{\slshape} 
\setkomafont{captionlabel}{\normalfont\bfseries} 

\newcommand{\corot}{CoRoT}

\newcommand{\sun}{\ensuremath{\odot}}

\newcommand{\Teff}{\ensuremath{T_{\mathrm{eff}}}}
\newcommand{\MJup}{\ensuremath{M_{\mathrm{Jup}}}}
\newcommand\vek[1]{\mathbf{#1}}
\newcommand{\AU}{\ensuremath{AU}}

\renewcommand{\deg}{\ensuremath{^{\circ}}}

\begin{document}
\definecolor{xdxdff}{rgb}{0.49,0.49,1}
\definecolor{qqqqff}{rgb}{0,0,1}
\definecolor{ffffqq}{rgb}{1,1,0}
\definecolor{uququq}{rgb}{0.25,0.25,0.25}
\definecolor{cqcqcq}{rgb}{0.75,0.75,0.75}
\definecolor{qqqqcc}{rgb}{0,0,0.8}
\definecolor{ffzzqq}{rgb}{1,0.6,0}
\definecolor{darkred}{rgb}{0.5,0,0}

\begin{titlepage}
\centering \textsf{
\linespread{1}\Large Mag.~rer.~nat.~J\"org Weingrill\\[25mm] 
\LARGE\textbf{Extrasolar Planets Orbiting Active Stars}\\[20mm]
\large \textbf{Dissertation}\\[5mm]
zur Erlangung des akademischen Grades eines\\
Doktors\\
an der Naturwissenschaftlichen Fakult\"at der\\
Karl-Franzens-Universit\"at Graz\\[30mm]
Begutachter: o. Univ.-Prof. Dr. Arnold Hanslmeier\\
Institut f\"ur Physik\\
Institutsbereich Geophysik, Astrophysik und Meteorologie\\[40mm]
\today}
\end{titlepage}

\pagenumbering{roman}

\chapter*{Vorwort}

Ich erinnere mich noch gut daran, wie ich als Achtjähriger mit meinem Großvater durch sein Fernglas die Plejaden, den Jupiter und seine Monde beobachtet habe. Der Auslöser für mein Interesse für Astronomie war damals das Buch von \citet{Ruekl1979}, das ich von meinen Eltern geschenkt bekommen hatte. Fast zwanzig Jahre später habe ich den Merkurtransit am 7.~Mai 2003 sowie den Venustransit am 8.~Juni bereits durch mein eigenes Teleskop beobachtet. Diese Ereignisse haben mir das vermittelt, was schon Johannes Kepler mit seinen \textit{Harmonices mundi}, den Weltenharmonien ausgedrückt hat.

Ich danke hiermit meiner Familie Sigrid und Christopher für ihre endlose Geduld auf unseren gemeinsamen Urlauben, die ich am Laptop verbracht habe, sowie Lissi, die mich mit einem guten Kaffee und einer Portion Ruhe bewirtet hat.

I have also to thank the CoRoT Exoplanet science team for the fruitful discussions, especially G\"unther Wuchterl, who encouraged me to focus my interests in the new field of transiting exoplanets and also Tsevi Mazeh for the long discussions about statistics and mathematics of eclipsing binaries and transiting exoplanets.

Mein Dank gilt auch der Dissertantengruppe, die von einer Runde Doktoratsstudenten gegründet wurde indem wir den Rat von \citet{Bolker1998} befolgt haben. Meine Kollegen Hannes Gr\"oller, Martin Leitzinger und Petra Odert waren ein große Hilfe bei der Korrektur der Arbeit, bei Diskussionen über den Fortschritt (und auch Rückschritt) oder einfach durch die gegenseitige Motivation.
Letztendlich danke ich auch dem Institut für Weltraumforschung für die Bereitstellung der Resourcen, die gute Zusammenarbeit und die Förderung in den letzten Jahren.\\

Diese Arbeit wurde mit \LaTeX\ unter Verwendung von Texmaker gesetzt. Das Layout liegt der Dokumentenklasse \textit{scrbook} von Markus Kohm zugrunde. Die Abbildungen wurden mit ITTVIS IDL, GeoGebra und InkScape erzeugt.
\tableofcontents
\newpage \setcounter{page}{0} \pagenumbering{arabic}
\selectlanguage{naustrian}
\chapter*{Zusammenfassung}
\section*{Kontext}
Wöchentlich werden neue Entdeckungen von Planeten publiziert, die mittels Transitmethode gewonnen wurden. Irdische Beobachtungen sowie Weltraummissionen wie CoRoT und Kepler sind dafür verantwortlich, dass wir die statistischen Lücken mit Planeten von entfernten Sternensystemen füllen. Das ultimative Ziel ist die Entdeckung eines habitablen Planeten, vielleicht einer zweiten Erde.

\section*{Ziele}
Ich möchte die stellare Aktivität und ihren Einfluss auf die Entdeckung von extrasolaren Planeten erörtern. Bis heute bilden CoRoT-7b und Kepler-10b die beiden Ausnahmen als kleine Gesteinsplaneten genannt ,,Super-Erden''. Die Frage wirft sich auf, warum unter den über 500~entdeckten und verifizierten Planeten die Anzahl der kleinen Planeten derart gering ist. Unsere Statistik die wir bis heute über die Verteilung der planetaren Massen und Durchmesser gesammelt haben, ist durch die Beobachtung verzerrt. Ein anderer Grund dafür könnte, abgesehen von der Schwierigkeit kleine Planeten zu verifizieren, der hohe Level an stellarer Aktivität sein, der bisher beobachtet wurde.

Stellare Aktivität verläuft auf unterschiedlichen Zeitskalen, von langjährigen Strahlungsänderungen wie dem bekannten Sonnenzyklus, über die stellare Rotation im Bereich von mehreren Tagen bis hin zur Beobachtung von akustischen Moden im Bereich von Minuten. Aber auch aperiodische Vorgänge wie Flares oder das Aktivitätssignal der Granulation kann die Entdeckung eines Planetentransits verhindern.

\section*{Methoden}
Ich möchte unterschiedliche Methoden zur Detektion von transit-ähnlichen Signalen beschreiben, die unterschiedlichen Einflüsse des Sterns, des beobachtenden Instruments und deren Auswirkung auf den Erfolg der Methodik. Unterschiedliche Filter-Methoden werden diskutiert um die Möglichkeit einer Transitdetektion zu vergrößern und um das Transitsignal vom intrinsischen stellaren Signal zu trennen.

Letztendlich werden unterschiedliche mathematische Modelle und Approximationen von Transitfunktionen auf deren Empfindlichkeit auf stellare Aktivität hin untersucht.

\section*{Ergebnisse}
Ein statistischer Überblick über die stellare Aktivität im CoRoT-Datenarchiv wird präsentiert. Der Einfluss stellarer Aktivität auf unterschiedliche Transitplaneten wird bei  CoRoT-2b, CoRoT-4b und CoRoT-6b untersucht.

\section*{Schlussfolgerung}
Stellare Aktivität kann die erfolgreiche Entdeckung eines Planetentransits verhindern, wobei CoRoT-7b sicherlich die Grenze markiert. Zukünftige Missionen wie \textsc{Plato} sind notwendig um langjährige Beobachtungen mit einer Genauigkeit im mmag Bereich zu liefern um die Einschränkungen zu überwinden, die durch die aktiven Sterne in unserer galaktischen Nachbarschaft gegeben sind.

\selectlanguage{british}
\chapter*{Summary}
\section*{Context}
New discoveries of transiting extrasolar planets are reported weekly. Ground based surveys as well as space borne observatories like CoRoT and Kepler are responsible for filling the statistical voids of planets on distant stellar systems. The ultimate goal is the discovery of a habitable planet and maybe a second Earth.

\section*{Aims}
I want to discuss the stellar activity and its impact on the discovery of extrasolar planets. Up to now the discovery of small rocky planets called ``Super-Earths'' like CoRoT-7b and Kepler-10b are the only exceptions. The question arises, why among over 500~detected and verified planets the amount of smaller planets is strikingly low. Our statistics, we collected so far on the distribution of planetary masses and radii, is obviously biased by observational effects. Another explanation besides that the verification of small planets is an intriguing task, is the high level of stellar activity that has been observed.

Stellar activity can be observed at different time-scales from long term irradiance variations similar to the well known solar cycle, over stellar rotation in the regime of days, down to the observations of acoustic modes in the domain of minutes. But also non periodic events like flares or the activity signal of the granulation can prevent the detection of a transiting Earth sized planet.

\section*{Methods}
I will describe different methods to detect transit-like signals in stellar photometric data, the different influences introduced by the star, the observer and their impact on the success. Different filtering techniques will be discussed to improve the transit detection capability and to separate the transit signal from the intrinsic stellar signal.

Finally different mathematical models and approximations of transit signals will be examined on their sensibility of stellar activity.
\section*{Results}
We present a statistical overview of stellar activity in the CoRoT dataset.
The influence of stellar activity will be analysed on different transiting planets: CoRoT-2b, CoRoT-4b und CoRoT-6b.

\section*{Conclusions}
Stellar activity can prevent the successful detection of a transiting planet, where CoRoT-7b marks the borderline. Future missions like Plato will be required to provide long-term observations with mmag precision to overcome the limitations set by active stars in our Galactic neighbourhood.

\chapter{Introduction}
With the invention of the telescope the last unknown planets of our solar system were detected by Herschel in 1781, who discovered Uranus, and Le Verrier (1845), who predicted the position of Neptune based on observations of Uranus. It took again about 150~years for the detection of the first extrasolar planet (or ``exoplanet'' for short) around the main sequence star  51~Pegasi by \citet{Mayor1995}. The first detection of a transiting planet came shortly after by \citet{Charbonneau2000}.

Since then the number of detected exoplanets has been increasing rapidly over the past years. This fact was also supported by dedicated space missions like MOST \citep{Rowe2006}, CoRoT \citep{Baglin2009} and Kepler \citep{Borucki2011} as well as ground based surveys like the HAT-network \citep{Bakos2002} or WASP and Super-WASP \citep{Pollacco2006}. The technological progress has also improved the detection capabilities of radial velocity searches by e.\,g. the HARPS spectrograph \citep{Pepe2004} and SOPHIE \citep[e.\,g.][]{Bouchy2009}. The large observatories VLT and the upcoming E-ELT will show us the next step of direct imaging \citep{Kalas2009}. 

One the long path to direct observation we are still tied to transit- and radial-velocity-measurements of the extrasolar planets to derive the important parameters like mass and radius. Upcoming missions like \textsc{Plato} \citep{Catala2009} will observe with unprecedented precision, revealing even more details, but also dealing with new problems. Even then the stellar signal becomes more and more relevant. The characterization and the understanding of the stellar activity on planet hosting stars will help to explain the planetary evolution and its habitability.

Spectro-polarimetric observations of transits will be required to answer the questions of habitability. The interaction of the planets' atmosphere with the stellar wind of the host star can only be observed in wavelengths of the respective atmospheric species.

By combination of the transit method with the observation in different wavelengths, the transiting planet with its opaque disk ``samples'' the surface of the star. This will reveal the host star's activity in its photosphere, chromosphere and even corona. If we imagine an ultra-high precision photometry of a small transiting body, we can even think of observing the granulation or corona of a distant star.

So far no working definition for an extrasolar planet has been agreed upon. The latest definition of planets dates back to 2006 in the IAU Resolution 5A 2006\footnote{\url{http://www.iau.org/static/archives/releases/pdf/iau0603.pdf}}:
\begin{quotation}
	
\textbf{RESOLUTION 5A}\\
The IAU therefore resolves that planets and other bodies in our Solar System, except satellites, be defined into three distinct categories in the following way:
\begin{enumerate}
\item A \textit{planet} is a celestial body that 
  \begin{enumerate}
  \item is in orbit around the Sun, 
  \item has sufficient mass for its self-gravity to overcome rigid body forces so that it assumes a hydrostatic equilibrium (nearly round) shape, and 
  \item has cleared the neighbourhood around its orbit.
  \end{enumerate}
\item A \textit{dwarf planet} is a celestial body that 
  \begin{enumerate}
  
  \item is in orbit around the Sun, 
  \item has sufficient mass for its self-gravity to overcome rigid body forces so that it assumes a hydrostatic equilibrium (nearly round) shape,     
  \item has not cleared the neighbourhood around its orbit, and 
  \item is not a satellite.
  \item All other objects , except satellites, orbiting the Sun shall be referred to collectively as \textit{Small Solar-System Bodies}.
  \end{enumerate}
\end{enumerate}
\end{quotation}

This resolution does not apply to extrasolar planets and has been refined in the working definition 2006 \citep{Boss2007}: 
\begin{quotation}
[A]n \textit{exoplanet}
\begin{itemize}
\item is an object with a true mass below the limiting mass for thermonuclear fusion of deuterium (currently calculated to be 13~Jupiter masses for objects of solar metallicity),
\item is in orbit around a star or stellar remnant,
\item has a mass and/or size that is superior to the one used as a limit for a planet in our Solar System.
\end{itemize}
\end{quotation}

We see that the limitations regarding the mass and the size are rather weakly defined. Especially the limiting mass for thermonuclear fusion of deuterium is estimated at $\approx13.0\pm0.8\MJup$, but depends on the pre-conditions like metallicity and can vary from $\approx 11.0\MJup$ to $16.3\MJup$\citep{Spiegel2011}. The definition also includes objects orbiting Pulsars, but excludes free floating planets \citep[and references therein]{Han2006}.

\section{Stellar and Solar Activity}
\label{sec:stellarsolaractivity}
Stellar activity in the context of transiting exoplanets has been investigated by \citet{Alapini2009} for the radial velocity method (see Section~\ref{sec:radialvelocity}).

Stellar activity happens on different scales and has many varieties. We are interested in activity signals that have an impact in extrasolar planet detection and have an implication on the evolution and the environment of those planets.

At first we take a look at the solar activity that has been studied for centuries since the detection of the first sunspots. Once again the sun is an ideal template -- and in frame of the work -- for the stars hosting exoplanets. Attempts to deduce stellar activity levels from solar variations have been carried out by e.\,g. \citet{carpano2008}. The main drawback is the fact that the solar activity levels only correspond to a stellar analogue that matches the spectral type and the age of the sun. So far transit search programs only cover a fraction of a solar/stellar cycle. The star HD~49933 \citep{Garcia2010} shows evidence for magnetic activity that might vary over longer time scales.

\subsection{Solar Activity}
\label{sec:solaractivity}
The solar activity has been known since the invention of the telescope 400~years ago. Spots on the sun have been observed even longer. The solar rotation as observed by the virtual motion of the solar spots across the solar surface is a signature of the magnetic activity of our central star. As the occurrence varies over the solar cycle, other non periodic events like flares occur along with it.

\begin{figure}[htb]
	\centering
\begin{tikzpicture}[mindmap,concept color=orange, font=\sf, text=white]
   \tikzstyle{level 1 concept}+=[font=\sf]
   \tikzstyle{level 2 concept}+=[font=\sf \small]
   \tikzstyle{level 3 concept}+=[font=\sf \tiny]
    \node[concept] {solar activity}
    [clockwise from=0]
    child[concept color=red] { node[concept] {oscillations} 
      [clockwise from=90]
      child { node[concept] {p-mode} }
      child { node[concept] {g-mode} }
      child { node[concept] {f-mode} }
      child { node[concept] {160-min cycle} }
     }
    child[concept color=black] { node[concept] {sunspots} }
    child[concept color=green] { node[concept] {solar cycle} }
    child[concept color=blue] { node[concept] {solar irradiance} }
    child[concept color=yellow] { node[concept] {flares} };
\end{tikzpicture}
	\caption{Mindmap of solar activity.}\label{fig:solaractivity}
\end{figure}
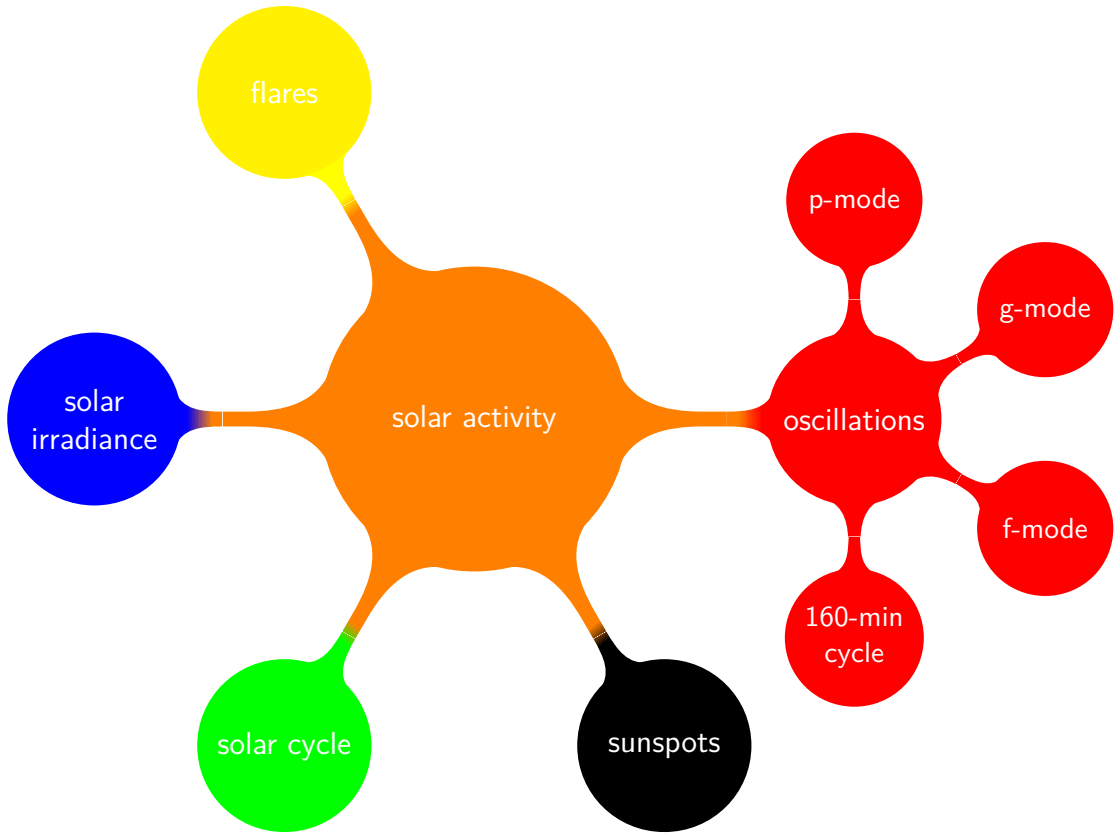

The solar oscillation namely the p-modes were discovered on the sun, revealing a new method to actually probe the interior of our sun. An overview of the solar activity is shown in Figure~\ref{fig:solaractivity}.\\

Virgo/SoHo-data has been used in the past to simulate realistic lightcurves for transit based observations by \citet{Aigrain2003}, where different time-scales $\tau$ of solar activity were found (see Table~\ref{tab:activitysources}).\\

\begin{table}
\centering
\captionabove{Activity sources and their respective timescales $\tau$ taken from \citet{Aigrain2003} and frequencies $\nu$.\label{tab:activitysources}}
\begin{tabular}{lrr}
\hline \hline
activity source & timescale $\tau$ & frequency $\nu$\\
\hline
active regions & $1 \dots 3 \times 10^5$\,s & $3\dots10\,\upmu$Hz\\ 
super-granulation & $3 \dots 7 \times 10^4$\,s & $14\dots33\,\upmu$Hz\\
meso-granulation & 8000\,s & 125\,$\upmu$Hz\\
granulation & 200 \dots 500\,s & $2\dots5$\,mHz\\
bright points & 70\,s& 14\,mHz\\
\end{tabular}
\end{table}

The power-spectrum $P(\nu)$ of the solar flux can be described by power-laws for non-periodic 
\begin{equation}
P(\nu) = \frac{A}{1+(2\pi\nu\tau)^b}
\end{equation}
and periodic components
\begin{equation}
P(\nu) = A\left(\frac{\nu}{\nu_0}\right)^c \left(\frac{\Gamma^2}{(\nu-\nu_0)^2+\Gamma^2}\right)^b,\label{eqn:harvey}
\end{equation}
where $A$ is the amplitude, $b$ is the decay time and $c$ represents the frequency trend for a central frequency $\nu_0$ \citep{Harvey1993}.\\

The known solar activity signals can be separated in two large groups,  periodic and non-periodic or eruptive solar activity.

\subsubsection{p-mode Oscillations}
A small scale periodic variability are the 5-minutes oscillations, which were later called solar p-modes and were firstly observed in wide band photometry with the ACRIM instrument \citep{Woodard1983}. The explanation for the observation is the change in temperature of the photosphere \citep{Renaud1999}. The discovery of  these acoustic modes goes back to 1960 \citep{Stix2002}, where periodic changes were visible in ''Doppler plates''. Their maximum amplitude is in the regime of $2\dots5$~mHz which roughly corresponds to 3\dots8~minutes.

\subsubsection{Solar Irradiance}
\begin{figure}[htb]
\centering
  \includegraphics[width=\textwidth]{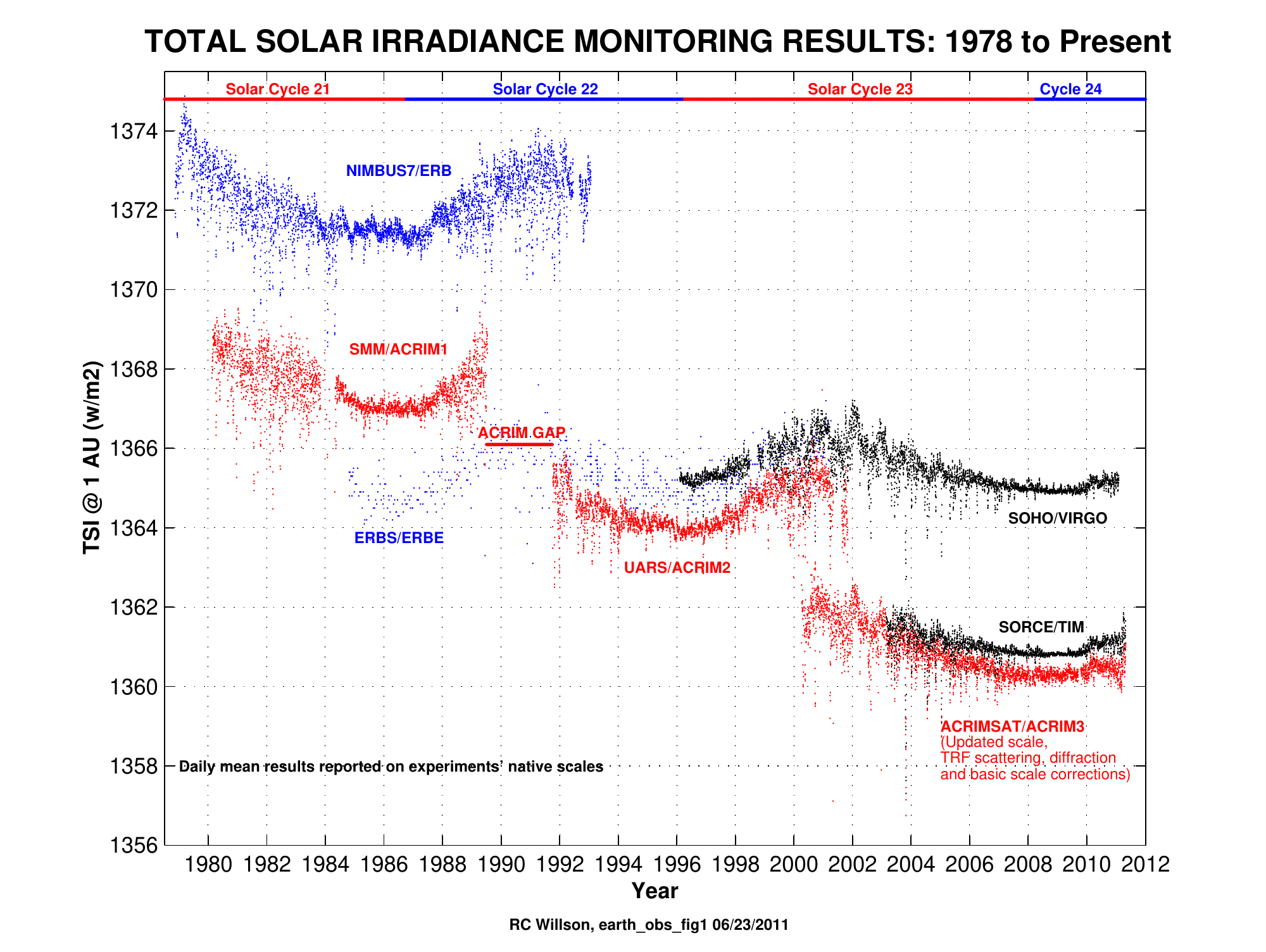}
  \caption{Total Solar Irradiance measured with different instruments. The variation of the activity levels during the solar cycle is clearly visible. The offsets between the different missions arise from unknown instrumental origin.}\label{fig:totalsolarirradiance}
\end{figure}

The Total Solar Irradiance (TSI) has been measured with a high precision, 10~ppm according to \citet{Willson1999} over two cycles. We need continuous space based observations to compare the solar activity with the stellar activity on the corresponding time scales. The variation with the solar cycles can be seen in Figure~\ref{fig:totalsolarirradiance}.\footnote{Data taken from \url{http://www.acrim.com/}}\\

TSI data is taken from the SORCE instrument\footnote{\url{http://lasp.colorado.edu/sorce/data/tsi_data.htm}}. The total irradiance is modulated by the solar rotation, being visible due to the active regions virtually moving across the solar disk and the magnetic cycle, that triggers the appearance of solar activity.

\subsubsection{Solar Rotation and Sunspots}
As mentioned in \citet{Willson1999} the spectral power at 27~days is evident, which corresponds to the solar rotation period. The inclination angle $i$ was initially defined by \citet{Carrington1863} [sic] and measured at Kanzelh\"ohe, Austria $i=7.137^\circ\pm0.017^\circ$ \citep{Stix2002}.

\subsubsection{Solar Flares and Active Regions}
Solar flares also contribute to the TSI. Only the most energetic flares are visible in the continuum, otherwise the most energy in the visible spectrum is emitted in H$\alpha$. The most prominent example of a white light flare was the Haloween event on 28 Oct. 2003 \citep{Kopp2004,Woods2005}.

\subsection{Stellar Activity}
Taking the sun as a template for other stars hosting planets, we might expect to observe the same phenomena as on the sun we described in Section~\ref{sec:solaractivity}. The main difference is the big variety, that we have to expect in the observation of numerous targets. Planet hosting stars will have different spectral types than the sun, but also different evolutionary stages and by that different levels of activity \citep{Guinan2003}. The size of the star as listed in Table~\ref{tab:cox2000}, has a big influence on the detected transit depth, which will be discussed in Section~\ref{sec:transitphysics}.
The inclination axis of the host star plays an important role, but especially for transit observations, where the observed inclination of the system is close to $90\deg$, the expected inclination of the host star's rotation axis can be assumed to have a similar value.

Stellar activity is the limiting factor in the detection and verification of planetary candidates \citep{Dumusque2011}.
If we want to derive stellar activity levels from the sun we need to scale at least with the stellar radii \citep{Aigrain2004}.
\begin{table}[ht]
\centering
\captionabove{Spectral types of dwarf stars taken from \citet{Cox2000} with their respective masses, radii and effective temperatures.\label{tab:cox2000}}
\begin{tabular}{lrrr}
\hline \hline
Spectral Type & $M_\ast$ & $R_\ast$ & \Teff \\ \hline
O5V &   60  & 12 & 42\,000\\
B0V & 17.5  & 7.4 & 30\,000\\
B5V &  5.9  & 3.9 & 15\,200\\
B8V &  $3.8$  & 3.0 & 11\,400\\
A0V &  2.9  & 2.4 & 9\,790\\
A5V &  2.0  & 1.7 & 8\,180\\
F0V &  1.6  & 1.5 & 7\,300\\
F5V &  1.4  & 1.3 & 6\,650\\
G0V &  1.05 & 1.1 & 5\,940\\
G2V &  1.0  & 1.0 & 5\,790\\
G5V &  0.92 & 0.92 & 5\,560\\
K0V &  0.79 & 0.85 & 5\,150\\
K5V &  0.67 & 0.72 & 4\,410\\
M0V &  0.51 & 0.60 & 3\,840\\
M2V &  0.40 & 0.50 & 3\,520\\
M5V &  0.21 & 0.27 & 3\,170\\
\end{tabular}
\end{table}

Stellar activity can be characterized in the power spectrum of the light curve as a `generalized' lorentzian profile
\begin{equation}
P(\nu) = \sum_{i=1}^N\frac{A_i}{1+\left(\frac{\nu}{\Delta \nu_i}\right)^{\alpha_i}}+B, \label{eqn:lorentzian}
\end{equation}
where $P(\nu)$ is the spectral power, $\alpha_i = 2$ if a temporal phenomenon is observed, $A$ is the amplitude and $1/\Delta \nu_i$ is the characteristic timescale \citep[and references therein]{Hulot2011}. Equation \ref{eqn:lorentzian} is closely related to Equation~\ref{eqn:harvey}.

\subsubsection{Oscillations}
Solar like oscillations in stars are difficult to detect photometrically, they have been at least proven by the radial velocity measurements \citep{Teixeira2008, Eggenberger2004}. The detection of p-modes on Procyon is controversial \citep{Matthews2004,Regulo2005}. Stellar oscillations have been measured by different space-based missions e.\,g. CoRoT \citep{Michel2008} and Kepler \citep{Gilliland2010}.

Even observations of p-modes have been reported on planet hosting stars \citep{Ballot2011,Gilliland2010}. Low-degree p-modes were successfully observed by CoRoT on the star HD~49933, which is of F5V type for 60~days\citep{Appourchaux2008}. The maximum amplitude of $3.75\pm0.23$\,ppm was found at $1760\,\upmu$Hz. 

As reported by \citet{Baudin2011} the average amplitude of the oscillations is around 75\,ppm with maximum amplitudes up to several 100\,ppm. The maximum amplitude is correlated with $\nu_\mathrm{max}$, where the lowest frequencies about 20\,$\upmu$Hz corresponds to amplitudes of 150\,ppm. The highest frequency at $\nu_\mathrm{max}\approx60$\,$\upmu$Hz has the lowest amplitude of 40\,ppm.

The F5\,V star HD~49933 shows solar like oscillations that have been measured by CoRoT \citep{Appourchaux2008}. Also the Kepler mission was able to observe acoustic modes \citep{Mathur2011}. The amplitudes can reach 200~ppm in a range of 20 to 60\,$\upmu$Hz\citep{Baudin2011}.

\subsubsection{Stellar Rotation}
Stellar rotation is similar to the solar rotation. It can be identified due to the movement of active areas across the stellar disk modulating the total flux that can be measured. But different inclination angles have to be taken into account. The inclination angles are uniformly distributed, so we also have to expect top-on observations.
\begin{figure}[htb]
\centering
  \includegraphics[width=0.75\textwidth]{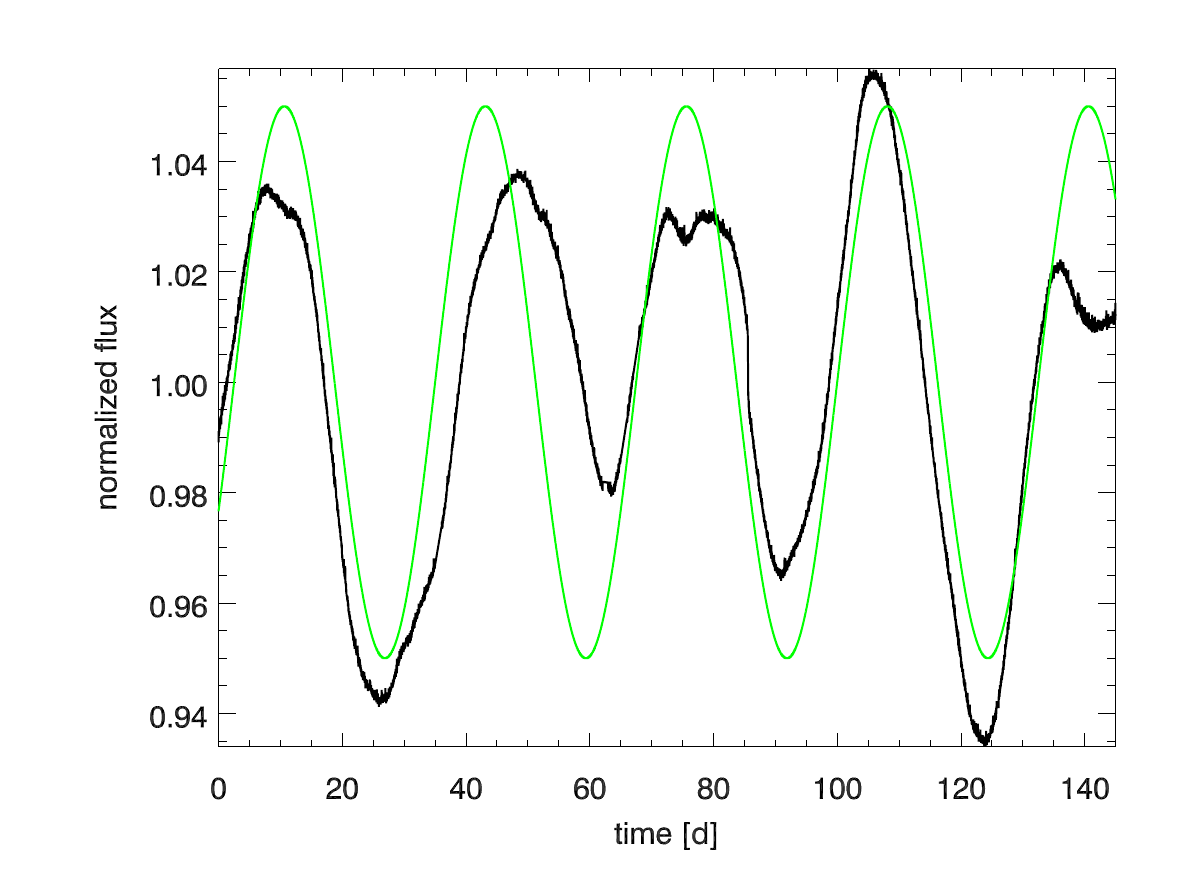}
  \caption{Stellar rotation as a sinusodial modulation of the flux. The highest amplitude is shown in green.}\label{fig:stellarrotation}
\end{figure}

\subsubsection{Flare Stars}
About 1101 active dwarf stars of spectral type F, G and K are known at the time being (July 2011)\footnote{\url{http://simbad.u-strasbg.fr/simbad/sim-sam}}. Among them only 57~flare stars are known.
For example the star \textit{V*~V1041 Tau} belongs to the same spectral class G2V as the sun and is a member of the Pleiades field. According to \cite{Gorlova2006} this star has a spectral type of G9. Nevertheless this object seems to be an interesting target. The star's properties are listed in table \ref{tab:V1041Tau}.

\begin{table}[ht]
\captionabove{Properties of V*~V1041 Tau as an example for a solar like flare star.\label{tab:V1041Tau}}
\centering
\begin{tabular}{lr} \hline
Property & Value\\ \hline
Identifier & V*~V1014 Tau\\
	& HII~0738\\
	& 2MASS J03453940+2345154\\
Spectral Type & G2V (G9)\\
B &	13.42\\
V & 12.26\\
J & 9.767\\
H & 9.224\\
K & 9.007\\
$(V-K)_0$ & 2.30\\ 
\hline
\end{tabular}
\end{table}

So far only little evident is known for flare activity at transiting planets. One of the few exceptions is a flare that has been observed at OGLE-TR10b by \citet{Bentley2009}.

\subsubsection{Blend Scenarios}
Since binary stars are very common throughout the galactic neighbourhood, we must expect them in transit searches. More or less they are closely related to transiting exoplanet systems, where only the secondary body is of smaller size. Brown dwarfs, however fall into this regime. The binarity is no intrinsic stellar activity, but definitely obfuscating the transit detection.

\begin{figure}[htb]
	\centering
		\includegraphics[width=0.9\textwidth]{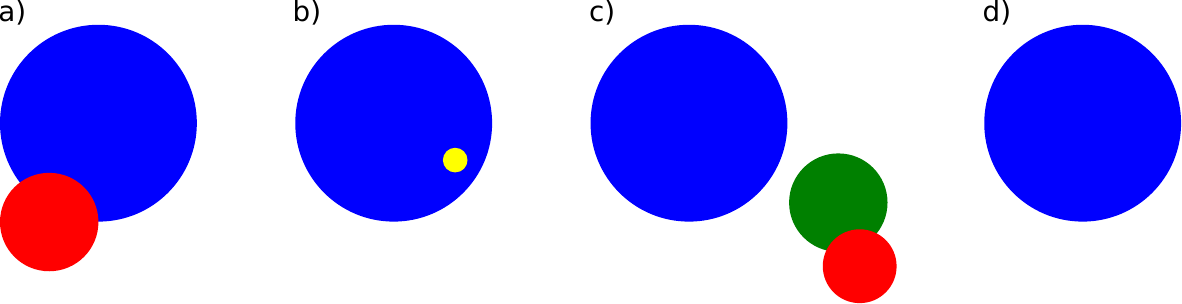}
	\caption{Blend scenarios adopted from \citep{Moutou2006}. a) Grazing binary; b) small stellar companion; c) eclipsing binary in a triple system; false positive caused by stellar activity or instrumental noise.}\label{fig:Blend}
\end{figure}

There are many different cases that can mimic a transiting planet. Most of them are caused by binary stars \citep[e.g.][]{Moutou2006}. These cases are illustrated in Figure~\ref{fig:Blend}.
The triple system does not have to be physically connected. Usually a background binary contaminates the PSF. These systems can be identified with high resolution imaging or spectroscopy.

\section{Detection of Exoplanets}
\label{sec:detectionmethods}
\begin{figure}[htb]
  \centering
  \includegraphics{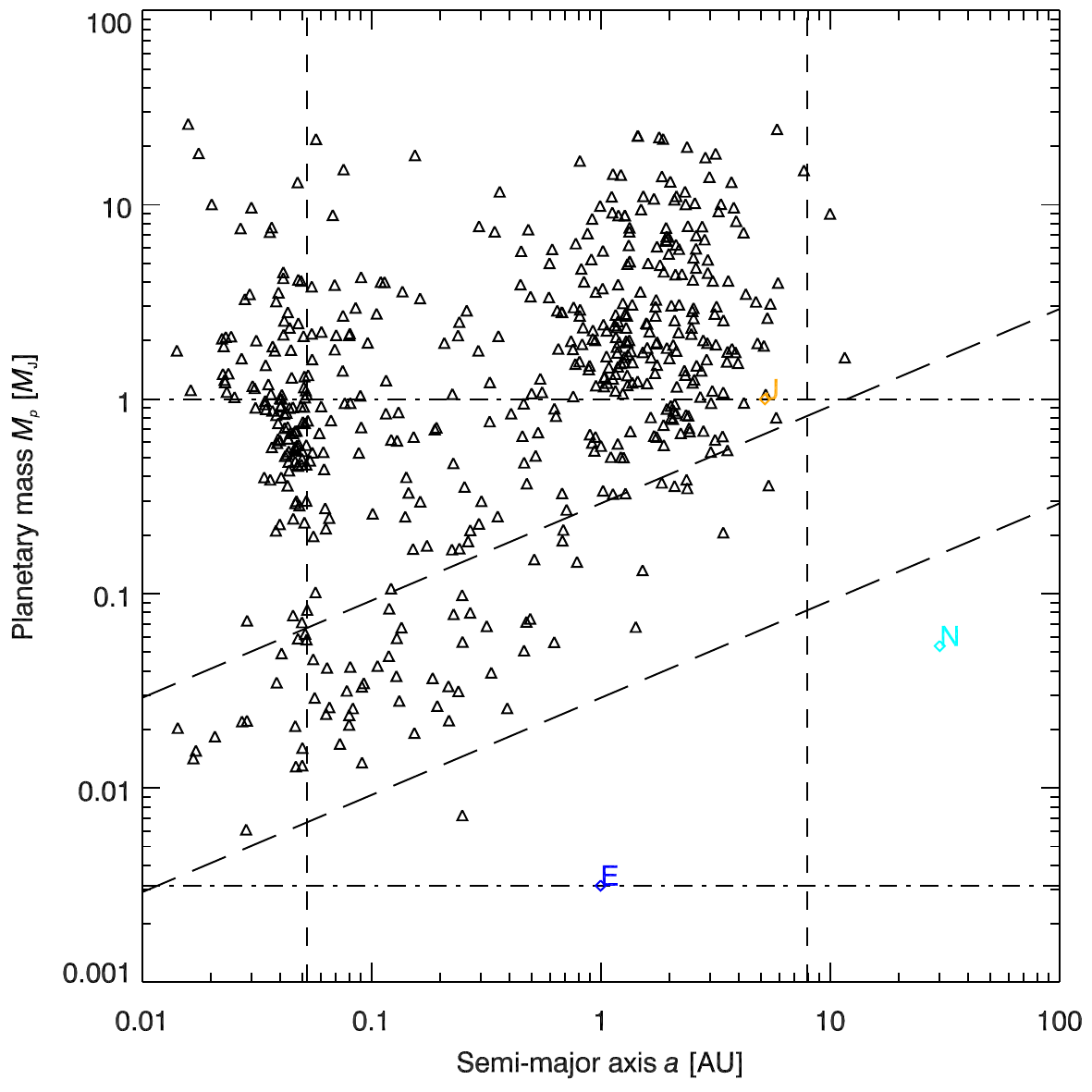}
  \caption{Semi-major axis in Astronomical Units ($1 \mathrm{AU} = 1.49598\times 10^{11} \mathrm{m}$) versus planetary mass in Jupiter masses ($M_\mathrm{Jup} =1.8986\times 10^{27} \mathrm{kg} $. The horizontal dashed-dotted lines represent the transit depth for an Earth- and Jupiter-sized planet, the dashed lines show the detection limits for radial velocity at $1 \mathrm{m/s}$ and $10 \mathrm{m/s}$ respectively. The vertical lines mark the maximum observable semi-major axis CoRoT (150~days) and Kepler (12 years lifetime assumed).  The empty domain between $a = 0.01$ and $a = 0.04$ is called \textit{the Neptune void}.
  }\label{fig:statistics}
\end{figure}

Many different ways of detecting extrasolar planets have been developed through the years, among them the radial-velocity or Doppler method that initially came from the investigation of close binaries. The transit method that evolved on the ground with several surveys finally went to space leading to successful mission like CoRoT and Kepler. A current distribution of confirmed exoplanets\footnote{Data taken from \url{http://exoplanet.eu} as of July 2011.} can be seen in Figure~\ref{fig:statistics} \citep{Schneider2011}.

There are various methods to detect an extrasolar planet. A detailed overview of all the methods can be found e.\,g. in \citet{Doyle2008}. 
\begin{figure}[htb]
\centering
\includegraphics{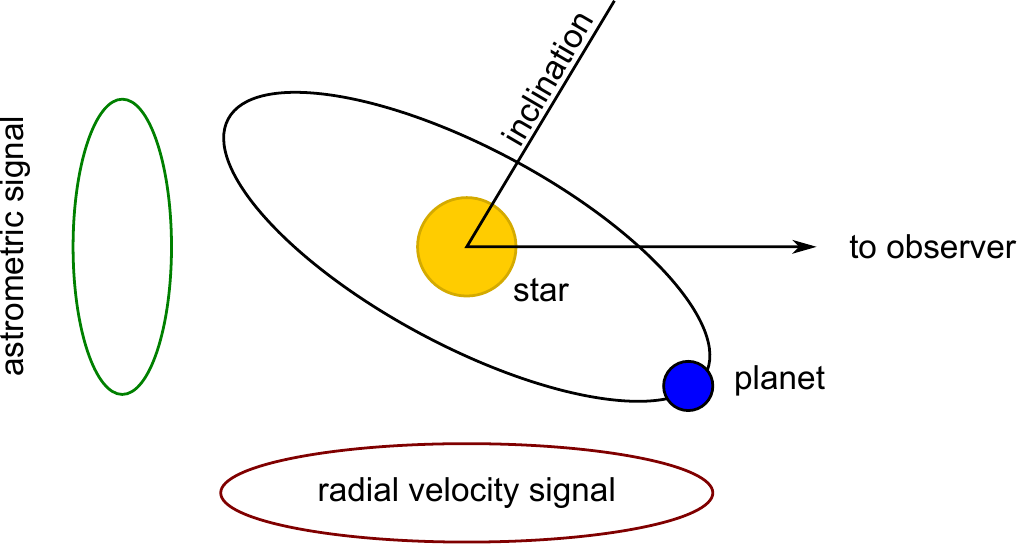}
\caption{The Doppler and astrometric measurements are complementary, depending on the inclination angle as seen by the observer (adapted from \citet{Brown2008}). A transit can only be observed in a small cone with an inclination $i\approx90$.}
\end{figure}

\subsection{Direct imaging}
The easiest way to detect an exoplanet would be the direct observation. This effective method has to deal with the problem that the light of the star is several magnitudes brighter than the planet. The brightness relation depends on two facts:\\
\paragraph{The albedo} of the planet. Since planets are not emitting light by themselves, except thermal emissions \citep{Charbonneau2005}, the only light source is the host star. A high albedo is required to measure reflected light from the planet. We may separate two classes of planets: rocky planets and Jupiter like planets. Taking Mercury as a template for a rocky planet one might expect a geometric albedo of 0.142 \citep{Mallama2002} and 0.52 for a Jupiter like planet\footnote{\url{http://nssdc.gsfc.nasa.gov/planetary/factsheet/jupiterfact.html}} . The albedo of an extrasolar planet can be directly measured with the observation of the secondary transit a.\,k.\,a. occultation of the planet by its host star \citep{Alonso2009}. We must be aware that our solar system is a rather bad template, and we might observe higher \citep[e.\,g.][]{Demory2011} or lower albedos \citep[e.\,g.][]{Rowe2008}.\\

\paragraph{The semi-major axis} of the planet. The albedo of the planet has to be multiplied with the semi-major axis $a$, because the measured intensity of the planet varies with its distance from the host star.\\

\begin{figure}[htb]
\centering
\includegraphics[width=\textwidth]{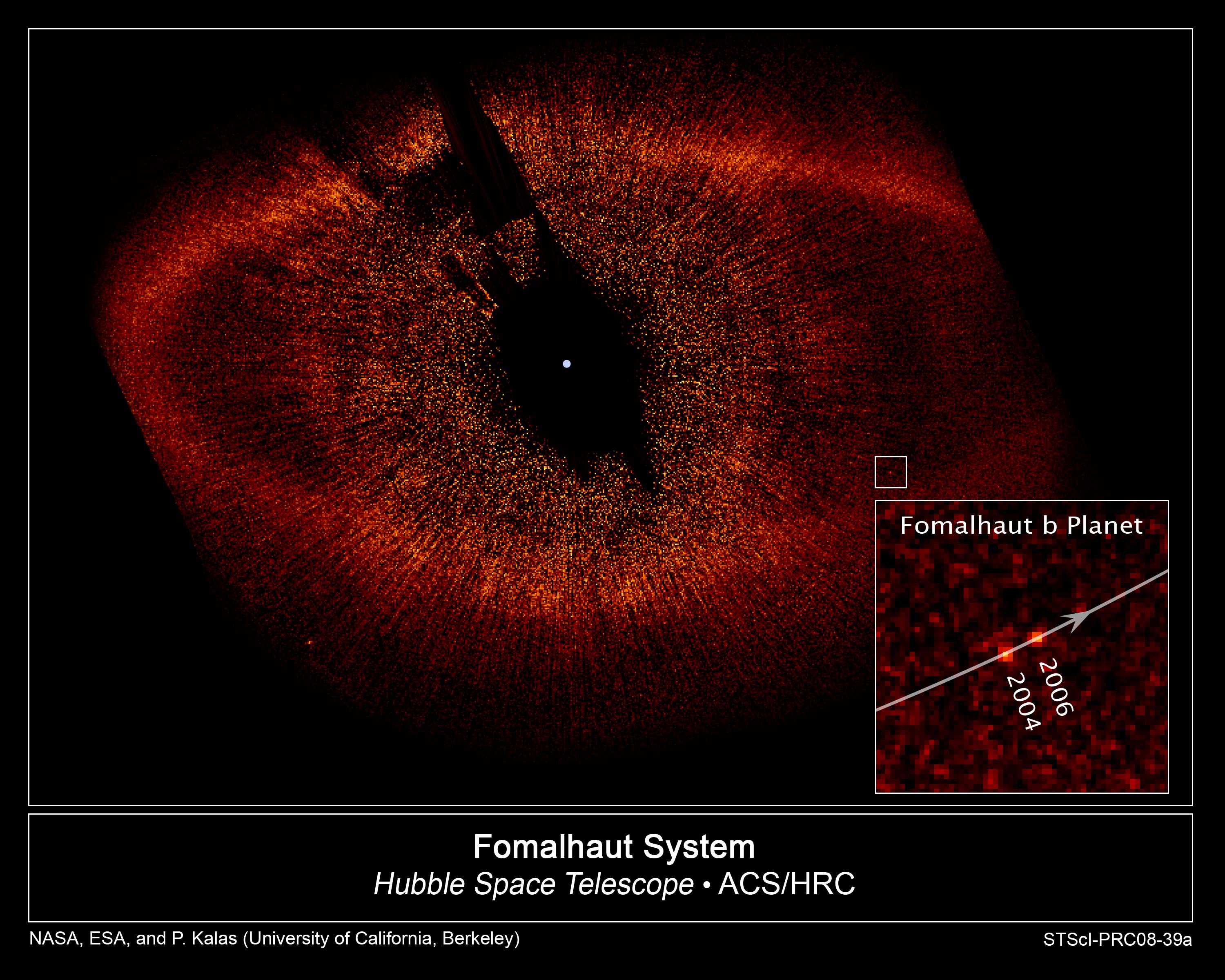}
\caption{Image of Formalhaut~b. Credits: \href{http://www.nasa.gov}{NASA}, \href{http://www.esa.int}{ESA}, P.~Kalas, J.~Graham, E.~Chiang, E.~Kite (University of California, Berkeley), M.~Clampin (NASA Goddard Space Flight Center), M.~Fitzgerald (Lawrence Livermore National Laboratory), and K.~Stapelfeldt and J.~Krist (NASA Jet Propulsion Laboratory)\label{fig:formalhaut}}
\end{figure}

The direct observation can be improved by using a coronal mask, where the light of the host star is reduced, or using a nulling interferometer, where the light of the host star is cancelled out or choosing an appropriate wavelength, like the far infrared, where the planet is relatively bright. This has been successfully accomplished with the Hubble Space Telescope on Fomalhaut by \citet{Kalas2009}, see Figure~\ref{fig:formalhaut}.

The biggest problem is the necessity for high resolution, and therefore for large apertures of interferometers. The next generation of observatories like the E-ELT or interferometric missions for the discovery of extrasolar planets are on the right way to overcome those limitations.

\subsection{The radial-velocity method}\label{sec:radialvelocity}
Since the discovery of 51~Pegasi by \citet{Mayor1995}, the discovery by radial-velocity (RV) has brought the majority of extrasolar planets. The mass of the extrasolar planet attracts the host star throughout the circular orbit and implies a sinusodial signal on the spectral lines of the host star. These lines are either blue- or red-shifted.

The velocity signal $V$ of the host star is made up by the host star's relative motion $\vek{V}_0$ in respect to the observer and the movement of its barycentre $\vek{v}_\ast$. It is given by
\begin{equation}\label{eq:radialvelocity}
V(t) = V_{0,z} + \frac{2\pi a M_\mathrm{p}\sin i}{(M_\mathrm{p} + M_\ast)P\sqrt{1-e^2}} (\cos(\theta(t) + \omega_\mathrm{OP}) + e\cos\omega_\mathrm{OP}),
\end{equation}
where $a$ is the semi-major axis, $M_P$ and $M_\ast$ are the masses of the planet and the star, $i$ is the inclination of the orbital plane, $P$ is the orbital period, $e$ is the orbital eccentricity, $\theta$ is the true anomaly and $\omega_\mathrm{OP}$ is the angle between the pericenter and the orbital plane. The $z$-axis  represents the vector from the observer to the star, wheras the $x$- and $y$-axes span the sky-plane. The expected amplitude
\begin{equation}
A_\mathrm{RV}=\frac{2\pi a M_\mathrm{p}\sin i}{(M_\mathrm{p} + M_\ast)P\sqrt{1-e^2}}
\end{equation}
can be calculated from Equation~(\ref{eq:radialvelocity}).

Up to date all discoveries are verified by radial-velocity. Especially in combination with the transit method (see Section~\ref{sec:Transitdetection}), the mass of the planet can be fixed. Having no other constraints on the planet, the radial-velocity suffers from the uncertainty of the inclination $i$ implying an additional factor of $\sin i$ on the mass of the planet. This means an amplification of $1.58$ on average.

Stellar activity puts a limit on the detection of exoplanets \citep{Boisse2011}. Low mass planets can not be detected due to stellar RV jitter. For example the verification of Corot-7b took more than 80~nights on HARPS. The weak mass-determination of CoRoT-7b is caused by the very active host star \citep{Hatzes2011}.

\subsection{Spectroscopic measurement of atmospheric features}
Another spectroscopic method similar to direct imaging would be the measurement of atmospheric species or molecules on the surface of the planet. Again, large apertures are required, to observe directly e.g. water or even chlorophyll on an illuminated planet.

This topic is also related to transit spectroscopy, where the absorption of hydrogen, sodium or water molecules can be detected. This kind of observation, additionally provides information on the evolution of the planet and especially its atmosphere \citep{Lammer2011}.

\subsection{Astrometric method}
\begin{figure}[htb]
	\centering
		\includegraphics[width=0.5\textwidth]{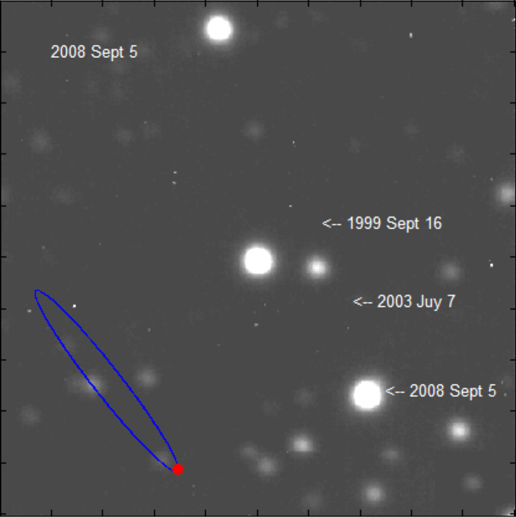}
		\caption{The planet VB10b induces a wobble of 6~mas on its central star.}\label{fig:vb10}
\end{figure}

Since the star and the planet orbit a common center of mass, a perturbation in the stars motion can be detected with astrometric methods. The planet must be sufficiently large and the inclination close to zero to detect a variation in the stellar motion. This obeys the limit of detecting Jupiter class planets in very tight orbits. The astrometric method has been probed by \citet{Lazorenko2011} on VB~10 as seen in Figure~\ref{fig:vb10}\footnote{\url{http://www.nasa.gov/topics/universe/features/C-VB10b-20090528.html}}.

The astrometric wobble $w$, where
\begin{equation}
w = \frac{M_\mathrm{P}a_\mathrm{P}}{M_\ast d},
\end{equation}
and $M_\mathrm{P}$ is the mass of the planet, $M_\ast$ is the mass of the star observed at a distance $d$.

\subsection{Pulsar-timing}
Pulsars are known to emit periodic radio-pulses. This effect is caused by a strong magnetic field which poles are not in alignment with the rotation axis of the stellar remnant. Electromagnetic radiation is emitted, and the beam emerges into space. In rare occasions the pulsar is aligned in a way, that the beam hits the Earth.

A massive body orbiting a pulsar causes variations of these radio emissions. Since habitable planets are not assumed around pulsars, they are no longer included in the exoplanet search. The first detection of this kind was a Jupiter-like planet orbiting the binary system PSR~B1620-26 detected by \citet{Backer1993}.

\subsection{Detection due to Microlensing}
As a planetary system passes the line of sight to a distant star, the passing objects act as a gravitational lens according to their masses. The effect of the lens is, that the light is focused and hence intensified to the observer. These events occur only once per system but the duration of microlensing events is rather long. Due to the nature of the event, only Jupiter mass planets at several AU can be detected, which are definitely out of range for a confirmation by radial-velocity.

\subsection{Detection of Transiting Exoplanets}\label{sec:Transitdetection}
We assume a planet that is orbiting a star. The inclination of the orbit is in the line of sight, so that the projected disk of the planet covers the star. This transit can be detected as a change in the light curve. The transits of Mercury or Venus, that can be observed from Earth in the solar system demonstrate this principle. To this day we are far from observing transits from Earth-sized planets. 

Observing a solar eclipse is a special case of a transit, that is caused by the finite distance of the observer to the transiting object. As for stars we can always assume an infinite distance, that has no impact on the projected size of the planet on the host star's disk. We observe $D\gg a$, where the distance $D$ is in the order of light-years and the semi-major axis $a$ does not exceed several hundred AUs. A general exception from this rule are free floating planets, which are not discussed in the frame of this work.

\begin{figure}[htb]
  \centering
  \includegraphics[width=\textwidth]{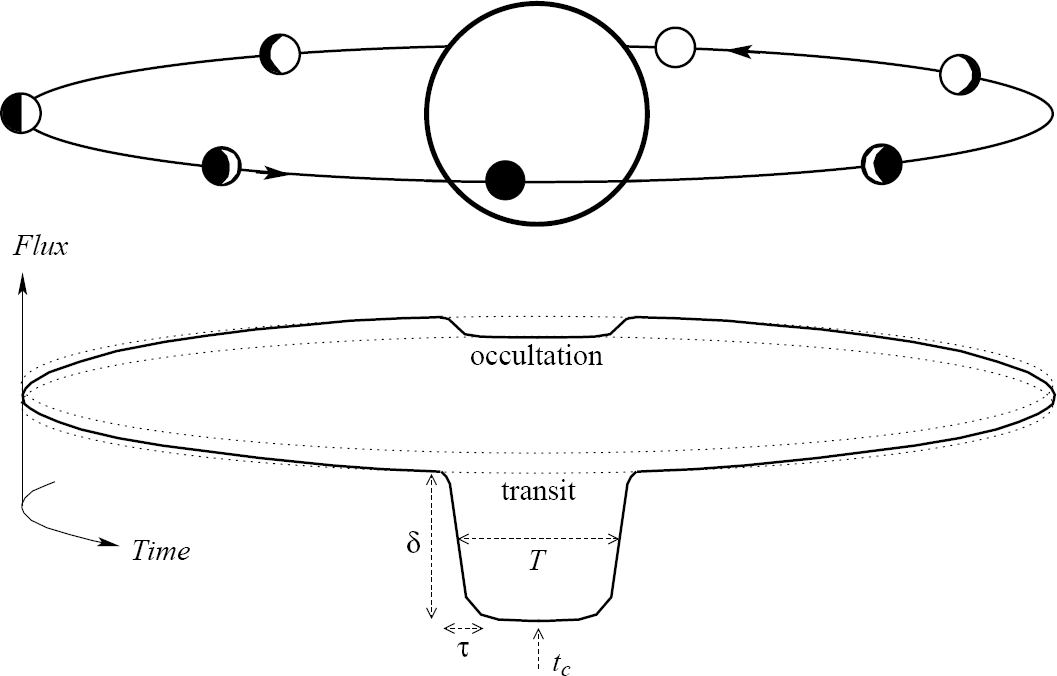}
  \caption{A planet transiting its host star, with the (primary) transit and the secondary transit (occulation). Figure taken from \citet{Winn2010}.} 
\end{figure}

\section{The Physics of Transits}
\label{sec:transitphysics}
A detailed discussion on the transit parameters can be found e.g. in \citet{Seager2003} or \citet{Winn2010}. I want to follow some considerations here:

\begin{figure}[htb]
\centering
\begin{tikzpicture}[line cap=round,line join=round,>=triangle 45,x=1.4cm,y=1.4cm]
\clip(-2.78,-3.57) rectangle (2.88,2.16);
\draw [color=ffffqq,fill=ffffqq,fill opacity=0.5] (0,0) circle (2.8cm);
\draw [fill=black,fill opacity=0.5] (-1.5,1) circle (0.27cm);
\draw [fill=black,fill opacity=0.5] (-1.96,1) circle (0.28cm);
\draw [fill=black,fill opacity=0.5] (1.5,1) circle (0.27cm);
\draw [fill=black,fill opacity=0.5] (1.96,1) circle (0.28cm);
\draw [line width=1.2pt] (-1.96,-2.5)-- (-1.51,-3.01);
\draw [line width=1.2pt] (-1.51,-3.01)-- (1.5,-3.01);
\draw [line width=1.2pt] (1.5,-3.01)-- (1.96,-2.5);
\draw [dash pattern=on 4pt off 4pt] (-1.96,1)-- (-1.96,-2.5);
\draw [dash pattern=on 4pt off 4pt] (-1.5,1)-- (-1.51,-3.01);
\draw [dash pattern=on 4pt off 4pt] (1.5,1)-- (1.5,-3.01);
\draw [dash pattern=on 4pt off 4pt] (1.96,1)-- (1.96,-2.5);
\draw (0,0)-- (-1.18,-1.62);
\draw (0,0)-- (0,1);
\draw [line width=1.2pt] (-2.61,-2.5)-- (-1.96,-2.5);
\draw [line width=1.2pt] (1.96,-2.5)-- (2.61,-2.5);
\draw [dotted] (-2.73,1)-- (2.88,0.99);
\draw [dotted] (-1.96,-2.5)-- (1.96,-2.5);
\draw[color=ffffqq] (-0.97,1.57) node {c};
\fill [color=uququq] (-1.96,-2.5) circle (1.5pt);
\draw[color=uququq] (-1.89,-2.72) node {$t_1$};
\fill [color=uququq] (-1.51,-3.01) circle (1.5pt);
\draw[color=uququq] (-1.45,-3.18) node {$t_2$};
\fill [color=uququq] (1.5,-3.01) circle (1.5pt);
\draw[color=uququq] (1.58,-3.21) node {$t_3$};
\draw[color=black] (0.07,-3.18) node {$t_F$};
\fill [color=uququq] (1.96,-2.5) circle (1.5pt);
\draw[color=uququq] (2.04,-2.7) node {$t_4$};
\draw[color=black] (-0.62,-0.62) node {$R_\ast$};
\draw[color=black] (0.32,0.59) node {$bR_\ast$};
\draw[color=black] (0.08,-2.68) node {$t_T$};
\end{tikzpicture}
\caption{Sketch of a transiting planet shown in gray in front of a star in yellow. $R_\ast$ is the radius of the star and $b$ is the impactfactor. The total transit time $t_T$ is the time between $t_1$ and $t_4$. The ``flat part'' $t_F$ is the time between $t_2$ and $t_3$.\label{fig:transitsketch}}
\end{figure}
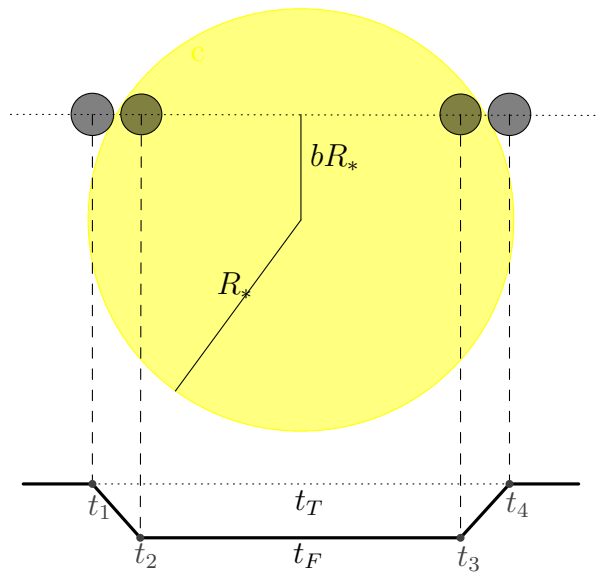

The depth of a transit $\Delta F$ is the relative change in flux
\begin{equation}
\Delta F\approx\left(\frac{R_\mathrm{p}}{R_\ast}\right)^2\label{flux}
\end{equation}
depending on the radius of the planet $R_\mathrm{p}$ and its host star $R_\ast$ as shown in Figure \ref{fig:transitsketch}. This is the simple geometric relation assuming an opaque planetary disk and uniform illuminating stellar disk. The planet itself does not emit any radiation either. Even in the far infrared the light emitted from a planet as thermal emission is several magnitudes fainter than the host star.

The impact parameter $b$ is defined as
\begin{equation}
b \equiv \frac{a}{R\ast}\cos i = \sqrt{ \frac{(1-\sqrt{\Delta F})^2-(t_F/t_T)^2(1+\sqrt{\Delta F})}{1-(t_F/t_T)^2} }
\end{equation}
depends on the semi-major axis $a$ and the radius of the star $R_\ast$ and the inclination $i$. In other words, the impact parameter is the distance between the transit path across the stellar disk and the center of the star as shown in Figure~\ref{fig:transitsketch}. It can be calculated from the measured lightcurve using the change in flux $\Delta F$, the total transit duration $t_T$ and the duration of the flat part of the transit $t_F$. High impact factors or inclinations respectively lead to a grazing transit. This type of transit is V-shaped and has to be distinguished from other blend scenarios as shown in Figure~\ref{fig:Blend}.

The transit duration $t_T$ with an impact factor $b=0$ is
\begin{equation}
t_T = \frac P \pi \sin^{-1}\left(\frac{R_\ast}{a}\right).\label{eqn:transitduration}
\end{equation}
By introducing
\begin{equation}
l = \sqrt{(R_\ast + R_\mathrm{p})^2 - a^2\cos^2i},
\end{equation}
we get
\begin{equation}
t_T = \frac{P}{\pi}\sin^{-1}\left( \frac{l}{a}\right).
\end{equation}
The transit duration can be approximated under the assumptions $ R_\mathrm{p} \ll R_\ast \ll a$ with
\begin{equation}
  t_T \approx 2 R_\ast\sqrt{\frac a {GM_\ast}},
\end{equation}
where $G = 6.67300 \times 10^{-11} \mathrm{m}^3 \mathrm{kg}^{-1} \mathrm{s}^{-2}$ is the gravitational constant.\\

The semi-major axis $a$ depends on the measured period $P$ using Kepler's third law
\begin{equation}
a^3=\frac{P^ 2G(M_\ast + M_\mathrm{p})}{4\pi^2},\label{eqn:kepler}
\end{equation}
and can be simplified under the assumption $M_\mathrm{p} \ll M_\ast$. Other parameters can be directly derived from the lightcurve
\begin{align}
\frac{a}{R_\ast} &= \frac{2P}{\pi}\frac{\Delta F^{1/4}}{\sqrt{t^2_T-t^2_F}},\\
\rho_\ast &= \frac{32P}{G\pi}\frac{\Delta F^{3/4}}{\sqrt[3]{t^2_T-t^2_F}}.
\end{align}
A more detailed derivation of these parameters and can be found in \citet{Perryman2011} and references therein. 

Additionally by the stellar mass-radius relation
\begin{equation}
R_\ast = kM^x_\ast,
\end{equation}
and by taking the values $k = 1$ and $x \approx 0.8$ from \citet{Cox2000} for main sequence stars we obtain
\begin{equation}
\frac{R_\mathrm{p}}{R_\sun} = \frac{R_\ast}{R_\sun}\sqrt{\Delta F} = \left(k^{1/x}\frac{\rho_\ast}{\rho_\sun}\right)^{x/1-3x}\sqrt{\Delta F}.
\end{equation}

A less physical parameter is the probability of detecting a transit
\begin{equation}
  P=0.0045\frac{1\AU}{a}\times\frac{R_\mathrm{p}+R_\ast}{R_\sun}\times\frac{1+e\cos(\omega/2)}{1-e^2}.
\end{equation}

It is assumed that the planets' orbits are distributed normally. Current calculations predict one planet for each star on the average or at least one planet among four solar like stars \cite{borde2003}.

To simulate a transit numerically the following parameters have to be taken into account:
\begin{itemize}
\item{the radius of the star}
\item{the radius of the planet}
\item{the distance between the star and the planet}
\item{the distance between the observer and the star}
\item{the center to limb variation of the star's atmosphere}
\item{inclination of the planet's orbit}
\end{itemize}

Since not all parameters can be observed or measured respectively, we have to reduce them to something useful:
\begin{itemize}
\item{fractional radius of the planet}
\item{the impact factor of the planet}
\item{the center to limb variation of the star's atmosphere}
\end{itemize}
There are several levels of accuracy to simulate a transit of an exoplanet. The intensity $I$ of the star can be normalized to $1.0$.

\subsection{Simple Transit Modelling}
The most simple approach separates the transit of a planet with radius $R_p$ across a star with radius $R_\ast$ into two distinct cases:
\begin{eqnarray}
I =\begin{cases} 2\pi R_\ast^2&\text{for $d>R_\ast$}\\
2\pi R_\ast^2-2\pi R_\mathrm{p}^2&\text{else}
\end{cases}
\end{eqnarray}
The variable $d$ represents the distance between the center of the two discs. Needless to say this intensity can be normalized as stated above to:
\begin{eqnarray}
I =\begin{cases} 1&\text{for $d>R_\ast$}\\
\frac{2\pi R_\ast^2-2\pi R_\mathrm{p}^2}{2\pi R_\ast^2}&\text{else}
\end{cases}
\end{eqnarray}
A discontinuity arises at the point where $d=R_\ast$.

\subsection{The Continuous Transit Model}
To provide a continuous function that approximates the observation, a geometric representation for this problem has to be found. 

\begin{figure}[htb]
	\centering
\begin{tikzpicture}[line cap=round,line join=round,>=triangle 45,x=1.0cm,y=1.0cm]
\clip(-3.24,-3.27) rectangle (6.31,3.18);
\draw(0,0) circle (3cm);
\draw(4,0) circle (2cm);
\draw [line width=1.2pt,color=green] (2.63,1.45)-- (2.63,-1.45);
\draw [line width=1.2pt,color=red] (2,0)-- (2.63,0);
\draw [line width=1.2pt,color=blue] (3,0)-- (2.63,0);
\draw (0,0)-- (2.63,1.45);
\draw (4,0)-- (2.63,1.45);
\draw [dash pattern=on 1pt off 2pt on 5pt off 4pt,domain=-3.24:6.31] plot(\x,{(-0-0*\x)/4});
\draw[color=green] (2.49,0.66) node {l};
\draw[color=red] (2.43,-0.23) node {$h_1$};
\draw[color=qqqqff] (2.89,-0.21) node {$h_2$};
\draw[color=black] (1.65,0.71) node {$R_\ast$};
\draw[color=black] (3.52,0.95) node {$R_p$};
\end{tikzpicture}  \caption{Geometrical illustration of a transiting object. The star with radius $R_\ast$ is partially covered by the planet $R_p$. The intersecting circles are separated by the line $l$. The heights of each segment is $h_1$ and $h_2$ respectively. The size of the planet has been scaled for better visualization.}\label{fig:transit_geom}
\end{figure}
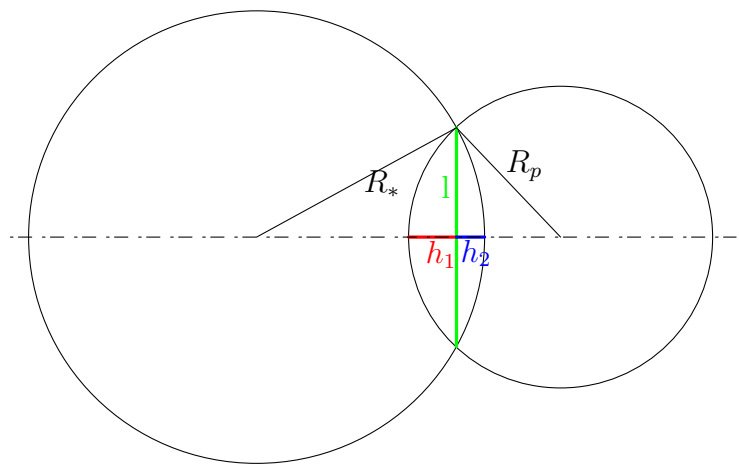

We have to introduce some variables to calculate the overlapping disc segments $A_1$ and $A_2$:

\begin{align}
l &= R_\mathrm{p}\sin\arccos\left(\frac{R_\ast^2-d^2-R_\mathrm{p}^2}{2dR_\mathrm{p}}\right),\\
h_1 &= R_\mathrm{p} - \sqrt{R_\mathrm{p}^2-l^2},\\
h_2 &= R_\ast - \sqrt{R_\ast^2-l^2},\\
A_1 &= f(R_\mathrm{p},h_1),\\
A_2 &= f(R_\ast,h_2).
\end{align}

Where $l$ ist the half length of the intersecting line and $h_1$, $h_2$ are the heights of the stellar and the planetary segment, see Figure~(\ref{fig:transit_geom}) for details. The formula for the area of a disc segment with radius $R_\mathrm{p}$ and height $h$ is

\begin{equation}
A(h,R_\mathrm{p})=R_\mathrm{p}^2\arcsin\left(\frac{a}{2R_\mathrm{p}}\right)-(R_\mathrm{p}-h)\sqrt{2hR_\mathrm{p}-h^2}.
\end{equation}

We can now calculate the intensity
\begin{eqnarray}
I(d, R_\ast,R_\mathrm{p}) =\begin{cases} 2\pi  R_\ast^2&\text{\dots $d\geq R_\ast+R_\mathrm{p}$}\\
2\pi R_\ast^2-2\pi R_\mathrm{p}^2&\text{\dots $d<R_\ast-R_\mathrm{p}$}\\
\pi R_\ast^2-A_1-A_2&\text{\dots $d^2<R_\ast^2+R_\mathrm{p}^2$}\\
\pi R_\ast^2-\pi R_\mathrm{p}^2-A_2+A_1&\text{\dots $d^2>R_\ast^2+R_\mathrm{p}^2$}\label{intensity}
\end{cases}
\end{eqnarray}
The function (\ref{intensity}) is analytic in $d$ within the whole interval $[-\infty,\infty]$ for given $R_\ast$ and $R_\mathrm{p}$.

\subsection{Stellar Limb Darkening}
\label{sec:limbdarkening}
To introduce the impact factor $b$ where $0\leq b\leq R_\ast$ or $0\leq b\leq 1$ respectivly the distance can be expressed as

\begin{equation}
d = \sqrt{x^2+b^2}\label{distance}
\end{equation}

where $x$ is a function of time $t$ and can be assumed linear. Inclination exceeding this boundary will lead to partial eclipses. To approximate the observations we need to introduce stellar parameters, at least the center-to-limb variation \citep{Claret2000} which can be described by

\begin{equation}
  \frac{I(0,\theta)}{I(0,0)}=\frac{2}{5}\left(1+\frac{3}{2}\cos\theta\right)\label{eqn:clv}
\end{equation}

which can be derived from the Eddington approximation. Since we need a function not dependent of an angle but as a function of the
radius or the distance, we modify equation \ref{eqn:clv} to 

\begin{equation}
I(d)=\frac 2 5\left(1+\frac 3 2\cos\arcsin(d)\right)
\end{equation}

Usually quadratic limb darkening is used where
\begin{equation}
\mu = \cos \theta = \sqrt{1-r^2}, 0\leq r\leq 1
\end{equation}
The intensity $I$ is then defined as
\begin{equation}
I(r) = 1-\gamma_1(1-\mu)-\gamma_2(1-\mu)^2,
\end{equation}
where $\gamma_1+\gamma_2<1$ are the limb darkening coefficients.
\subsection{Eclipse of a Uniform Source}
An analytic description of planetary transits is given by \citep{Mandel2002}, where I want to summarize some considerations. The model describes a transit of a opaque spherical planet on a spherical star. We are interested in the relation between unobscured to obscured flux $F^e(p,z)=1-\lambda^e(p,z)$ with neglecting the limb-darkening effect, we get
\begin{eqnarray}
\lambda^e(p,z)=\begin{cases}
0 & 1+p < z\\
\frac{1}{\pi}\left[p^2\kappa_0+\kappa_1-\sqrt{\frac{4z^2-(1+z^2-p^2)^2}{4}}\right] & |1-p|<z \leq 1+p\\
p^2 & z\leq 1-p\\
1 & z \leq p-1\\
\end{cases}
\end{eqnarray}
where $d$ is the center-to-center distance, between the star with radius $R_\ast$ and a planet with radius $R_\mathrm{p}$. The normalized separation of centers $z=d/R_\ast$ and $p=R_\mathrm{p}/R_\ast$ the size ratio with
\begin{equation}
 \kappa_1=\cos^{-1}\frac{1-p^2+z^2}{2z}
\end{equation}
and
\begin{equation}
 \kappa_0=\cos^{-1}\frac{p^2+z^2-1}{2pz}.
\end{equation}


\section{Overview}
The different detection methods will be explained briefly in Section~\ref{sec:detectionmethods}. We will discuss the transit detection method in detail and focus on the parameters that can be derived from transit observations of exoplanets.

We will briefly describe the data reduction pipeline used for observations at the observatory Lustb\"uhel. An overview of the most important and successful transit detection algorithms is given in Section~\ref{sec:detectionalgorithms}. Different transit modelling codes including codes that initially were used to describe stellar binaries will be discussed in Section~\ref{sec:transitmodelling}.

Finally the effects introduced by ground- and space-based observations are discussed in Section~\ref{sec:ObservationalEffects}. Several active stars hosting transiting planets are discussed in Section~\ref{sec:exoplanetsandstars}.


\chapter{Methods}
Transit detection in the frame of active stars does not only depend on fitting the right model to the data, it affects the whole process from data acquisition to the resulting model.

The process of transit observation is described in Section~\ref{sec:observation}. Since transit depths are in the domain of a few mmags, care has to be taken  for observational planning as well as for the data-reduction.

Usually one in thousand stars or even less have a transiting exoplanet. This implies a large amount of stars to be observed simultaneously to ensure a detection. This principle is carried out on space-based and on ground-based observatories. Transit detection algorithms that can handle the data gaps caused e.\,g. by the night-and-day cycle are necessary to sieve the large number of targets. A selection is presented in Section~\ref{sec:detectionalgorithms} including the de-facto standard by \citet{Kovacs2002}. I present my own considerations on Section~\ref{sec:detectionalgorithm}.

Various filtering methods are discussed to increase the signal-to-noise ratio in Section~\ref{sec:filteringmethods}. Special care is taken on the analysis of periodic signals in Section~\ref{sec:analysis}.

\section{Observation of Transiting Exoplanets at the Observatory Lustbühel}
\label{sec:observation}
I want to give a short description, how transiting exoplanets are observed at the Observatory Lustbühel Graz, and describe in detail the data reduction pipeline. Both telescopes, the 40\,cm Schmidt-Cassegrain and the 30\,cm Zeiss refractor are capable of observing transits on brighter targets.
\subsection{Target Selection}
The targets have to fulfil a set of criteria to be observed. Usually the target is selected from a list provided by the Exoplanet Transit Database (ETD)\footnote{\url{http://var2.astro.cz/ETD/index.php}}
\begin{itemize}
\item The transit depth must be at least $0.01$~mag to ensure a detection of the transit at a reasonable signal-to-noise ratio.
\item The V~magnitude must be brighter than 15~magnitudes.
\item The airmass of the object must be less than $1.8$.
\item The lunar phase must be less than 50\% and the angular distance between the target and the moon should be at least $90°$.
\item The target should be in an azimuth range between 60 and 120 degrees to avoid regions with light-pollution near Graz.
\end{itemize}
\subsection{Observation}
The observation starts at least 15~minutes before the transit to measure the level of stellar activity out-of-transit.\\
A set of at least 20 dark frames is taken after the observation with the same exposure time as the light frames. Also a corresponding set of bias frames is taken at the same temperature to ensure the reduction. Flat fields are usually taken from a library.

\subsection{Image Registration Techniques}
Inaccuracies in tracking the target as well as guiding errors lead to image displacements. Higher airmasses usually imply higher guiding errors. To measure the flux of a target star and compare it to comparison stars, a constant position of the target is mandatory throughout the whole image stack. This can be basically accomplished by two different methods, either by keeping the image coordinates in pixel constant, or by calibrating the image to calculate the absolute positions on the sky in \textit{right ascension} and \textit{declination}. 
\subsubsection{Fourier Image Registration}
If we assume a displacement in $x$- and $y$-axes between an image $i$ and a reference $r$, the easiest way would be to probe all possible displacements starting from $(0,0)$ to $(w,h)$ where $w$ is the width of the image and $h$ is the height of the image. Actually the image shift can be assumed in the range $[-w/2,w/2]$ and $[-h/2,h/2]$. To calculate the match between the two images $i$ and $r$, we calculate $|i-r|$, which we want to minimize.
Using the Fourier cross-correlation
\begin{equation}
c \equiv \mathcal{F}^{-1}(\mathcal{F}(i)\cdot\mathcal{F}^\ast(r)), \label{eqn:crosscorrelation}
\end{equation}
between the both images $i$ and $r$, where $\mathcal{F}$ is the Fourier transform and $\mathcal{F}^{-1}$ its inverse and $\mathcal{F}^\ast$ the complex conjugate. 
The Fourier transform of a 2D function is calculated by
\begin{equation}
\mathcal{F}(u,v) = \iint f(x,y)e^{-2i\pi(ux+vy)}dxdy,
\end{equation}
where $u$ and $v$ are spatial frequencies and its inverse
\begin{equation}
\mathcal{F}^{-1}(u,v) \equiv f(x,y)= \iint\mathcal{F}(u,v)e^{2i\pi(ux+vy)}dudv.
\end{equation}
The shift theorem in Fourier-space 
\begin{equation}
\mathcal{I}[f(x-a)]=e^{-2i\pi ua}\mathcal{F}
\end{equation}
The coordinates of the maximum in the image-array $c$ in Equation~\ref{eqn:crosscorrelation} marks the displacement. This registration is only accurate by integer pixel values, which anyway is flux conserving. If we want to register the images in sub-pixel accuracy, the images have to be scaled up. So an upscale by the factor 2 will result in $1/2$-pixel accuracy.

If displacement and rotation have to be corrected the Fourier-Mellin transform has to be applied. An appropriate algorithm is described by \citet{Reddy1996}.
\subsubsection{Astrometric Registration}
Stellar-like sources can be easily identified and extracted with \textsc{SExtractor} by \citet{Bertin1996}. Each star is fitted with a two-dimensional Gaussian, where its photo-center can be determined numerically. This method provides sub-pixel accuracy down to 10~milliarcseconds. The limitations are given by the detector noise and by the distortion caused by imperfect optics.

By matching the stars to a precise catalogue, the exact transformation from pixel coordinates to celestial coordinates can be determined. In this way the target star is located by right ascension and declination and the stellar flux can be either measured from the gaussian fit done by \textsc{SExtractor} or by applying aperture photometry using the \textsc{DAOPhot} routines \citep{Landsman1995}. In latter case the stellar flux is measured within a circular area and the average sky-flux that has to be subtracted is measured in an annulus.

\subsection{Data Reduction}
In the whole dataset the target star $V$ and two comparison stars $C$ and $K$ are selected, both being known to be non-variable. One of the comparison stars should be dimmer, the other brighter than the target star. All three stars must be present throughout the whole dataset, otherwise the frames where one star is missing is omitted. The comparison stars should have a photometric precision better than 0.01~magnitudes. Usually the photometric magnitude is mapped to a R-,J- or K-magnitude in a catalogue regardless of the fact that the measurements were taken in the visual or R-band.

For each frame the observed values of the comparison stars are mapped to the known magnitudes resulting in a correcting linear function, where the precise photometric value for the target star is calculated by interpolation. This could be in principal be extended to a whole set of comparison stars by applying a linear regression. 

For high precision, the comparison stars should be of the same color temperature as the target star. This is generally not the case due to a limited field-of-view. Therefore as a last step the data has to be corrected for influences caused by the airmass.
\section{Detection Algorithms}
\label{sec:detectionalgorithms}
Various methods for the detection of transit signals have been developed. Most of them require a more or less ``clean'' dataset or \textit{lightcurve}. Usually the data contains gaps and other irregularities like jumps, hot pixels (see Section~\ref{sec:hotpixels}). The detection of transit like features can be achieved in different approaches:
\paragraph{Brute force}
The simplest idea would be, to scan the lightcurve for recurring transit-like shapes. The parameter space is spanned by periods, depths and different impactfactors.
\paragraph{Fold and Search}
The lightcurve can be folded with different periods, and the transit signal can be found in the folded data (see Section~\ref{sec:phasefolding}). 
\paragraph{Correlate with Model}
Depending on the spectral type of the star, a model can be generated in respect of the limb-darkening. This model lightcurve can be correlated with the dataset.

A description of a \textit{Monte Carlo Markov Chain} (MCMC) analysis of the parameters derived from the lightcurve can be found in e.\,g. \citep{Burke2007}. A good comparison of performance of the algorithms described in the following sections can be found in e.g. \citep{Tingley2003,Moutou2005}.

\subsection{The BAST algorithm for transit detection}
The BAST (Berlin Automatic Search for Transits) algorithm \citep{Renner2008} was developed in advance to the CoRoT mission. It could not be tested on real data but simulated lightcurves for CoRoT. The basic principle is the search for box shaped  signals in pre-filtered lightcurves. The filtering includes normalization, application of a low-pass filter and correction for variability. 

After filtering, the algorithm performs a search for periodic signals. The performance is similar to BLS (Section~\ref{sec:BLS}). Since BAST was only tested on simulated data, the results on real data are degraded. \citet{Renner2008} claims to be able to find transit depth down to $0.01\%$ corresponding to an Earth-sized planet around a solar-type star.

\subsection{A box-fitting algorithm in the search for periodic transits (BLS)}
\label{sec:BLS}
The algorithm by \citet{Kovacs2002} performs a box-fitting to the lightcurve. The transit is here assumed as a box-shaped function with two levels $H$ and $L$. The box-function is fitted to a folded time series at a trial period. The period is then permuted. The algorithm weights different levels $L$, giving higher levels a higher priority. 

A signal-to-noise-ratio of 6 is required to ensure a significant transit detection. The BLS algorithm is the current standard algorithm performed for transit searches. Its performance can be improved by filtering the lightcurve prior to processing. BLS is computational costly on the downside.

\subsection{TRUFAS, a wavelet-based algorithm for the rapid detection of planetary transits}
The \textsc{Trufas} algorithm by \citet{Regulo2007} is based on a continuous wavelet transform of the lightcurve, which has been detrended in a first step. In a second step the period is searched within the wavelet transformation. At least three transits are required to ensure a proper detection.

This algorithm had been developed on simulated CoRoT-data as the BAST algorithm. With \textsc{Trufas} the preprocessing is not crucial. \citet{Regulo2007} claims that the algorithm is well suited for large data volumes.

\section{Filtering Methods}
\label{sec:filteringmethods}
An attempt of filtering of lightcurves in the presence of stellar activity has been demonstrated on one of the \corot Blind Tests by \citet{Alapini2009}. Filtering is necessary to remove spurious signals from the lightcurve and therefore to increase the signal-to-noise ratio (SNR).

In our case the signal is the transit signal and the noise is the standard deviation of the out-of-transit data. All measurements show intrinsically Gaussian noise as a first approximation. For faint stars the photon noise becomes relevant, which is proportional to $\sqrt{N}$, where $N$ is the number of photons \citep[see][]{Weingrill2011,Aigrain2009}. Usually lightcurves are poised with red noise caused by the finite observational length and transient stellar activity features.

Usually the model-fit of the transit has to be better than $3\sigma$ in the unfiltered folded data. In rare occasions a marginal detection of $1\sigma$ is accepted, when the lightcurve is noisy or one or more transits are missing. Assuming a model $\mu$ for the transit (\ref{sec:transitmodelling}) with $n_x$ data-points in transit and performing a reduced chi-squared fit
\begin{equation}
\chi^2_N\equiv\frac{1}{n_x - n_p}\sum\left(\frac{x_i - \mu_i}{\sigma_i}\right)^2,
\end{equation}
of $n_p$ parameters with a variation of $\sigma_i$ out-of-transit.
\subsection{Median Filter}
The median filter is a low pass filter, where a datapoint $x_i$ is replaced by the median of its surrounding data $x_{i-k},\dots,x_{i+k}$. The length of the filter corresponds to the characteristic timescale $\tau$. Usually high frequency content with $\tau < 0.5$\,days is removed with this filter.

By subtracting the median filtered data from the unfiltered original, a high pass filter is realised. It works ideally to remove long term variations caused by stellar activity.

\subsection{Wiener Filtering}
The Wiener filter operates in Fourier space under the assumption that the noise $N(f)$ can be estimated in a power spectrum. So the optimal Wiener filter $W(f)$ for a measured signal $S(f)$ is defined as
\begin{equation}
W(f) \equiv \frac{|S(f)|^2}{|S(f)|^2 + |N(f)|^2}
\end{equation}
A priori we have to estimate the noise background independent of the frequency $f$, so the noise is a constant background in the power spectrum.

In the case of planetary transit searches the instrumental effects and the stellar activity show generally higher levels at the lower frequency end of the power spectrum which is called \textit{red noise}. Wiener filtering is not suitable for eliminating stellar activity since it fails to estimate the noise background.

\subsection{Butterworth Filter}
The Butterworth filter consists of a flat part that leaves the wanted frequencies unchanged and a slope of variable steepness to block the unwanted frequencies. The gain $G(\omega)$ is defined as
\begin{equation}
G^2(\omega) = \frac{G_0^2}{1+\left(\frac{\omega}{\omega_c}\right)^{2n}},
\end{equation}
where n is the order of the filter determining the steepness of the slope, $\omega_c$ is the cutoff frequency sets in and $G_0$ is the gain at zero frequency, which is usually unity.

This filter can be applied as a substitute to the median filter, as a high- and a low-pass filter.
\subsection{Savitzky-Golay Filter}
The filter by \citet{Savitzky1964} is generally speaking a smoothing filter for spikes. For that purpose the adjacent points $x_{i-k},\dots,x_{i-1},x_{i+1},\dots,x_{i+k}$ of each datapoint $x_i$ is interpolated by a low order polynomial. The value of $x_i$ is replaced by the interpolated polynomial value.

As a digital finite impulse response filter, the data is usually convolved with the coefficients of the filter kernel. The filter works fast and removes single outliers in the data without affecting the shape of the transit. The filter is mostly used to remove instrumental noise, caused e.\,g. by spurious outliers from the Southern-Atlantic-Anomaly~(SAA).

\subsection{Pre-whitening}
\label{sec:prewhitening}
Stellar signals that are generated by rotation or pulsation can be approximated with a small set of sinusoidal functions. By determining the most prominent frequencies including their harmonics, those rotation signals can be eliminated. The filter works iteratively by performing the following steps:
\begin{enumerate}
\item Calculate the Fourier power spectrum and determine the maximum amplitude with the respective frequency.
\item Fit the function $A\cdot\cos(\omega t+\phi)$ with the amplitude $A$ the frequency $\omega$ and the phase $\phi$.
\item Evaluate the stop condition (e.\,g. remaining total frequency power) and repeat step 1.
\end{enumerate}

Usually fitting a cosine function (Section~\ref{sec:cosinefit})is more robust than a sine function. The fitting by minimizing the least squares can be either done using the AMOEBA or the Levenberg-Marquardt (LM) \citep{Press1992} algorithm. The same algorithm can be used to separate the noise from the signal, by iterating the removal of frequencies unless the amplitudes are in the order of $1\sigma$.

Pre-whitening is usually performed to clean radial-velocity measurements from unwanted frequencies \citep[e.\,g.][]{Queloz2009}. It can also be applied selectively to remove known instrumental frequencies.

\subsection{Transit Filter}
The basic idea for a specialized filter is that we are looking for transits. We know that the transits cover a distinct area in parameter space, periods $P=0.5\dots T/3$~days, where $T$ is the total length of observation, transit durations $t_T=1\dots20$~hours, depths $\Delta F = 0.01\dots0.1\%$.

\begin{figure}[htb]
\centering
\includegraphics{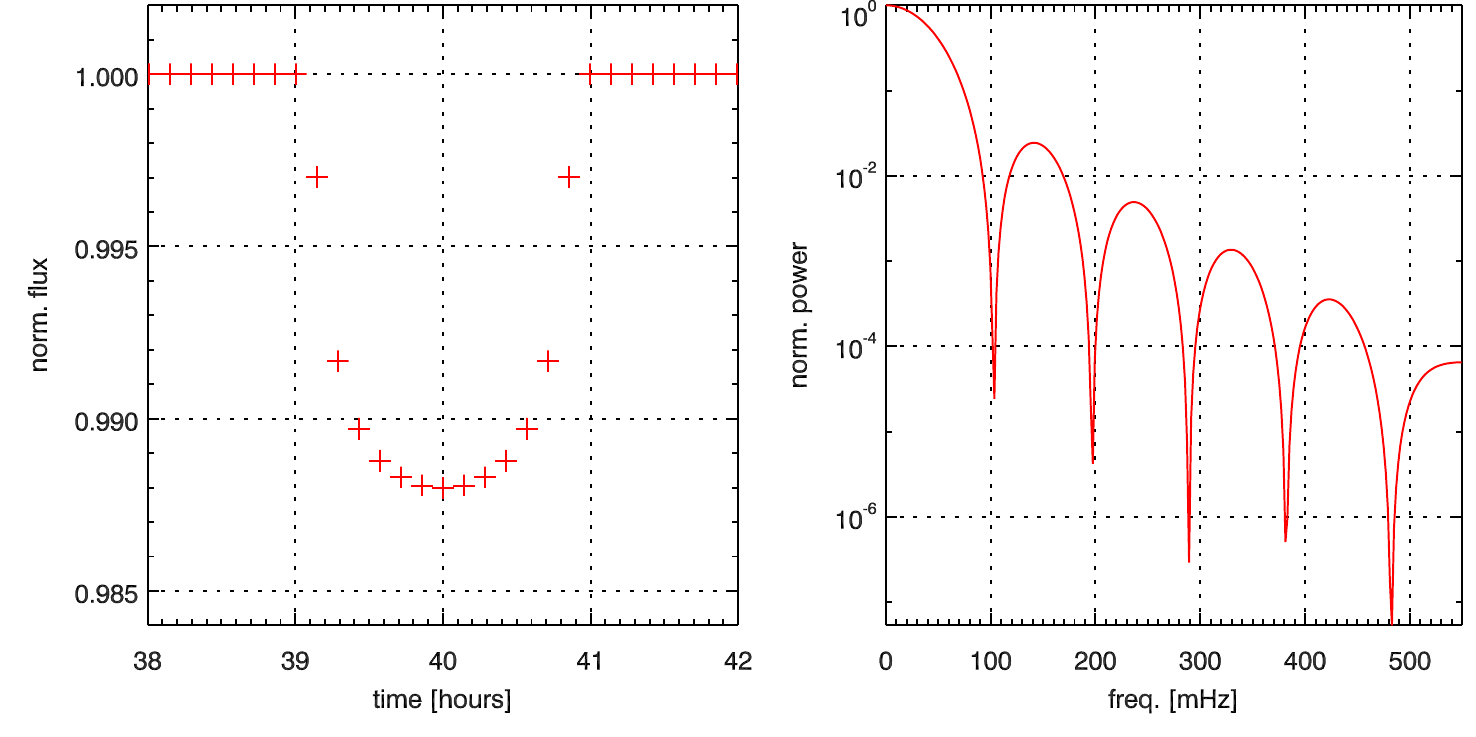}
\caption{\textit{left:} Transit of a Jupiter-sized object with a duration of 2~hours and a depth of $0.1\%$. The sampling was 512~s. \textit{right:} Power spectrum of the transit-shape.\label{fig:transitfilter}}
\end{figure}

Using a Fourier filter, it is possible to filter out just the signal, which we want to detect. Using the low-pass filter as seen in the right Figure~\ref{fig:transitfilter}, we are able to filter out transit shapes that look like Figure~\ref{fig:transitfilter} (left). The transit itself was synthesized using the \citet{Mandel2002} algorithm with limbdarkening coefficients $u_1=0.25$ and $u_2 = 0.75$.

This filter must be adapted to different transit durations, while the depth and the limb- darkening is uncritical. Usually a single filter functions covers a large volume in parameter space.

\subsection{The SARS Algorithm}
A filtering method that considers mostly instrumental effects has been developed for CoRoT by \citet{Ofir2010}. It is based on the assumption that all lightcurves are affected by instrumental noise on the CCD in the same way, hence showing slight correlations. The algorithm tries to minimize those influences by generating a photometric reference.

The SARS algorithm is necessary to reach the domain of ppm~precision. Using this kind of cleaning procedure faint transits in the regime of $0.01\%$ are visible. Nevertheless systematic errors can be introduced and transit signals are often below a $3\sigma$ confidence level.

\section{Transit Modelling}
\label{sec:transitmodelling}
Many parameters of a transiting system can be derived by analysing the observed lightcurve and comparing it to a calculated model. Most of the codes are descendants from binary star modelling tools. The most prominent codes that have been applied to exoplanets' transits are \textsc{Phoebe} by \citet{Prsa2007} and \textsc{JKTEBOP} by \citet{Southworth2009}. 

Stellar limb darkening coefficients are usually taken from \citet {Sing2010}, which are available from  \Teff from 3500 to 50000\,K including different surface gravities and metallicities.

Other codes have been developed to test transit search algorithms, like the \textsc{Universal Transit Modeller} (UTM) by \citet{Deeg2009}. One has to distinguish transit modelling from transit search algorithms. Transit models are usually specialized on many parameters like e.g. limb darkening coefficients, which themselves are dependent on the temperature of the host star. These are not suitable to detect transits in a set of lightcurves due to the mathematical complexity

Codes designed for binaries have the major drawback that the secondary is assumed to be a stellar object, which extrasolar planets are definitely not. As a workaround the surface temperature of the secondary is assumed to be 3500~K to satisfy the fitting of limb darkening coefficients. Nevertheless the results from these codes are questionable since other parameters like e.g. surface brightness, gravity brightening may falsify the output.

Analytic lightcurves as a base for searching transiting planets have been discussed by \citet{Mandel2002}, which is today the de-facto standard for transit models.

\subsection{Phoebe}
\label{sec:phoebe}
The \textsc{Phoebe} program by \citet{Prsa2007} uses the code by \citet{Wilson1971} for binary stars. It is capable of deriving model parameters for different types of binares using radial-velocity and photometric measurements. In the case of transiting planets the model for a detached binary is used and the surface temperature of the planet has to be assumed 3500~K.

This code has been applied by \citet{Poddany2008} successfully on Wasp-2b. The big advantage of Phoebe, using a scripted version to automatically find the best model parameters, can not be used due to the high mass ratio between primary and secondary object. Mass, radius and inclination as well as the ephemeris of the mid-transit can be estimated using Phoebe.

The program provides a graphical user interface (GUI), which greatly simplifies the manipulation of many parameters for fitting the parameters. The major drawback is the fact that the initial parameters for the fit have to be very close (within a few percent) to converge. Usually the fitting procedure halts after some steps without convergence.

\subsection{JKTEBOP}
The JKTEBOP code has been used by \citet{Southworth2009} to model several transits of transiting exoplanets. Similar to \textsc{Phoebe} (see Section~\ref{sec:phoebe}), this code incorporates the Levenberg-Marquardt optimisation algorithm \citep{Press1992} and advanced limb darkening parameters.

JKTEBOP requires several input parameters (\ref{tab:jktebop}) as a first estimate, but works more robust than Phoebe and in an automated way.

\begin{table}[ht]
\centering
\captionabove{Some input parameters for JKTEBOP used to calculate WASP-4.\label{tab:jktebop}}
\begin{tabular}{lr}
\hline \hline
parameter & value\\
\hline
Sum of the radii   & 0.21 \\
Ratio of the radii & 0.15 \\
Orbital inclination (deg) & 88.5 \\
Mass ratio of system &  0.0013\\
LD law type for star A & quad\\
LD law type for star B & lin\\
\hline
\end{tabular}
\end{table}

\subsection{Mandel \& Agol Algorithm}
\begin{figure}[htb]
\centering
\includegraphics{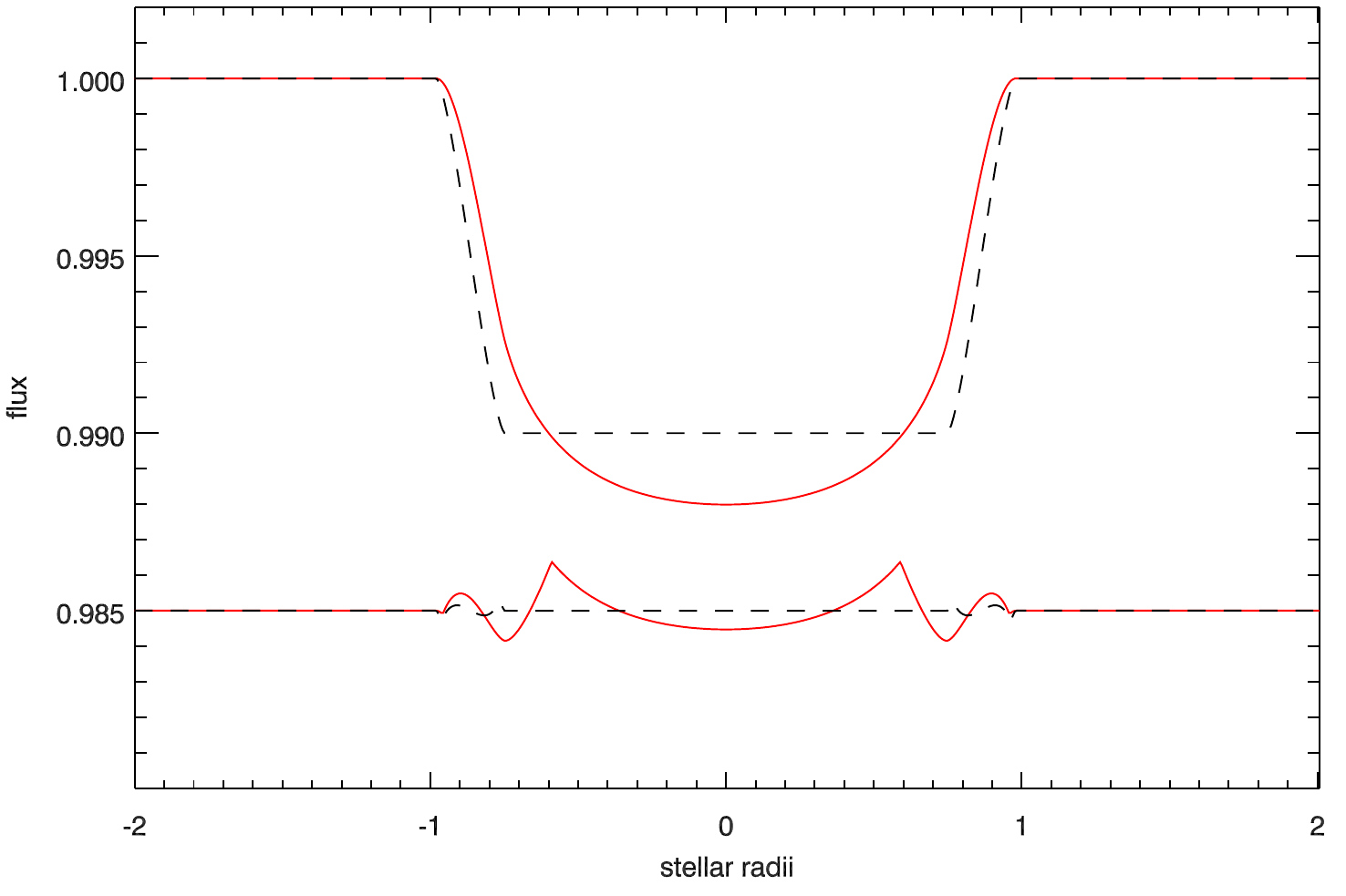}
\caption{Transit model after \citet{Mandel2002} with parameters taken from Table~\ref{tab:mandelagol}. The solid red line includes the limb-darkening coefficients $\gamma_1, \gamma_2$, whereas the dashed line is calculated for a uniform source. The residuals are plotted with an arbitrary offset at the same scale.\label{fig:mandelagol}}
\end{figure}

The core of the M\&A algorithm is used in the \textsc{Transit Analysis Package} (TAP) by \citet{Gazak2011}. With the given parameters from Table~\ref{tab:mandelagol} we get the result seen in Figure~\ref{fig:mandelagol}.

\begin{table}[ht]
\centering
\captionabove{Imput parameters for a transit model. For a explanetion of the limb darkening coefficients see Section~\ref{sec:limbdarkening}\label{tab:mandelagol}}
\begin{tabular}{llr}
\hline \hline
description & parameter & value\\
\hline
$p = R_\mathrm{p}/R_\ast$ & $p$ & 0.1\\
impact factor & $b$ & 0.5\\
limb-darkening coefficient & $\gamma_1$ & 0.25\\
limb-darkening coefficient & $\gamma_2$ & 0.75\\
\hline
\end{tabular}
\end{table}

The algorithm is computational intensive and is not suitable for transit detection. It is intended to perform transit fitting to derive the transit parameters $p$, $b$, $\gamma_1$ and $\gamma_2$.

By fitting a trapezoid using the Levenberg-Marquardt least squares algorithm implemented in IDL \citep{Markwardt2009}, we get the parameters $p = 0.107$, with a duration of $t_T = 1.92$ and the impact factor $b = 0.92$, which is highly overestimated. The $\chi^2 = 3.73\times10^{-4}$ is sufficient. The uniform source is generally better fitted with  $p = 0.1$,  $t_T = 1.93$  and $b = 0.585$ with a $\chi^2 = 4.3\times10^{-5}$ about one magnitude better.

\section{Analysis of periodic signals}
\label{sec:analysis}
%
One initial problem is finding the main period of a signal. In transit detection, at least three transit signatures are required to assume a transiting object. In rare cases, where a high signal-to-noise ratio is evident, two transits will suffice, bearing the risk of a binary or even a blend scenario. Having time-limited observation windows like for CoRoT 150~days at maximum, sometimes mono-transits are detected. These detections are candidates for follow-up observations to eventually detect further transits.

We define a signal as period, if a signature occurs at equidistant time intervals. Hence, if we fold the signal at the length of the given interval, and rebin the signal we get a minimal \textit{root-mean-square}, which is defined as
\begin{equation}
\mathrm{rms} \equiv L_2(x - \bar{x}).
\end{equation}

\subsubsection{Rebinning}
\label{sec:rebinning}
Each measurement of the flux of a star requires a certain time of integration on the CCD. For CoRoT this is typically 512s but will be changed to 32s if a target is of interest. For the detailed analysis, we are interested in constant time bins for the data. Hence we need to rearrange the data by conserving the flux. This process is called \textit{drizzling} and initially intended for images \footnote{\url{http://www.stsci.edu/ftp/science/hdf/combination/drizzle.html}}.

By rebinning the data to larger time intervals we get the standard deviation ($\sigma$) in each bin as an error estimation. This value can be later on used for fitting the data. If we rebin 32s cadence data to 512s, we get 16 measurements in each new bin. Data gaps smaller than 512s still leave at least one measurement in the bin, but show a higher $\sigma$. By looking for transit signals which last three hours, we might even consider to rebin the data to e.g. one hour bins, which condenses the data by a factor of seven. However, this is dangerous, since a single outlier in the large bin may cause an alteration of the data, in a way, such a transit search algorithm may not be able to find a transit or even detects a false positive.

The major disadvantage of the drizzling algorithm is the fact that bins with a high local derivative, e.g. in the ingress or egress of a transit show abnormal high $\sigma$-values.

\subsubsection{Folding}
\label{sec:folding}
For $n$ data-points $d_i$ and their time-indices $t_i$ we calculate 
\begin{equation}
t_i' = t_i \mathrm{mod} P,
\end{equation}
where $P$ is the period and which fulfils 
\begin{equation}
t_\mathrm{min} \leq t_i' < P
\end{equation}
by sorting the times $t_i'$ ascending and applying the sort to the data $d_i$, we get a new sorted dataset $d_i'$. This new dataset is now folded at the period $P$. If the period is not an integer multiple of the original time-interval we have unequal spaced data. Using the algorithm from the previous section \ref{sec:rebinning}, we rebin the data. The implementation can be found in Section~\ref{sec:foldedpro}.

\subsection{Finding the period by phase folding}
\label{sec:phasefolding}
The easiest method to find the main period of a signal is by iteratively trying possible periods, where we fold the data and look for the ``best fit'' in our case, where the maximum of all $\sigma_i$ for each bin $i$ is a minimum. We give an example on the binary SRa01\_E1\_0359. The domain, where we look for a periodic signal can be restricted between 0.5 days and 11 days for a total run length of 33~days. The lower boundary is given by physical limits. Planets or in our case a secondary star would be within the Roche-lobe of the host star or the primary respectively. The Roche limit $d$ is 
\begin{equation}
d = r_\ast\left(2\frac{\rho_\ast}{\rho_\mathrm{p}}\right)^{1/3},
\end{equation}
depending on the stellar radius $r_\ast$, the density $\rho_ast$ of the star and the density $\rho_\mathrm{p}$ of the planet. 

The higher boundary is a third of the observation period, which means that at least three transits can be observed. This limit is given by logical constraints: a single transit won't reveal a period, whereas two transits can not exclude a binary system. Hence for looking for an Earth-analogue at 1~AU we need to observe three years in total.

\begin{figure}[htb]
	\centering
		\includegraphics[width=0.9\textwidth]{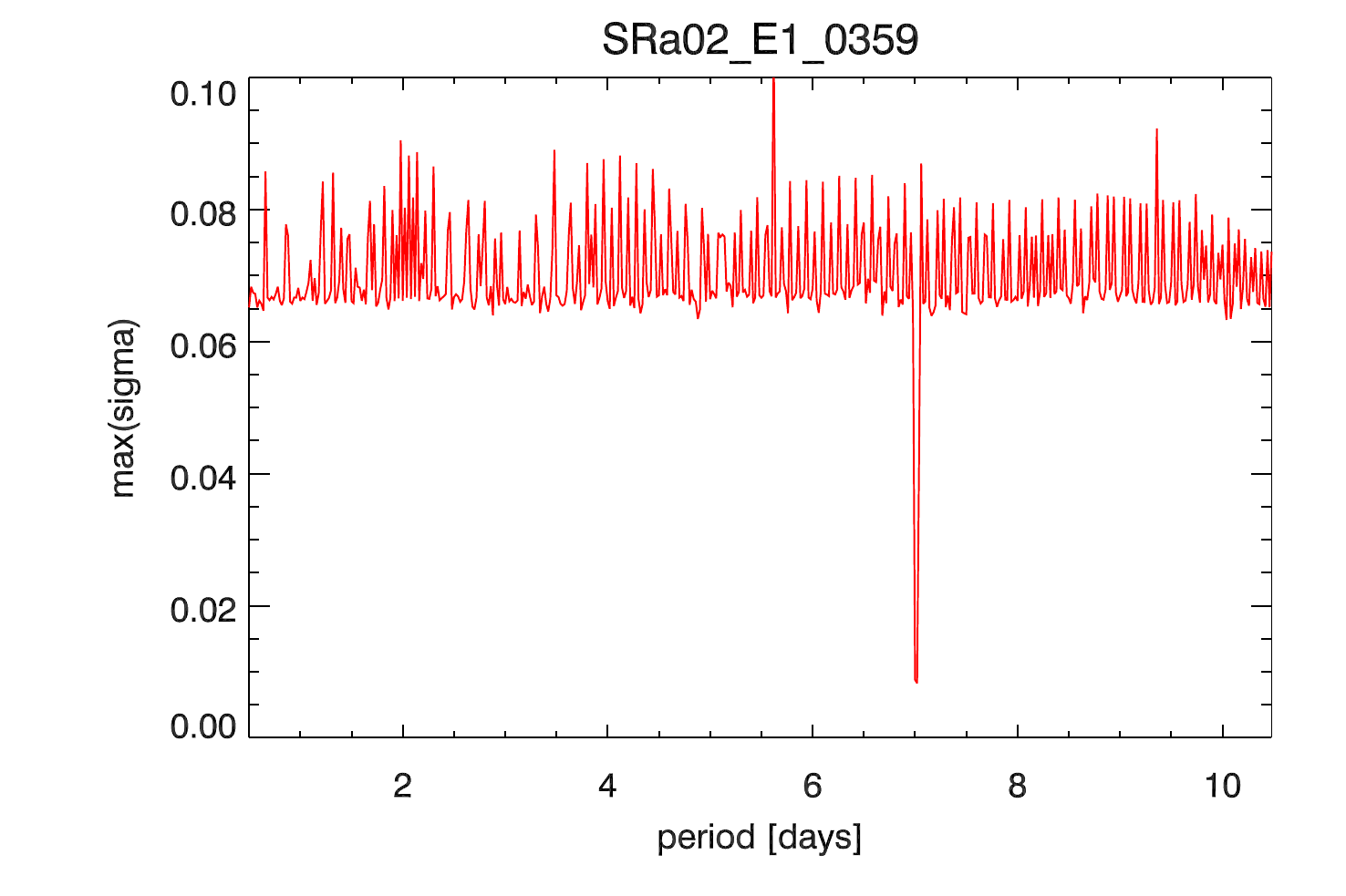}
	\caption{The maximum of the RMS of each bin of the binary SRa01\_E1\_0359. A total of 500~periods between 0.5~days and 11.5~days were probed, giving the best period of $7.02$~days.}\label{fig:periodtest}
\end{figure}

We start with a rough estimation of the period, which can be deduced from the second local maximum of the autocorrelation of the signal or the maximum in the power-spectrum of the signal. A proper algorithm is e.g.

\begin{enumerate}
\item initialize a function-minimizer e.g. AMOEBA with the estimated period $P$
\item fold the data with the given period $P$ as described in Section~\ref{sec:folding}
\item rebin the folded data to $d_i'$
\item calculate the $\mathrm{max}(\sigma_i)$ over all bins as the weight for the function.
\end{enumerate}
We reiterate from point 2. until a local minimum is found.

For the case of the binary SRa01\_E1\_0359, AMOEBA found $7.01408$~days.

Algorithms that minimize a function like AMOEBA or Levenberg-Marquardt however need good initial conditions. In our case, a good guess of the true period.

\subsection{Fourier Autocorrelation}
The Fourier cross-correlation of two functions $f$ and $g$ is defined as
\begin{equation}
  c \equiv \mathcal{F}^{-1}(\mathcal{F}(f)\cdot\mathcal{F}^\ast(g)),
\end{equation}
where $\mathcal{F}$ is the Fourier transform, $\mathcal{F}^{-1}$ its inverse and $\mathcal{F}^\ast$ is the complex conjugate.
The \textit{Fourier auto-correlation} of a function $f$ is defined as
\begin{equation}
 a \equiv \mathcal{F}^{-1}(\mathcal{F}(f)\cdot\mathcal{F}^\ast(f)),
\end{equation}
as the cross-correlation of the function with itself (see also Section~\ref{sec:acorrpro}).

By performing a sliding windowed auto-correlation over the data with the length twice a characteristic timescale (e.\,g. rotation period or orbit period), transient features can be followed. This method is superior in using a sliding Fourier window, since it ``locks'' on the most prominent feature without any presumptions.

Sliding auto-correlation may be applied to detect timing variations. It operates as a wavelet-filter in a specific frequency.

\section{Description of a Transit Detection Algorithm}
\label{sec:detectionalgorithm}
We want to give a short description of a transits detection algorithm, that has been developed in the frame of the CoRoT\citep{Baglin2006} detection pipeline. After an observation run of CoRoT the raw (N0) data is undergoing two pipelines (N1, N2) where the invalid data-points are flagged, e.g. for hot pixels and for crossings of the satellite over the South Atlantic Anomaly~(SAA). The SAA causes a semi-periodic alteration of the on-board electronics and therefore impacts the data.

\begin{figure}[htb]
  \centering
  \includegraphics{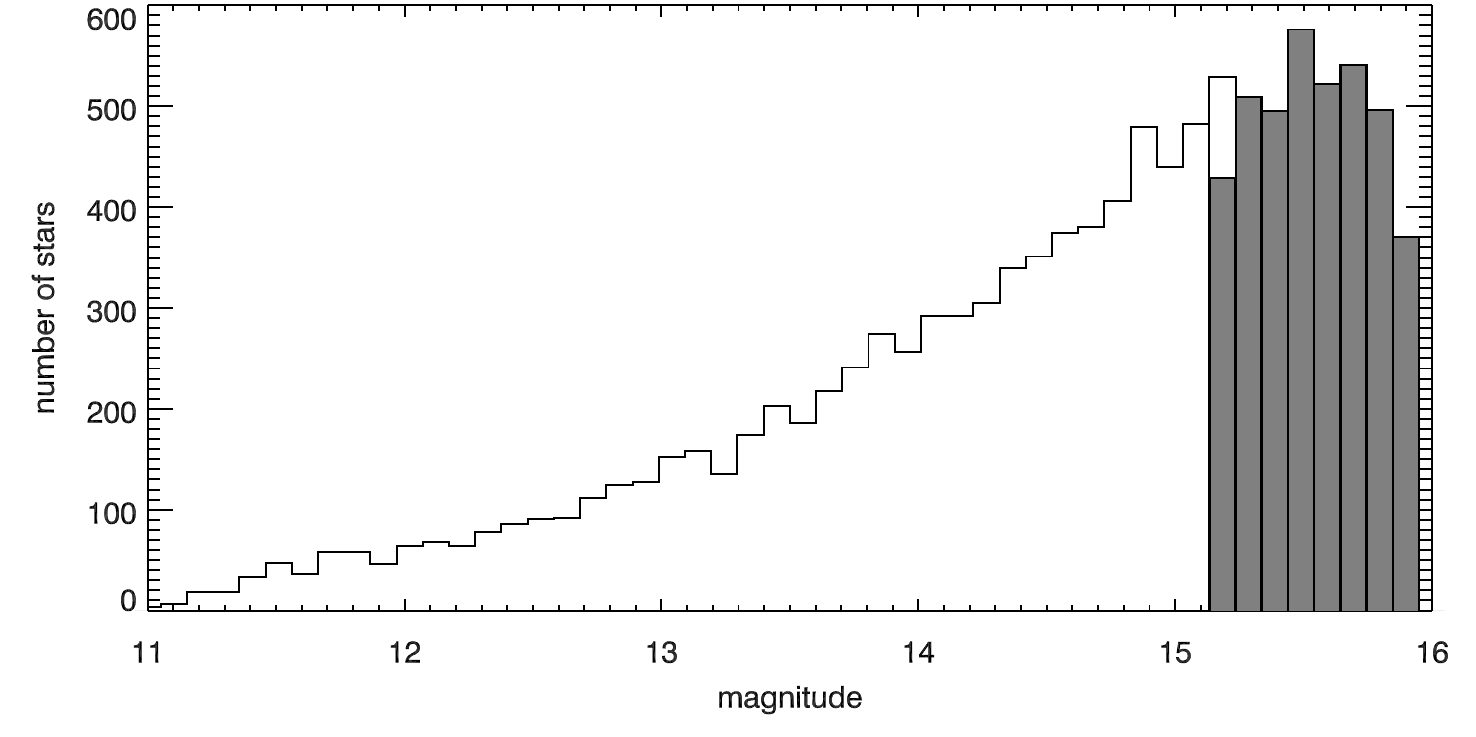}\label{fig:magnitude}
  \caption{Distribution of chromatic targets over magnitude for the CoRoT run LRa01. The gray bars show the amount of monochromatic targets.}
\end{figure}

The CoRoT data can be downloaded from the CoRoT Data Center at IAS archive\footnote{\url{http://idoc-corot.ias.u-psud.fr/}}. Each run in separated between chromatic and monochromatic targets. The chromatic targets are usually the bright targets (Figure~\ref{fig:magnitude}). 

\begin{figure}[htb]
  \centering
  \includegraphics[width=0.5\textwidth]{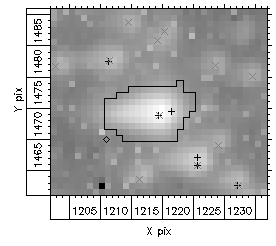}\label{fig:imagette}
  \caption{Imagette of the CoRoT target \#0102728404. The coordinates are given in absolute pixel of the CCD. The stars $\ast$ mark the position of other CoRoT targets, the crosses $\times$ indicate faint background stars and the plus-signs represent the true position on the sky. Imagette taken from Exo-Dat \citep{Deleuil2009}.}
\end{figure}

In front of the the CCDs E1 and E2 for the exo-fields a grism has been inserted to spread the point-spread-function (PSF) in its spectral colors. This enlarges the PSF to several arcseconds. The image of the PSF on the CCD is called \textit{imagette}, as seen in Figure~\ref{fig:imagette}. By selecting distinct columns in the imagette, the color information can be extracted. On fainter stars the whole intensity of an imagette is integrated.

\subsection{Data Preparation}
\label{sec:datapreparation}
Usually a CoRoT-lightcurve has a cadence of 512~seconds. The on-board cadence is 32~seconds which is kept for interesting targets and candidates in the so called Alarm-mode. Due to bandwidth limitations in the satellite downlink, only a few targets are available in high cadence. The onboard electronics downsamples the 32s-data to 512s. The timing of each datapoint is given in CoRoT Julian Date (CoRoTJD) and heliocentric Julian Date (HelJD). To use an absolute Julian date, we have to add a constant
\begin{equation}
\mathrm{JD} = \mathrm {HelJD} + 2451545.0
\end{equation}
where the constant corresponds to 1 January 2000.
The total number of datapoints $n$ in a dataset is calculated by 
\begin{equation}
n = \left\lceil 168.75(\max t - \min t) \right\rceil,
\end{equation}
where the constant is used to convert to samples per day.

For detection purposes the signal-to-noise ratio (snr) is at highest for the integrated flux combining all three color channels. To minimize the noise all data is oversampled to 512~seconds. This is done using the drizzling algorithm (see Section~\ref{sec:drizzlealgorithm} and Section~\ref{sec:rebinning}) which bins the data according to its timestamps. Then the mean of each bin is calculated and the rms in each bin is used as an error estimate. A better way to integrate one bin would be taking the median, which produces a better snr but the noise statistics differ from the on-board processing of CoRoT.

Since the drizzling may produce empty bins resulting from the flagged data, this data has to be substituted for following algorithms, that assume equidistant data-points. For that purpose the missing data is simply linearly interpolated. The linear interpolation has the major advantage, that the power spectrum is not severely affected by the introduction of the artificial data-points, whereas other interpolation methods like quadratic, or spline interpolation induce several artificial frequencies. The optimal interpolation function would be the Fourier interpolation (for an implementation see Section~\ref{sec:finterpolpro}) which minimizes the total energy in the power spectrum. This however is computational expensive.

\begin{figure}[htb]
  \centering
  \includegraphics{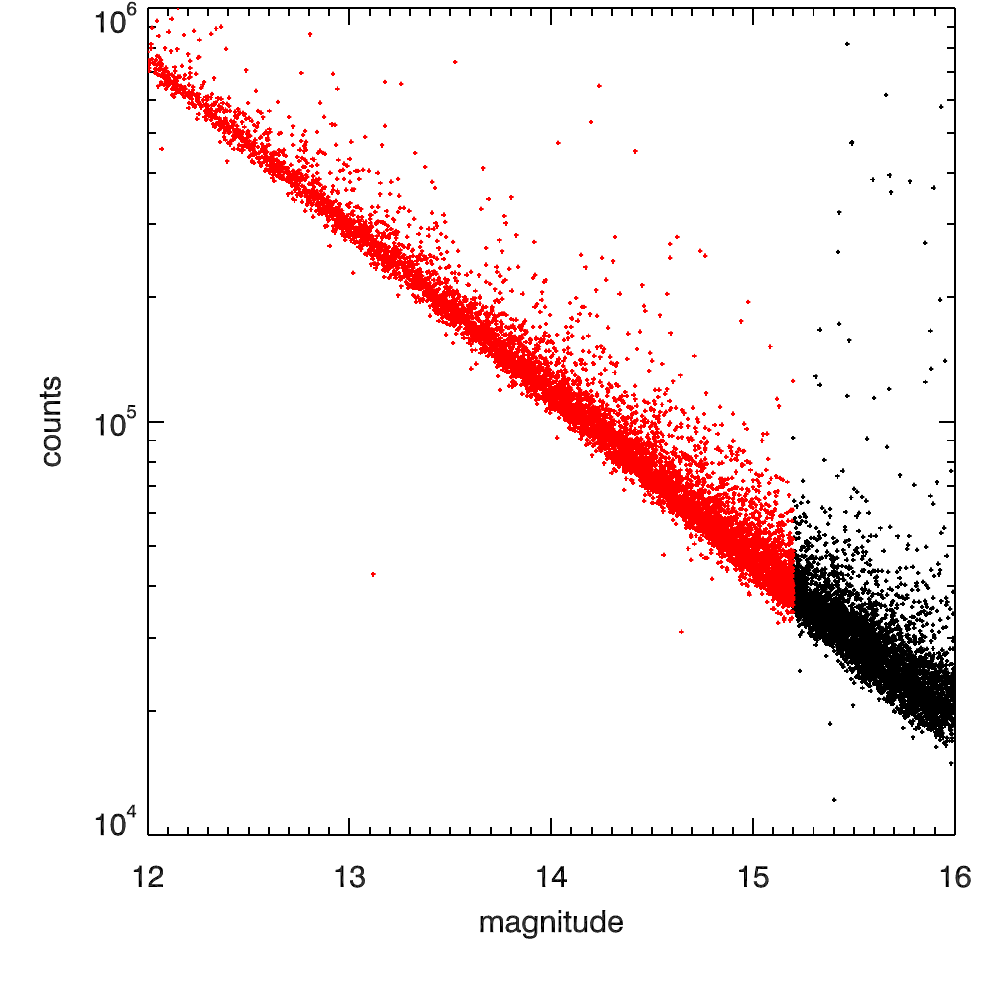}
  \caption{Mean R magnitude versus counts in a logarithmic scale for the run LRa01. The linear correlation is evident. Chromatic targets are shown in red, where the flux from all three channels was summed up.\label{fig:magmean}}
\end{figure}

After interpolation the mean flux of the lightcurve is normalized to 1. This is to overcome numerical problems, since using the original flux values which are in the order of $10^5$. An alternate method is the conversion to logarithmic values.
\begin{equation}
f' = -2.5x_1\log_{10}f + x_2,
\end{equation}
where $x_1 = 0.9674492$ and $x_2 = 26.3424511$. The value of $x_2$ marks the so called zero point. \citet{Mazeh2009} has shown that this is however dependent on instrumental effects. Both values were determined by applying a linear fit to  the magnitudes of all target-stars of the run LRa01 (Figure~\ref{fig:magmean}).

With $x_1 \approx 1$ and $x_2$ in the order of the galactic background the values look reasonable.

\subsection{Detrending}
Most of the light-curves show a clear more or less linear trend, that can be caused either by long term stellar activity or by instrumental effects. Especially for giants, the rotational periods are in the order of the length of one observation run. The most puzzling instrumental effect on the long runs of CoRoT is the aberration caused by the satellite's motion along the earth's orbit. The stellar positions on the CCD move in respect to the guide sensor, but as the imagette remains at a constant position, the flux varies between the color-channels.

Generally a second order polynomial in the form 
\begin{equation}
f'(t) = x_0 + x_1f(t) + x_2f(t)^2
\end{equation}
can be fitted and subtracted from the light-curve, where $f(t)$ is the flux. The fit must be applied on original data-points and not on those added by interpolation (see Section~\ref{sec:datapreparation}).

In rare cases, where the modulation by the stellar rotation is evident, a fit with a sinusoidal is applicable
\begin{equation}
f'(t) = a\cos\left(\frac{2\pi t}{\tau} +\phi\right)+k,
\end{equation}
with the period $\tau$, a phase shift $\phi$, an amplitude $a$ and an offset $k$. This fit can also be iteratively used for pre-whitening (Section~\ref{sec:prewhitening}). Using the cosine instead of a sine function has numerical advantages to avoid $\sin 0 = 0$, where the amplitude is undetermined.

\subsection{Filtering}
There are various approaches to separate the transit signal from any other signal caused by stellar activity or instrumental influences. For a run of 150~days we expect the transit signal to be shorter than 12~hours for a dwarf star. Therefore we apply a median filter that operates as a high-pass filter on the data. The width of the filter is 84~data-points where the median for a sliding window is calculated which contains 84~measurements. As a precondition, the measurements have to be equidistant in time. Otherwise, if large gaps occur, the median filter will be broader than necessary.

To remove spikes and single outliers the filter by \citet{Savitzky1964} is very efficient. Outliers will be replaced by a polynomial evaluated through the adjacent values. Usually 12 values (six before and six after the data-point) are interpolated by a second order polynomial. The filtering is applied by convolution of the filter kernel with the dataset.

\begin{figure}[htb]
\centering
\includegraphics{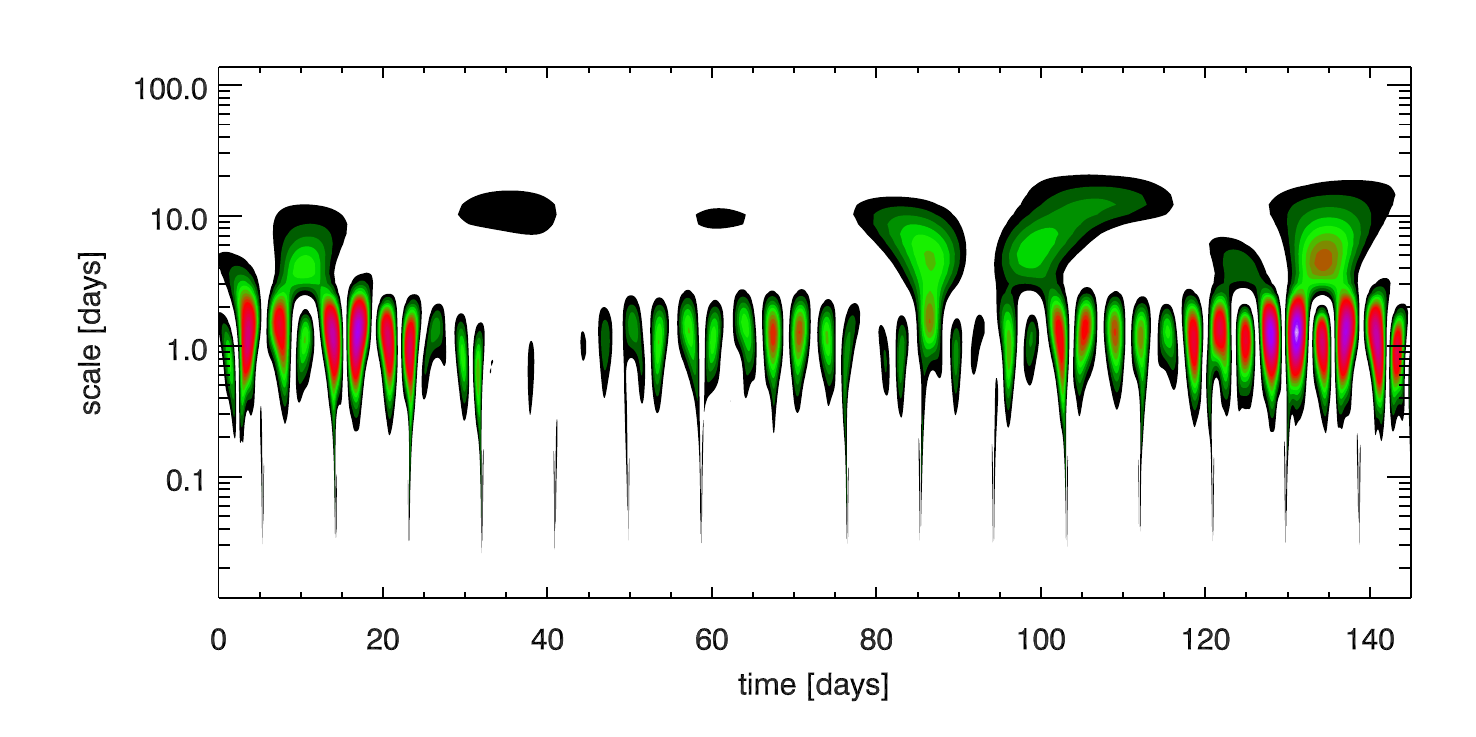}
\caption{Wavelet representation of CoRoT-6. The power of the signal is represented by differet colors. The wavelets in the domain from 0.05 to 0.2 represent the transit signatures.\label{fig:gaussianwavelet}}
\end{figure}

As a last optional filter stage, a low-pass Butterworth filter can be applied. Other filtering techniques are application of a Gaussian wavelet filter or a filter using the Haar-wavelet. An example of a wavelet representation can be seen in Figure~\ref{fig:gaussianwavelet}. A filter would cut a horizontal slice to isolate the transits.

Wavelets in principal are a powerful tool to remove noise. One should keep in mind that any filtering also deforms the transit signals, sometimes to an unknown extend.

\subsection{Transit Detection}
After the preparation and the filtering process the main signal that remains is (hopefully) transits with minor residual noise. The last task is the detection of transit-like signals. The most prominent parameter is the transit period (Section~\ref{sec:transitphysics}) that can be accomplished by calculating maximum of the autocorrelation function (Section~\ref{sec:analysis}).

At this stage it is sufficient to fit a trapezoid or a box-function to estimate the transit parameters like depth, duration and the impact parameter.

\section{Separation of Stellar Activity from Instrumental Effects}
\label{sec:ObservationalEffects}
While instrumental effects on ground based observations are not an issue, they have a big impact on space based missions. Most observational and instrumental effects on ground are the duty cycle of observation. Photometric measurements are interrupted by daylight, which imprints a spurious frequency signal corresponding to one per day or $11.5741\,\upmu$Hz. Other effects are the changing airmass during the observation, different seeing conditions and sporadic cosmic ray hits.

Space is a rougher environment for doing observations. The radiation level is much higher causing pixel alterations by cosmic ray hits as the main source of disturbances. Other variations derive from the satellites orbit and the altitude control.

\subsection{Instrumental Effects}
Instrumental effects can affect the whole image or specific pixels. We might separate effects on their impact on the measurements. For the whole discussion in this section we focus on space photometry, since ground based photometry suffers from other limitations like observational gaps due to weather conditions and or the day and night cycle. Even for observatories located at the South Pole \citep{Strassmeier2008} there are limits due to the length of the Arctic night and the stability of the conditions as well as local seeing conditions. 

Most important effects are 
\begin{itemize}
\item Spacecraft jitter
\item thermal noise
\item readout noise
\item cosmic ray hits
\end{itemize}

We must not forget optical effects either wanted as in CoRoT, where a prism extends the PSF to obtain color information, a Fresnel mask like in MOST to increase the signal-to-noise ratio by expanding the PSF. We also have unwanted or unavoidable effects like diffraction spikes from the spider holding the secondary mirror. The last one has a mayor impact on multi target photometry missions like Kepler and CoRoT. 
 
Bright stars in the observed field cause diffraction spikes affecting adjacent columns and rows of the star. This can lead to weird results where multiple targets share the identical period and epoch if the diluting star is a binary. Even the CCD-readout may produce correlated events along the readout direction.

\subsubsection{Jitter}
Jitter is caused by the spacecraft pointing. Since a satellite is a free floating object, its pointing has to be maintained by closed loop positioning system based on a star finder and reaction wheels. The pointing accuracy is in the order below one arcsecond. However the centroid of a star is slightly moved on the CCD where pixels do not have the same sensitivity. 

\subsubsection{Thermal noise}
The thermal noise is almost neglectable as it adds uniformly signal on the detector. Usually the detector is kept at a constant temperature. Ground based observations may suffer from temperature changes and introducing noise at different levels. The thermal noise is usually removed using a so-called dark-frame. It has the same exposure time as the image, but with the shutter closed.

Dark frames also include cosmic ray hits. Depending on the level of accuracy the number of dark frames is usually in the order of 20 or even higher for the mmag regime. For space based observatories dark-frames are taken in the calibration \& validation phase prior to the scientific part of the mission.

\subsubsection{Readout noise}
With todays electronics the readout noise is negligible. Generally the pixel-to-pixel variations have a larger impact on the data. Occasionally we can speak of pixel correlated noise where the signal is correlated with the position of the target on the CCD.

\subsubsection{Cosmic ray hits}
\begin{figure}[htb]
	\centering
		\includegraphics[width=0.9\textwidth]{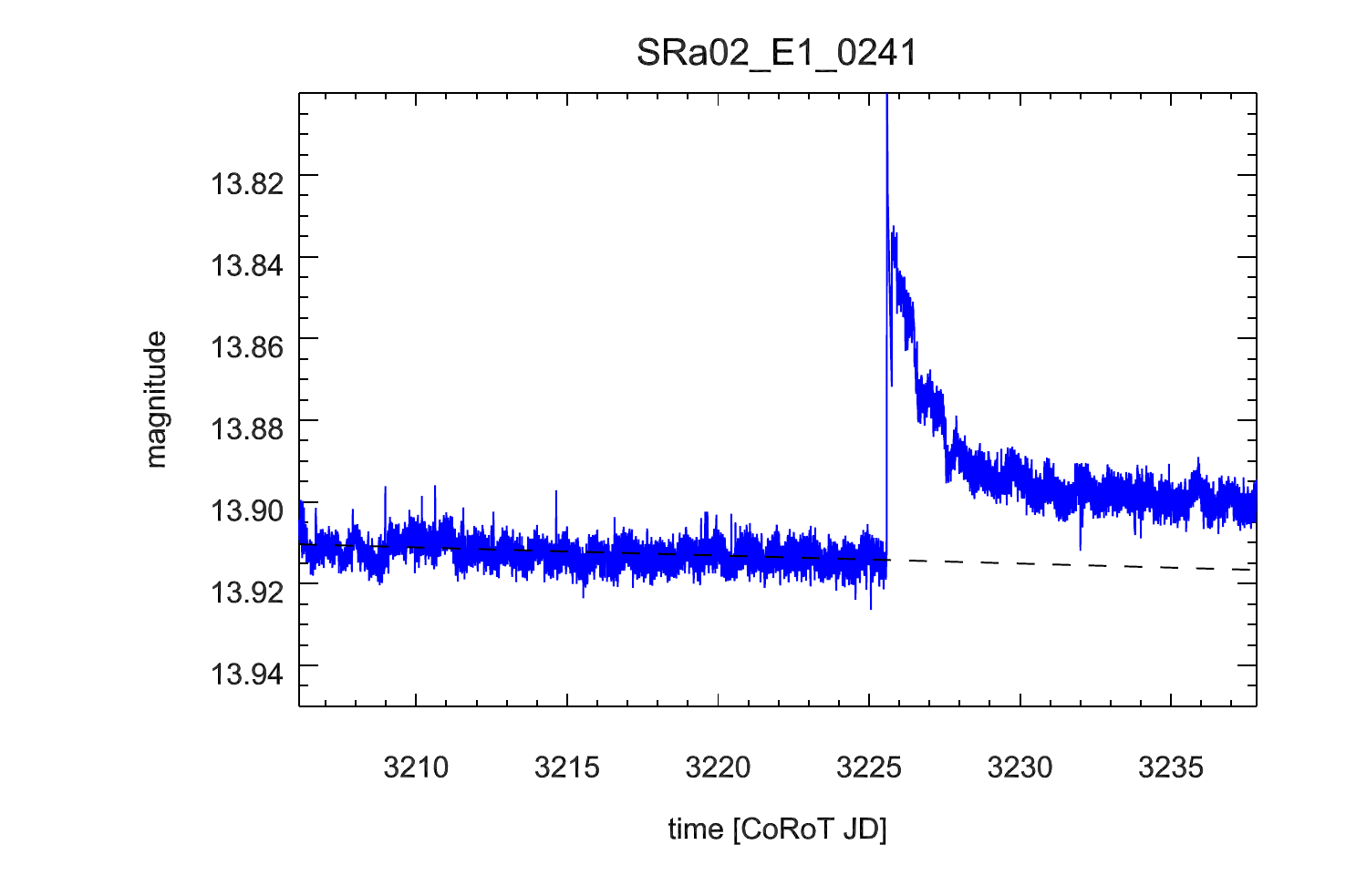}
	\caption{Cosmic ray hit in the blue channel of SRa02\_E1\_0241 at \corot JD $3225.62$. The other two channels remain unharmed. The linear trend (dashed line) shows an offset by almost $0.1$~mag after the hit, which indicates a permanent alteration of the CCD's physical structure. A typical exponential decay is visible.}
	\label{fig:cosmicray}
\end{figure}

Cosmic ray hits cause a local alteration in the CCD with different results. The most prominent effect is a sudden rise combined with a slow decay back to the previous level. The effect of the cosmic ray hit can be seen in Figure~\ref{fig:cosmicray}.

Unfortunately cosmic ray hits have the same shape as stellar flare, the only difference is its time scale. Usually the pixel alteration of a cosmic lasts longer than one day, whereas a flare lasts only hours. Also flares affect all color channels almost equally.

\subsubsection{Hot pixels}\label{sec:hotpixels}
Pixels on the detector with a constant (high) saturation level or a non-linear behaviour are called \textit{hot pixels}. If they occur on the border of a photometric mask, they will show abnormal activity due to spacecraft jitter which is often referred as \textit{telegraphic noise}.

\begin{figure}[htb]
	\centering
		\includegraphics[width=0.9\textwidth]{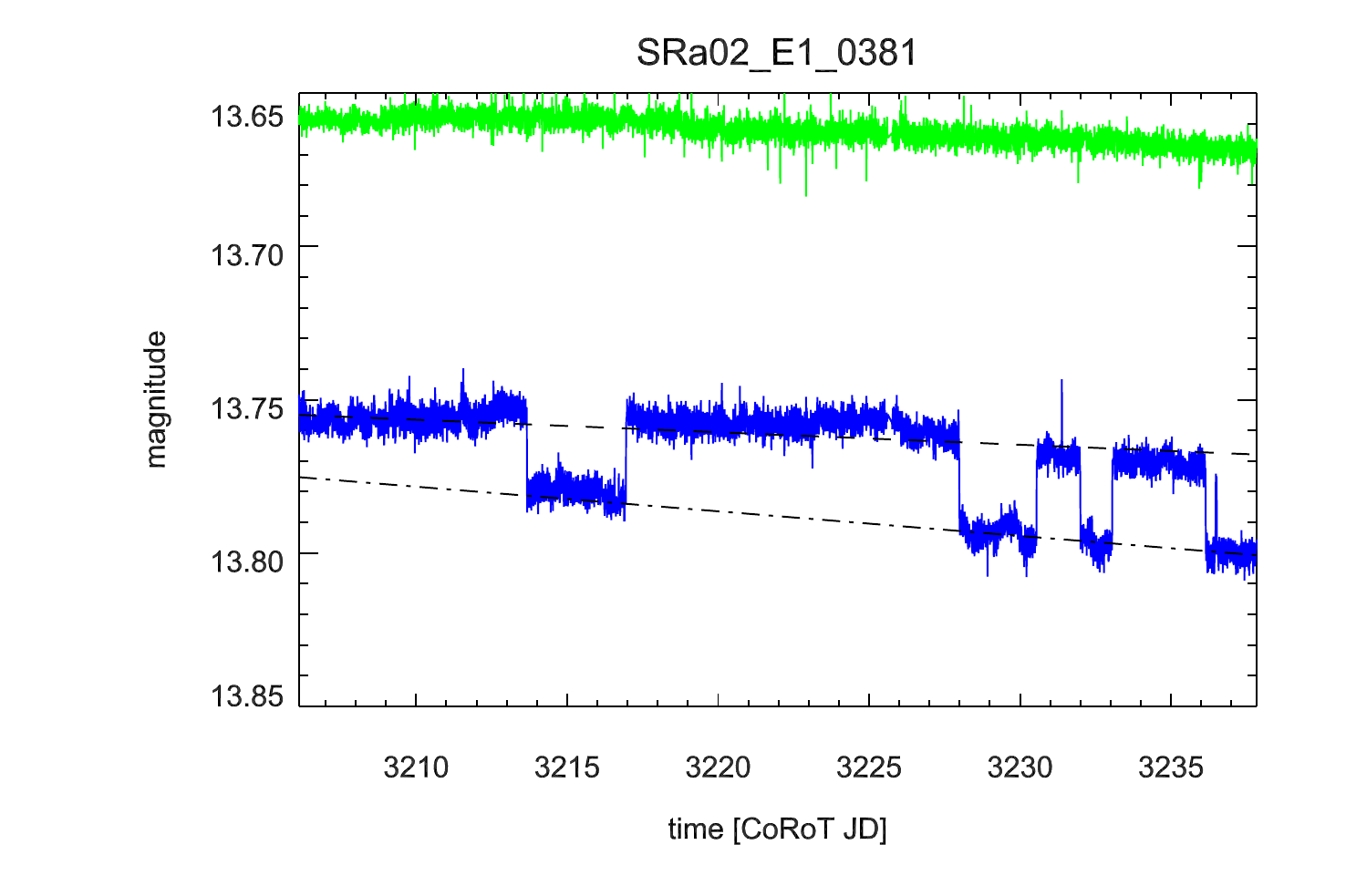}
	\caption{We see telegraphic noise occurring in the blue channel of the \corot lightcurve SRa02\_E1\_0381. The green channel remains unharmed. The jumps have a depth of 30~mmag. Spacecraft jitter causes constant jumps occurring quasi-periodic. In this case the photometric mask was moved over a hot pixel.}
	\label{fig:telegraphic}
\end{figure}

Telegraphic noise can occur quasi-periodic or even periodic and might be misinterpreted as a transit signal. At low signal-to-noise ratios, this is hard to distinguish. 

\begin{figure}[htb]
	\centering
		\includegraphics[width=0.9\textwidth]{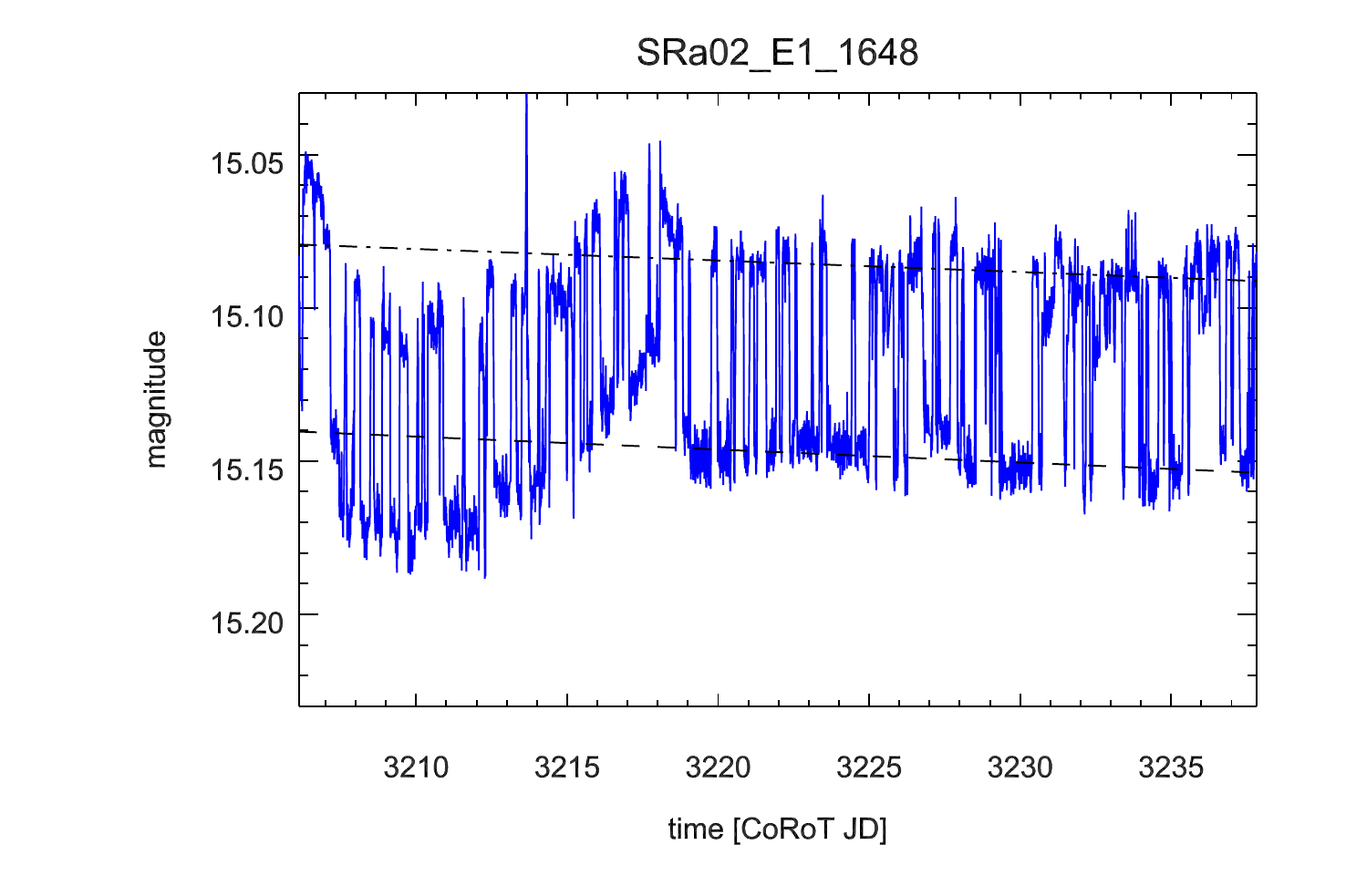}
	\caption{We see telegraphic noise occurring in the blue channel of the CoRoT lightcurve SRa02\_E1\_1648. The green and red channel are not affected. The jumps have a depth of 60~mmag. This lightcurve can not be repaired for the search for transit signals.}
	\label{fig:telegraphic2}
\end{figure}

We can categorize the effects in groups:
\begin{itemize}
\item affecting the whole image in the same manner by means of ADU and noise statistics.
\item signal depending on the pixel coordinates
\item random events like cosmic ray hits
\end{itemize}

\subsubsection{Image affecting noise}
The first one is also referred to as bias and can be corrected using calibration frames acquired by the spacecraft. But usually this frames are only taken at the calibration phase prior to the main mission. Calibration frames are not recorded during the ongoing mission since observation time is valuable on a spacecraft. The background signal is subtracted in die CoRoT N0-pipeline.

\subsubsection{Pixel correlated noise}
The second can be quantified using autocorrelation between stellar targets or correlation between pixel locations (stellar coordinates) and noise properties (Section~\ref{sec:noise}). The correlation may show itself in a small area or cluster of pixels on the CCD and therefore on stellar coordinates or correlate to the distance of a coordinate.

\subsubsection{Random pixel affecting events}
Events like cosmic ray hits cause a sudden change in the pixel ADU value. Whereas hot pixels cause a (almost) constant value.  

\subsection{How to quantify noise}
\label{sec:noise}
\begin{figure}[htb]
	\centering
		\includegraphics{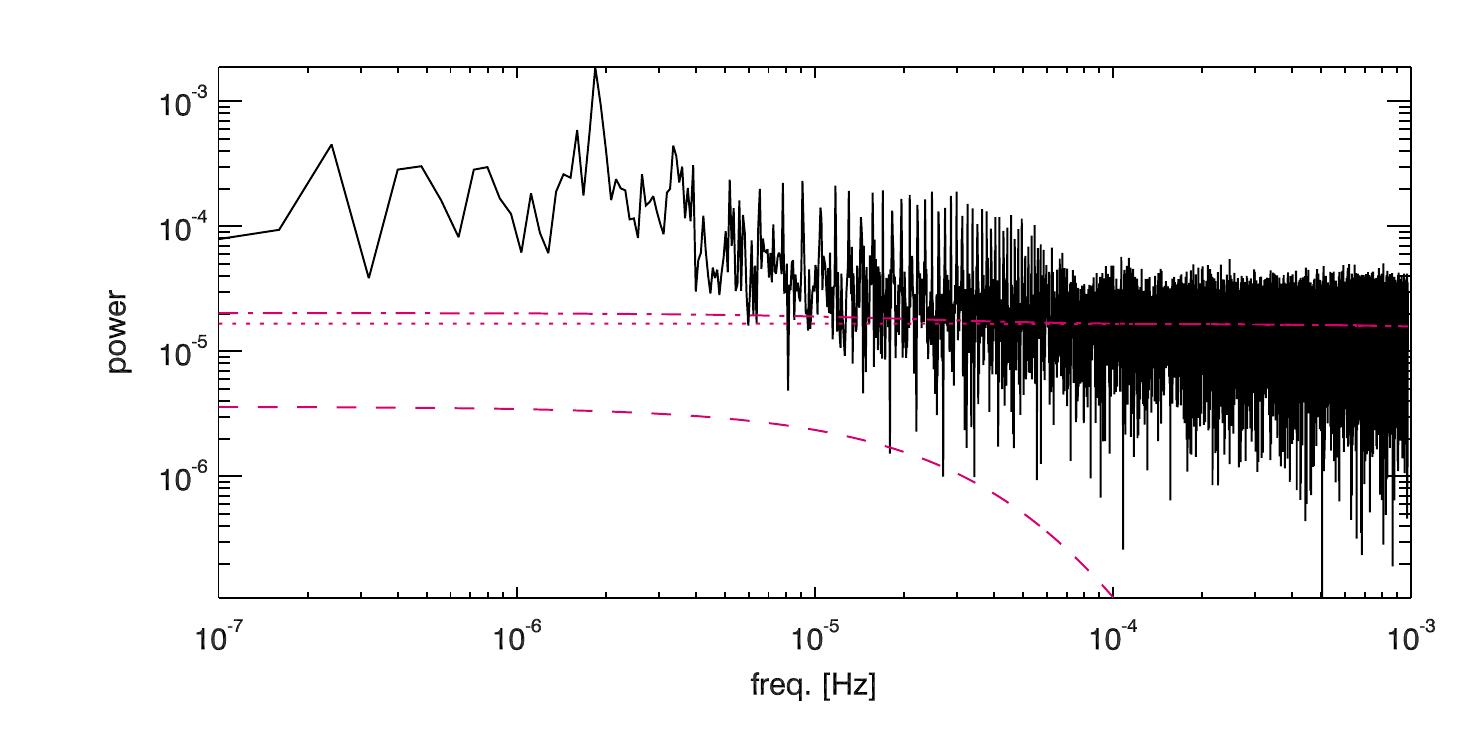}
	\caption{Power spectrum of CoRoT-6 with signal and noise levels estimated with a polynomial. The signal is indicated with a dashed line, the noise is indicated by a dotted line. Both signal plus noise are represented by the dashed dotted line. We see that the signal is underestimated, while the level of noise is somewhat accurate.\label{fig:wiener}}
	\label{fig:telegraphic}
\end{figure}

Usually noise is identified in the frequency spectrum as a constant background contributing to all frequencies equally. This is so called \textit{white noise}. It is easily identified and removed by Wiener filtering (Figure~\ref{fig:wiener}) or a Butterworth filter operating as a low-pass filter.

Other quantification methods are e.g. the \textit{mean absolute derivative} (mad) or median absolute derivative or the rms which exceeds a specific value to locally identify noisy events. Generally it is not easy to clearly identify noise. 

Stellar activity reveals itself as \textit{red noise} since activity is always limited to certain timescales like rotation periods, spot-lifetimes, granulation lifetimes, \dots. This amplifies low frequencies in different levels.

\subsection{Analysis of Stellar Signals}

Stellar variability can be separated into two main classes:
\begin{itemize}
\item intrinsic stellar variability, e.g. pulsation, flares, Novae
\item extrinsic stellar variability, e.g. binaries, transits of exoplanets, stellar rotation (spots).
\end{itemize}

Pulsation and rotation signatures are easily identified using either a Fourier spectrum or the autocorrelation function. Usually the latter is more robust. A sliding autocorrelation function with a certain window size (e.\,g. twice the stellar rotational period) can be used to observe transient features like stellar spots. The lifetime of spots and their appearance and disappearance causes a change in phase that might result in cancelling out the frequency of the true stellar rotation.

A semi-automatic approach to identify stellar variability classes in the HR-diagram has been developed and applied by \citet{Debosscher2009} on the CoRoT data.

Other non-periodic phenomena like flares could be identified by using a model which is cross-correlated with the data.

%
%
%
%
%
%
\chapter{Results and Discussions}
\section{Instrumental Noise}
\label{sec:instrumentalnoise}
\begin{figure}
\centering
\includegraphics{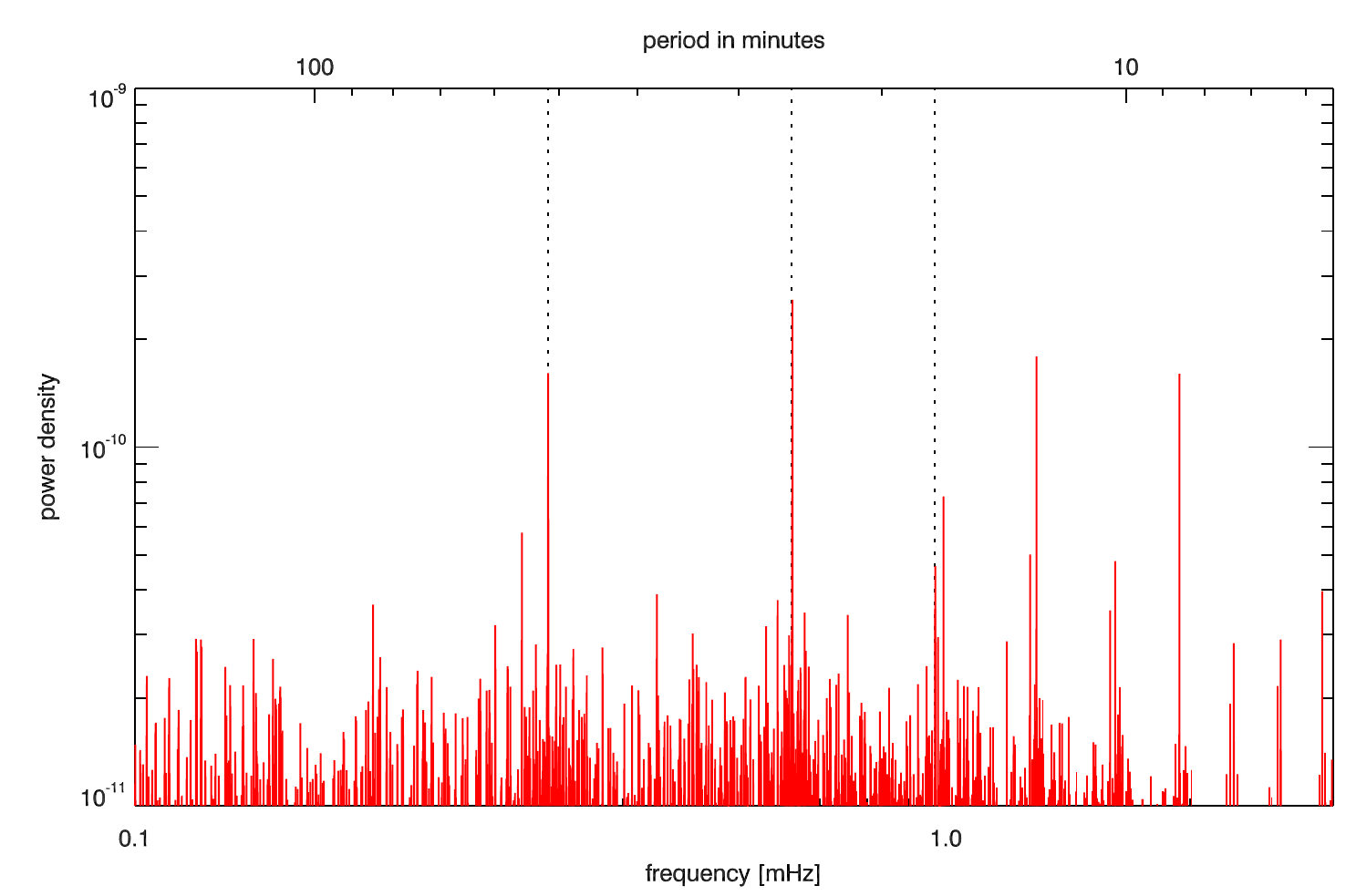}
\caption{High frequency part of the power spectrum of a F9V~star CoRoT LRa01\_E2\_0205. The orbital frequency of 0.323\,mHz and its harmonics are indicated by dashed vertical lines.\label{fig:orbfreq}}
\end{figure}

The CoRoT telescope was equipped with a unprecedented baffling system to avoid parasitic stray-light in the optical path \citep{Auvergne2009}. Nevertheless the orbital cycle of CoRoT and its disturbance in the SAA is visible in the data, which can be seen in Figure~\ref{fig:orbfreq}.

Though CoRoT is an exceptional example, the orbital cycle is not visible in general. It has been avoided by Kepler-mission in an Earth-trailing orbit. Smaller photometric satellites like MOST are more affected.

\section{Filtering}
\begin{figure}
\centering
\includegraphics{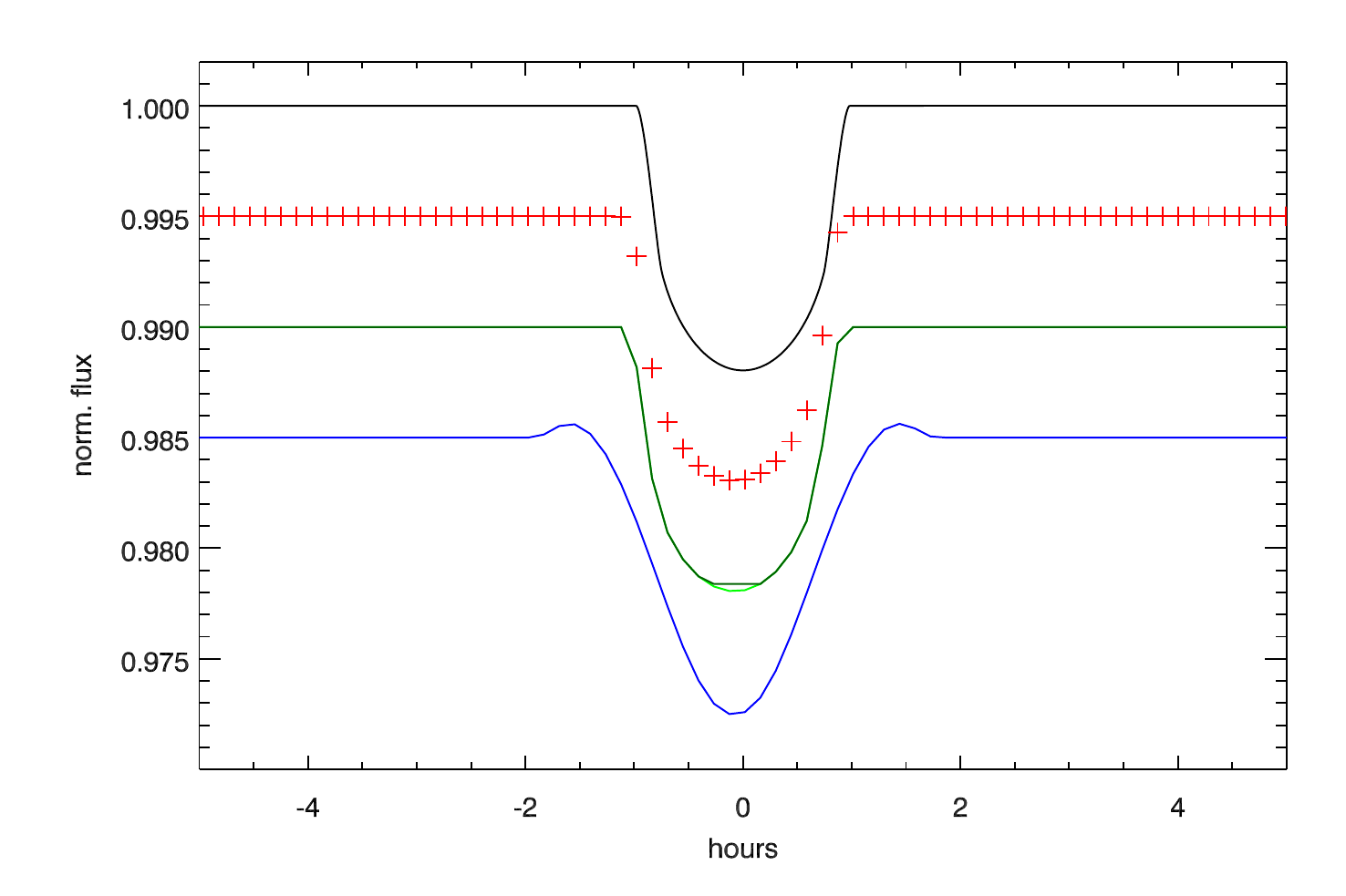}
\caption{A simulated transit (black solid) and the results of the filtering process. The red crosses indicate the 512\,second oversampled data, green after a median filter was applied and blue after applying \citet{Savitzky1964}.\label{fig:occfilter}}
\end{figure}

A variety of filters for enhancing the signal-to-noise ratio in lightcurves was discussed in Section~\label{sec:filteringmethods}. As stated earlier filtering has to be applied carefully without harming the transit-signal. A result of the filer-pipeline is shown in Figure~\ref{fig:occfilter}.

A planetary transit in a five day orbit was simulated with the transit duration in hours. The original sampling was 32~seconds corresponding to CoRoT's high cadence sampling rate. The original series consisted of 13\,500~points. Generally the filtering process does not depend on the shape of the transit itself, so an `average' transit was chosen with $p=0.1$, $b=0.5$ and $\gamma_1=0.75$, $\gamma_2=0$. The data was downsampled to 512~seconds leaving 844 points. The general shape of the transit remains unharmed. A slight shift appears since the datapoints are rebinned to the beginning of the sampling time. 

The median high-pass filter was applied with a width of 84 datapoints, which did not change the transit at all. Using the median filter as a low-pass, the flat part of the transit was a little bit more enhanced (dark green in Figure~\ref{fig:occfilter}).

At a last stage the filter by \citet{Savitzky1964} was applied. This filter is essentially to remove outliers that have not been flagged as invalid datapoints in the pipeline before. This filter has a severe impact on the transit transforming it into a Gaussian. Ingress and egress are slightly enhanced.

\section{Solar Activity}
The sun may act as a template for stellar activity, but we have to keep in mind that the sun, as we see and measure it now represents a star at the age of 4.5\,Gyr and a spectral type G2V. It is difficult to scale solar activity on photometric levels to stellar activity. An attempt so simulate transits at solar like stars has been conducted by \citet{carpano2008}.

Strong flare events that radiate in the continuum are called  ``white-light flares'' in contradiction to the usual flare event that can be observed in H$\alpha$. Even moderate C-class flares contribute to the TSI \citep{Kretzschmar2011}.

\begin{figure}
\centering
\includegraphics[width=0.75\textwidth]{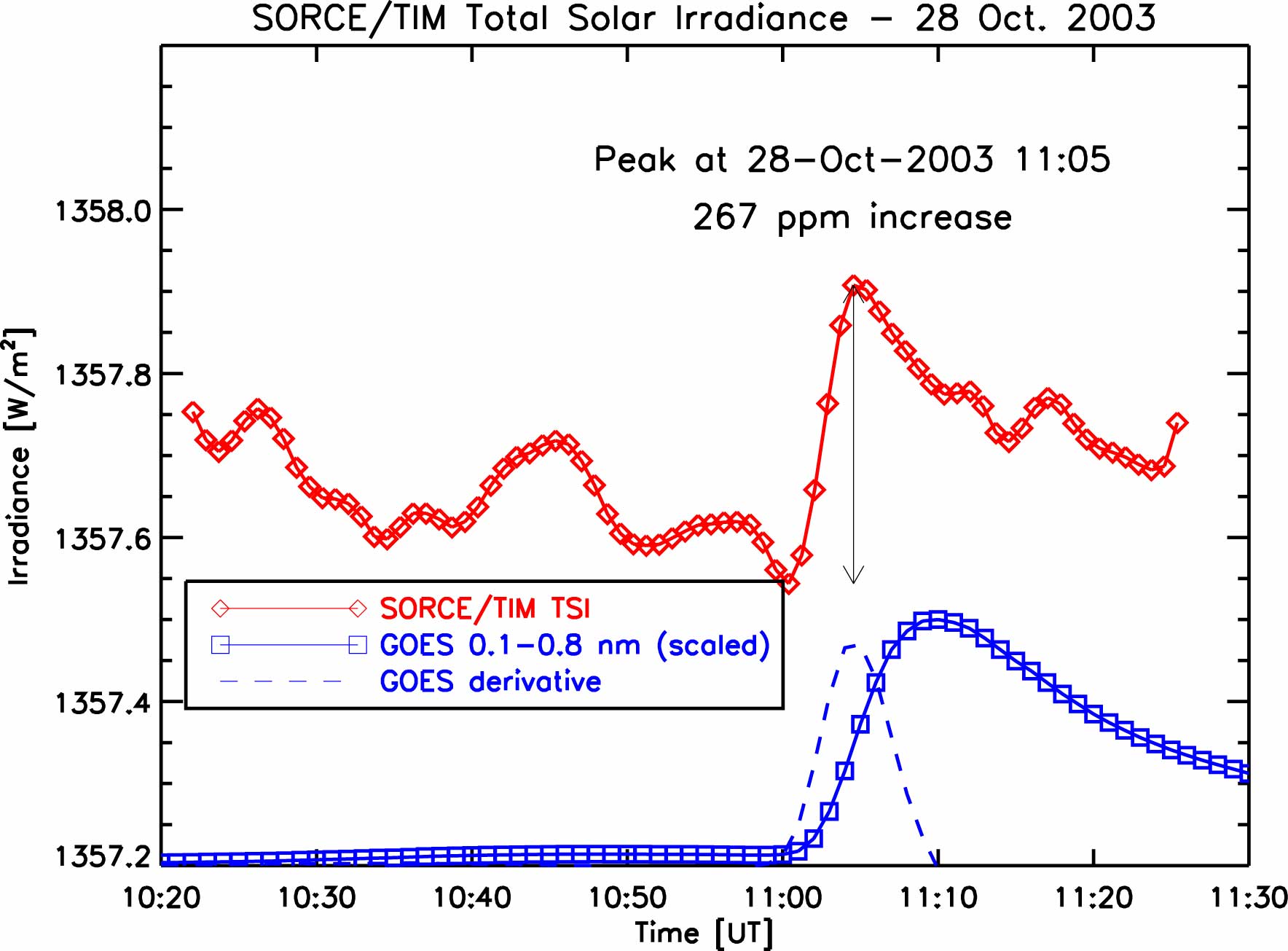}
\caption{First white-light flare observed with a TSI (red diamonds) instrument. The NOAA/GOES X-ray flux is shown for comparison (blue squares).\citep{Kopp2004,Woods2005}\label{fig:haloween}}
\end{figure}

The Haloween event on 28~October 2003 was an exceptional event with an X17-class flare. The flare, that was also visible in the continuum increased the TSI for 265\,ppm \citep{Kopp2004,Woods2005} which can be seen in Figure~\ref{fig:haloween}\footnote{Image source:~\url{http://spot.colorado.edu/~koppg/TSI/}}.

\subsection{Solar Irradiance}
\begin{figure}
\centering
\includegraphics{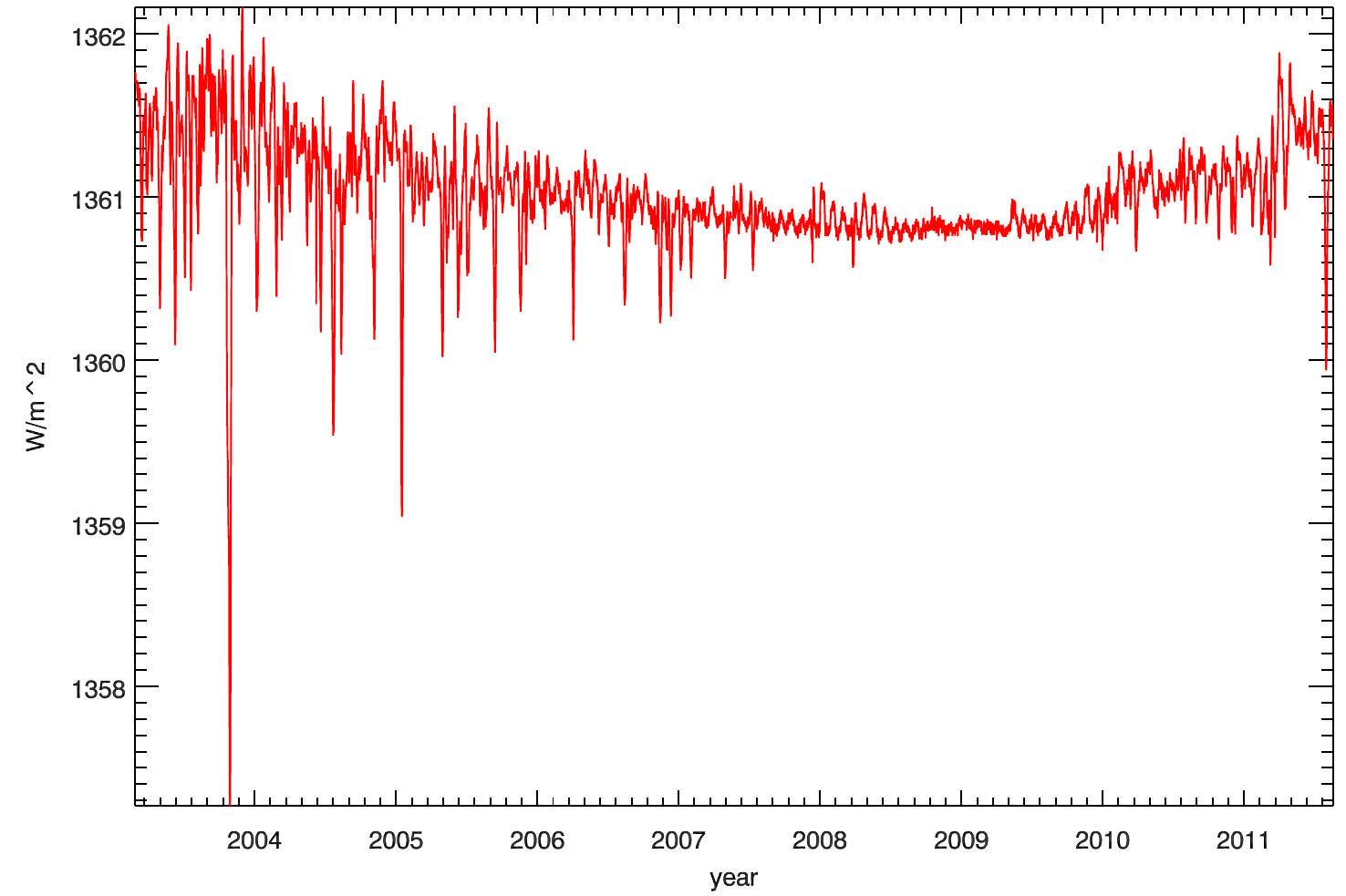}
\caption{Total Solar Irradiance measured with the TIM/SORCE instrument with a 6~hour cadence. The data covers a little less than a solar cycle. High activity signals correspond to a higher output of the sun. The largest event in 2003 was caused by a solar spot\label{fig:tsi}}
\end{figure}

\begin{figure}
\centering
\includegraphics{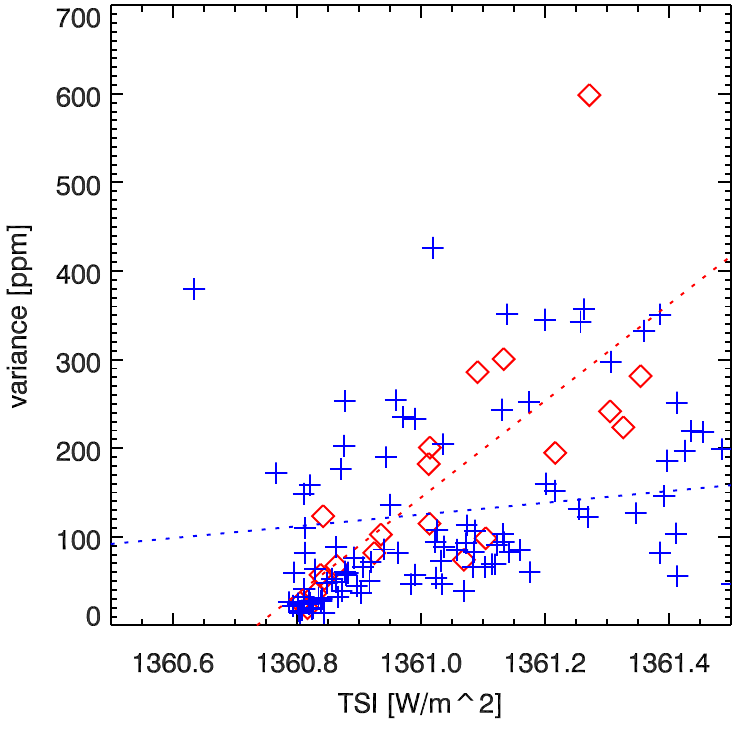}
\caption{Correlation between the Total Solar Irradiance measured with the TIM/SORCE instrument and its variance over a timescale of 150~days (blue) and over 30~days (red). The lengths have been chosen in accordance with the CoRoT observational run lengths.\label{fig:tsidev}}
\end{figure}

The Total Solar Irradiance (TSI) varies with the solar magnetic cycle (Figure~\ref{fig:tsi}). The output variation correlates with the total output and the phase of the solar cycle (Figure~\ref{fig:tsidev}). 

Space missions like CoRoT and Kepler don't have observational data that covers a full stellar cycle, other photometric missions like e.\,g. \textsc{Hipparcos} do not have sufficient datapoints to identify a periodic behaviour. Since the TSI is available with a cadence of 6~hours, the CoRoT-data can be resampled to be comparable with the solar activity. Since LRa01 is one of the longest runs observed by CoRoT, this run with its 150~days is the ideal testbed to compare to solar irradiance.

\subsection{Solar Rotation}
The mean synodic rotation period of solar sunspots is $27.2753$~days, which was defined by Carrington \citep[p.~278]{Stix2002}. The rotational signal is not clearly evident in the Fourier spectrum of the TSI. This fact is caused by the lifetime and the distribution of the solar spots.
\subsection{Solar Oscillations}
Solar low-degree p-modes are centered around 3\,mHz. Their energy does not exceed 2.5\,ppm \citep{Finsterle2001}. Generally they can be observed in the domain from 2.1\,mHz to 3.9\,mHz with lowest power at the limits between 0.3\,ppm and 0.1\,ppm \citep{Stix2002}.

\section{Stellar Activity}
\subsection{Stellar noise}
A detailed discussion of noise properties in the CoRoT data is discussed by \citet{Aigrain2009}, where the variability for some observed targets reaches $0.5$~mmag for hours timescales. This behaviour can be correlated to the activity of giant stars. On the other side dwarf stars show median correlated noise levels over 2~hours that reach 0.05~mmag at the bright end.

\begin{figure}
\centering
\includegraphics{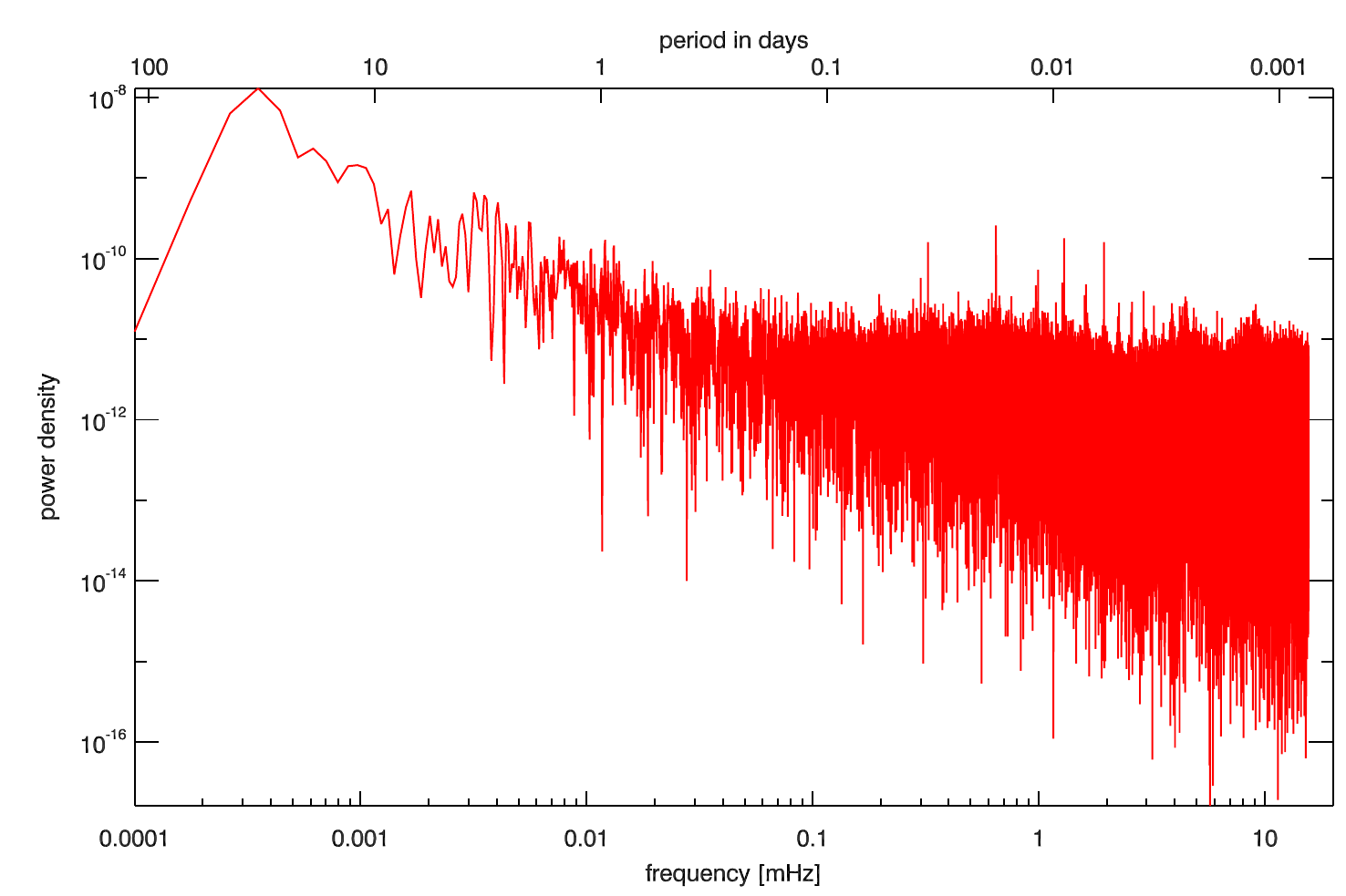}
\caption{Power spectrum of a F9V star CoRoT LRa01\_E2\_0205. The spectral power decreases from low frequencies to high frequencies. A small peak at the one day period is visible, as well as stronger peaks for the orbital period at 0.323\,mHz and its harmonics \citep[see also][]{Ballot2011}. The granulation noise is visible between 0.1 and 1\,mHz. The p-mode oscillations are centred around 4\,mHz.\label{fig:rednoise}}
\end{figure}

The noise level of the R-magnitude of dwarfs increases with the length of the mission, which is expected from sensor degradation. In general we observe red noise. An example of red noise is shown in Figure~\ref{fig:rednoise}.

In order to compare the stellar activity levels with the activity of the sun, the long run LRa01 was investigated. The rms levels of this run are comparable with other runs, with a slight increase caused by sensor degradation.

The light-curves for the run were rebinned to 6~hours using the drizzle algorithm (Section~{sec:rebinning}). The relative average standard deviation was calculated, that is naturally magnitude dependent since the photon noise contributes stronger on weaker fluxes. To get magnitude independent measurements the level of photon noise ($\sim1/\sqrt(N)$) was subtracted.

To get the noise properties for each spectral class the stars were grouped by their color temperature, putting five subclasses in each group. The minimum, median and maximum rms in each bin was calculated with the results shown in Table~\ref{tab:stellarrms}. The minimum noise level is $144\pm38$~ppm for all classes which contains granular noise and a not definable instrumental noise. This might explain the higher levels for late type stars, which are at the fainter end.

\begin{table}[ht]
\centering
\captionabove{Spectral types of dwarf stars taken from \citet{Cox2000} with their effective temperatures. The minimum and median rms are given in ppm, the maximum rms in percent. The values for the spectraltype K is poised by binary stars.\label{tab:stellarrms}}
\begin{tabular}{lrrrr}
\hline \hline
Spectral Type & \Teff & min. rms & median rms & max. rms\\ \hline
A5V & 8\,180 & 123 & 268 & 1.62\\
F0V & 7\,300 & 101 & 394 & 1.69\\
F5V & 6\,650 &  92 & 526 & 1.95\\
G0V & 5\,940 & 121 & 465 & 6.79\\
G2V & 5\,790 & 145 & 450 & 1.46\\
G5V & 5\,560 & 104 & 349 & 2.38\\
K0V & 5\,150 & 160 & --  & --\\
K5V & 4\,410 & 182 & --  & --\\
M0V & 3\,840 & 192 & 420 & 2.32\\
M2V & 3\,520 & 193 & 213 & 3.37\\
M5V & 3\,170 & 175 & 310 & 1.28\\
\end{tabular}
\end{table}

\begin{figure}
\centering
\includegraphics{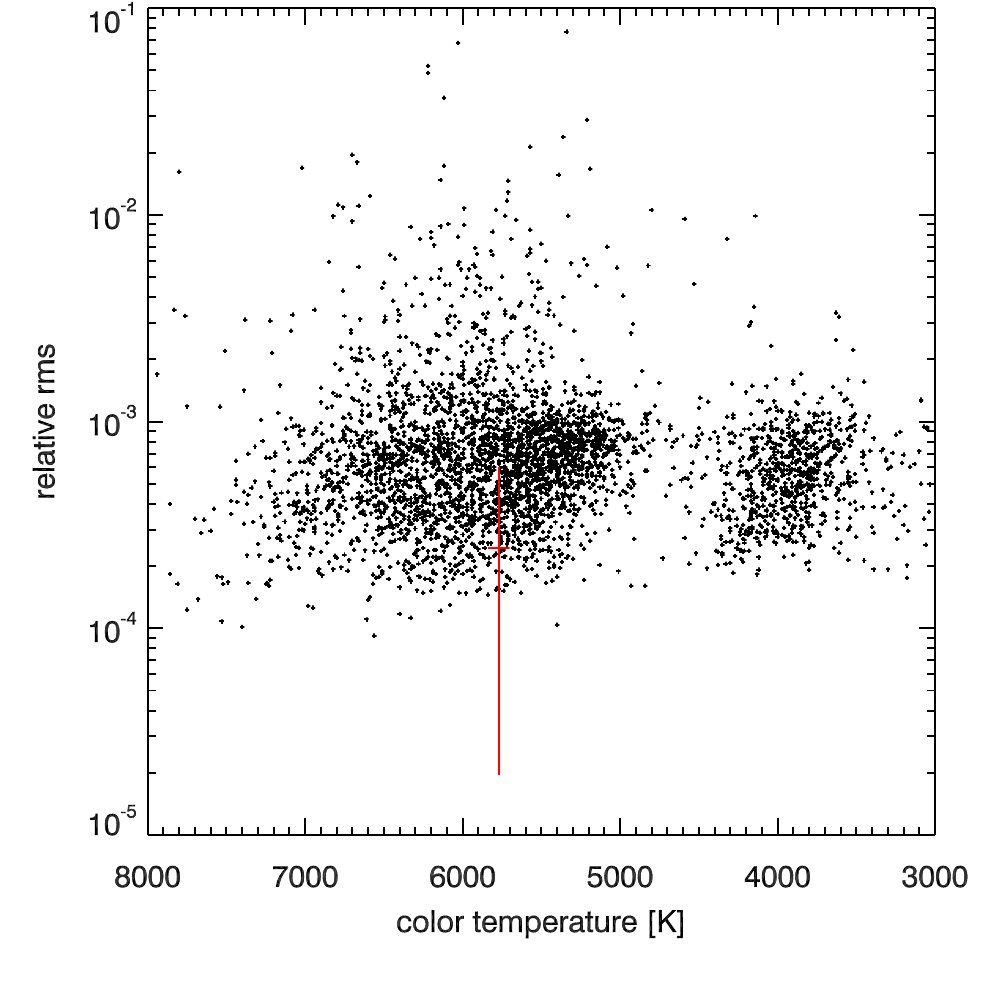}
\caption{Stellar noise levels within 6~hours on the CoRoT run LRa01. The solar noise level is shown in red indicating the minimum and maximum noise signal during the solar cycle.\label{fig:stellarrms}}
\end{figure}

The mean activity level of all spectral types is $377\pm100$~ppm, which is more than the average sun at $244.13$~ppm, which is indicated in Figure~\ref{fig:stellarrms}.

\subsection{Stellar Rotation}
\begin{figure}
\centering
\includegraphics[width=0.75\textwidth]{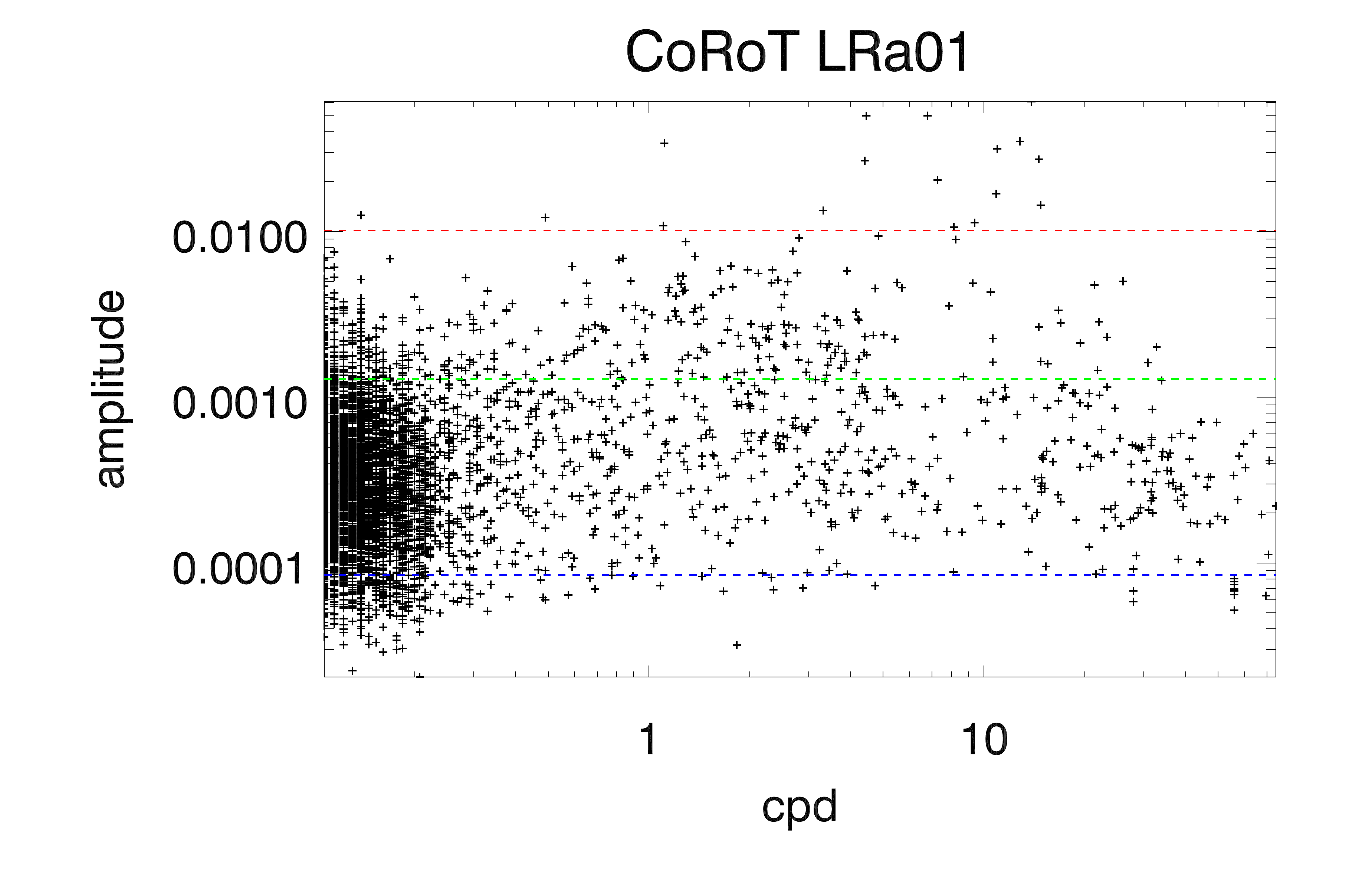}
\caption{Periods in cycles-per-day (cpd) versus amplitude of all stars in the CoRoT run LRa01. The red line corresponds to a transit depth of Jupiter, the green line corresponds to a Neptune and the blue line to an Earth-sized transit depth.\label{fig:LRa01cpds}}
\end{figure}

The long run of LRa01 with its 150~days provides an observational window of sufficient length to observe stellar rotation signals. In Figure~\ref{fig:LRa01cpds} all stars (including giants) were probed for ther dominant frequency in the Fourier spectrum which is assumed to be either from rotational or pulsational origin.

We see a pileup of rotations/pulsations for more than 3\,days ($\sim0.3$\,cpds). Generally we would expect the amount of pulsators among these stars to be rather low.

\subsection{Stellar Oscillations}
\begin{figure}
\centering
\includegraphics{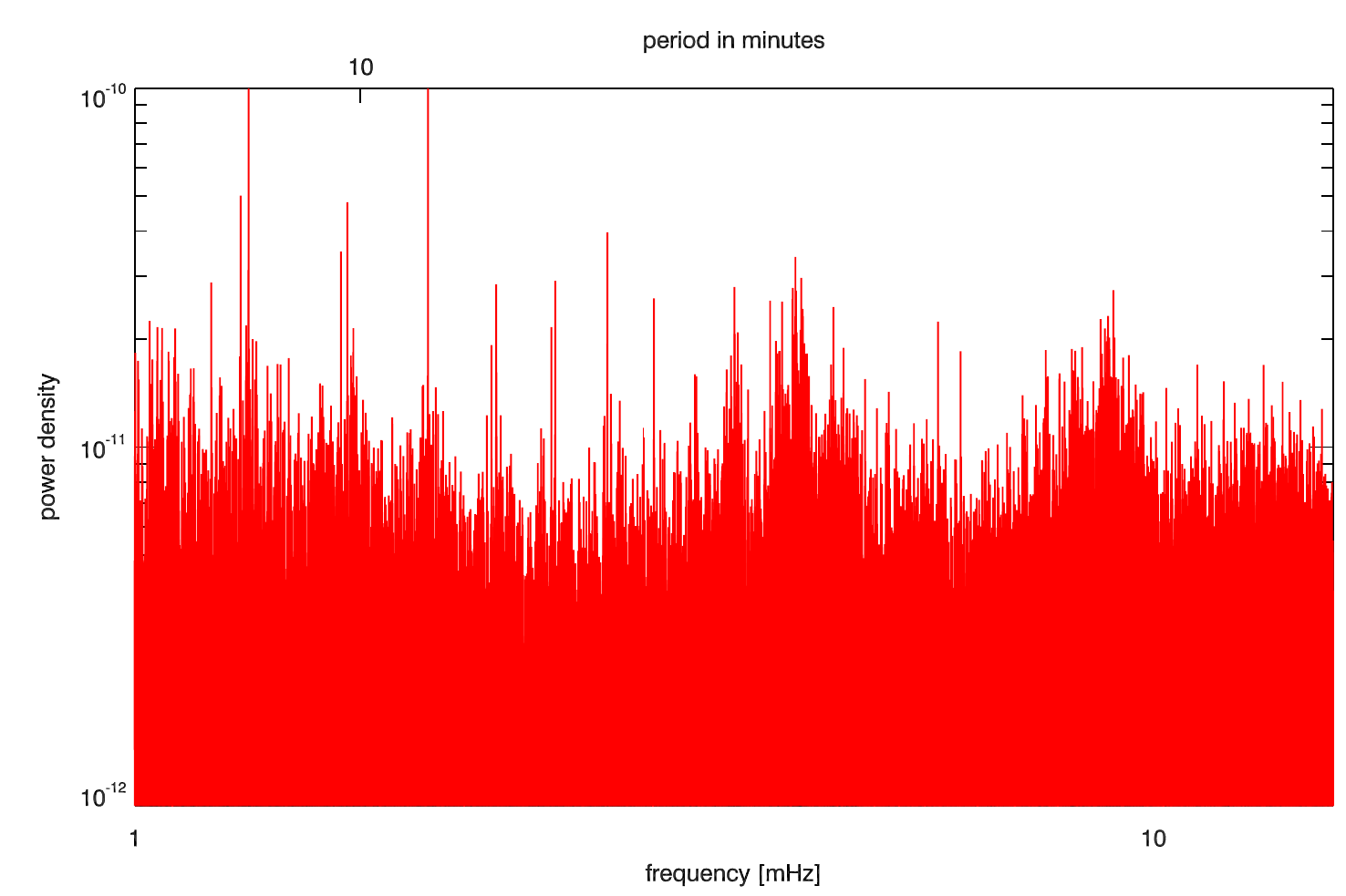}
\caption{Partial power spectrum of a F9V star CoRoT LRa01\_E2\_0205. The spectral power decreases from low frequencies to high frequencies. A small peak at the one day period is visible, as well as stronger peaks for the orbital period at 0.323\,mHz and its harmonics \citep[see also][]{Ballot2011}. The granulation noise is visible between 0.1 and 1\,mHz. The p-mode oscillations are centered around 4\,mHz\label{fig:pmodes}}
\end{figure}

Stellar oscillations of a F9V star can be seen in Figure~\ref{fig:pmodes}. We have to clarify that these measurements are form the so called exo-field of CoRoT which is not intended to be used for precise measurements of stellar oscillations. Nevertheless a sampling frequency of 
31.25\,mHz is sufficiently high to show first hints of stellar oscillation signatures.

\section{Exoplanets around Active Stars}
\label{sec:exoplanetsandstars}
In this section, case-studies on several discovered exoplanets that transit active stars are performed, as well as a hypothetical body with two Earth-radii transiting the sun will be investigated.

\subsection{A Super-Earth transiting the sun}
\begin{figure}[htb]
\centering
\includegraphics[width=0.5\textwidth]{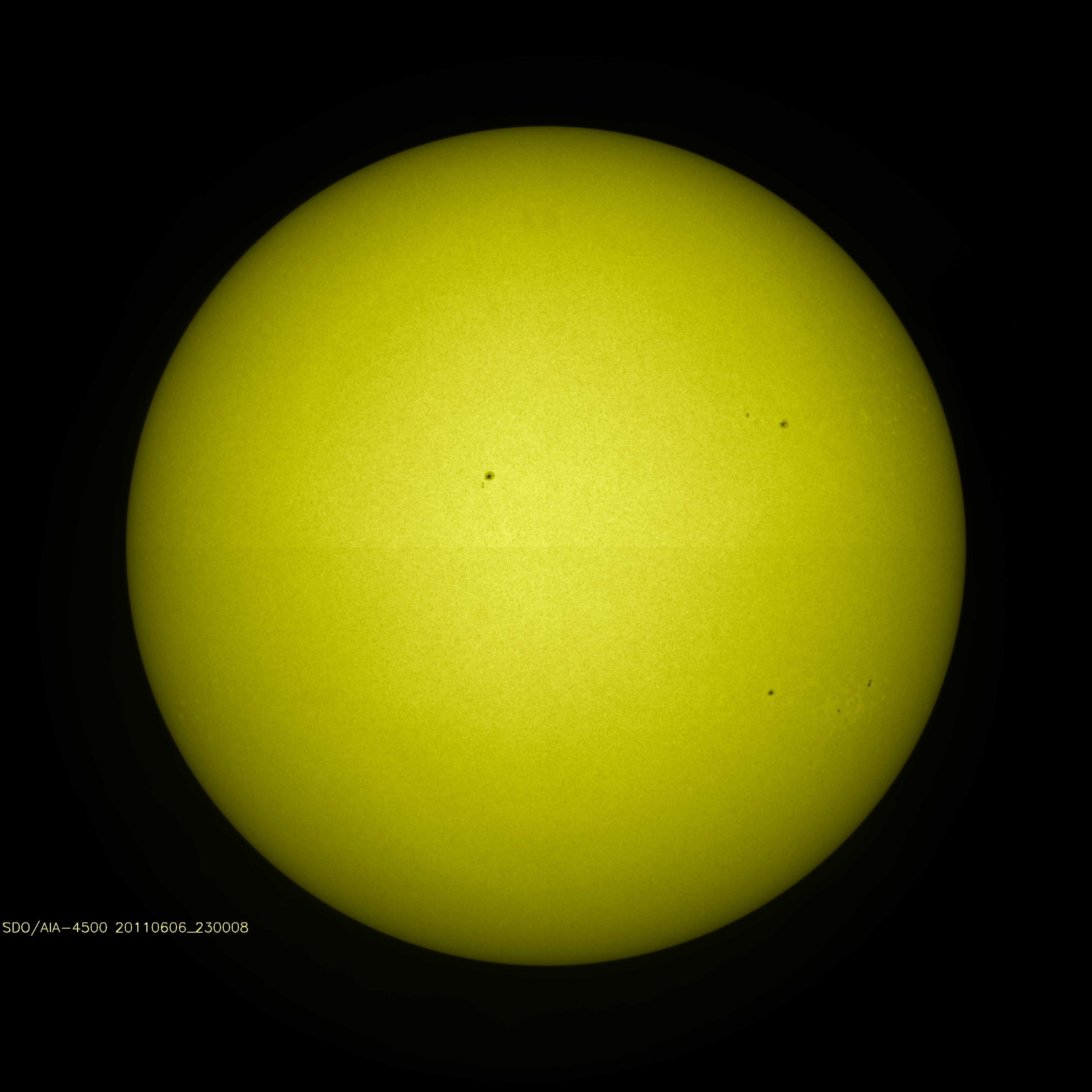}
\caption{Solar image form 6 June 2011 taken with the SDO/AIA instrument at a wavelength of 450\,nm. Four different quadrants are visible in the image, which might be caused by instrumental effects.\label{fig:sun4500}}
\end{figure}

The sun is the only star so far that can be spatially resolved. It is hence a valuable target to perform several simulations to gain knowledge about surface features. High resolution images from the NASA SDO data archive\footnote{\url{http://sdowww.lmsal.com/}} are used as the image source. The continuum wavelength of 450\,nm provides a substitute for a broadband white light image. The daily image of 6 June 2011 as shown in Figure~\ref{fig:sun4500} was chosen for the simulations. The solar image shows moderate activities.

The diameter of the sun was measured from the image with $3145.45$\,pixels. The total image has a height and a width of $4096 \times 4096$\,pixels. The radius in pixels of the transiting Super-Earth with $2R_\mathrm{Earth}$ was determined $28.81$\,pixels. All calculations and the transit-simulation were performed in subpixel accuracy.

\begin{figure}
\centering
\includegraphics{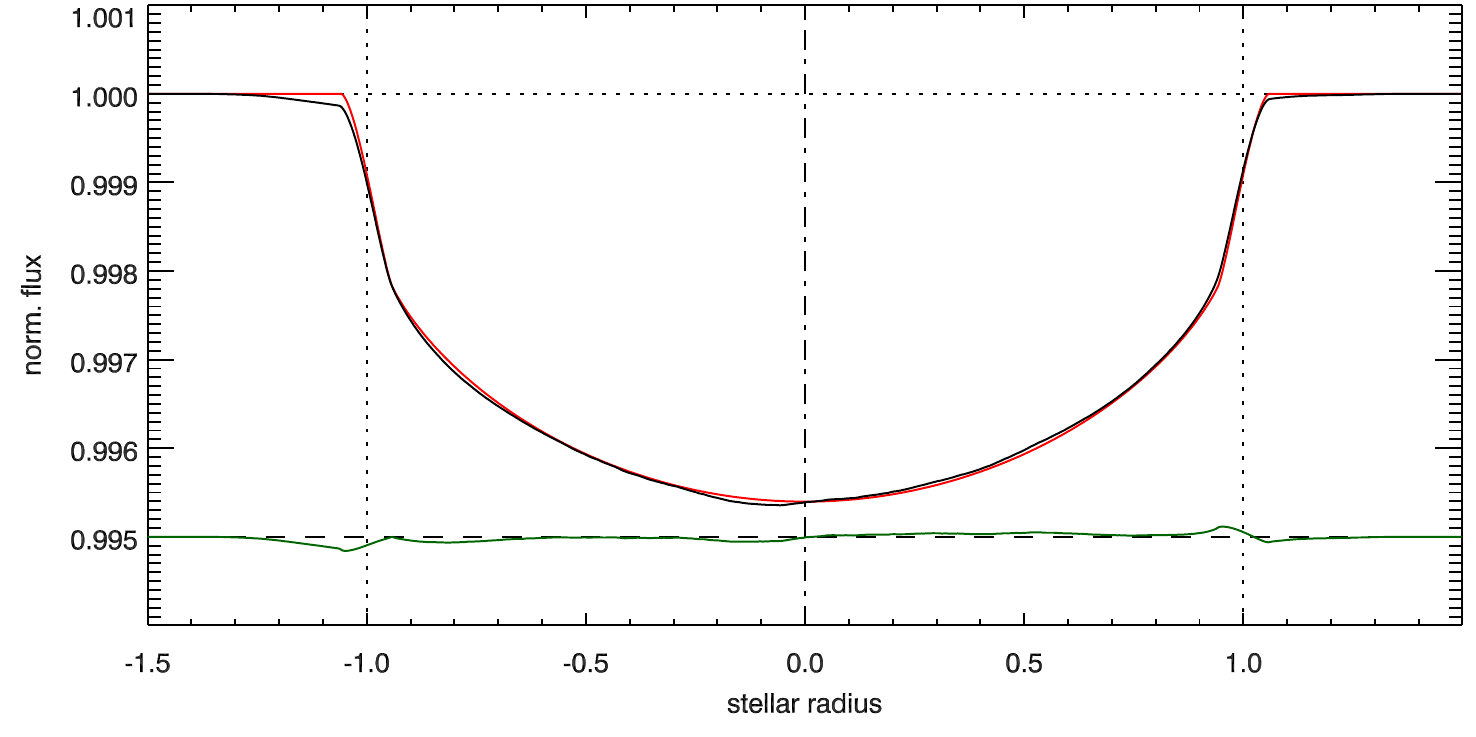}
\caption{Simulation of a transiting planet with two Earth radii. The black sold line is the simulated transit measurement. The red line is the fit of a model by \citet{Mandel2002}. The solid green line are the residuals of the fit.\label{fig:earth2}}
\end{figure}

For the transit-simulation itself, the planet was shifted pixel-by-pixel over the solar image, with the integral flux being measured and normalized with the integral flux of the empty image as seen in Figure~\ref{fig:sun4500}. The result of this simulation can be seen in Figure~\ref{fig:earth2}.

The fit was performed using the Levenberg-Marquard algorithm of the \citet{Mandel2002} algorithm. The parameters of the fit and its results can be seen in Table~\ref{tab:earth2fit}.

\begin{table}[ht]
\centering
\captionabove{Parameters for fitting the model by \citet{Mandel2002}. The initial values represent the a priori parameters for the simulation. The limb darkening parameters were taken from \citet[p.~162]{Stix2002}. The errors are the reduced $\chi^2$ values derived from the fit.\label{tab:earth2fit}}
\begin{tabular}{lrrrr}
\hline \hline
parameter & initial value & fit & error\\ \hline
b & 0.0 & $-0.00147472$ &  0.011713614\\
p & 0.068 & 0.0580131 & 1.4103004e-005\\
$\gamma_1$ & 0.6 & 0.890749 & 0.0042045146\\
$\gamma_2$ & 0.0 &$-0.165174$ & 0.0066625388\\
\hline
\end{tabular}
\end{table}

\begin{figure}
\centering
\includegraphics{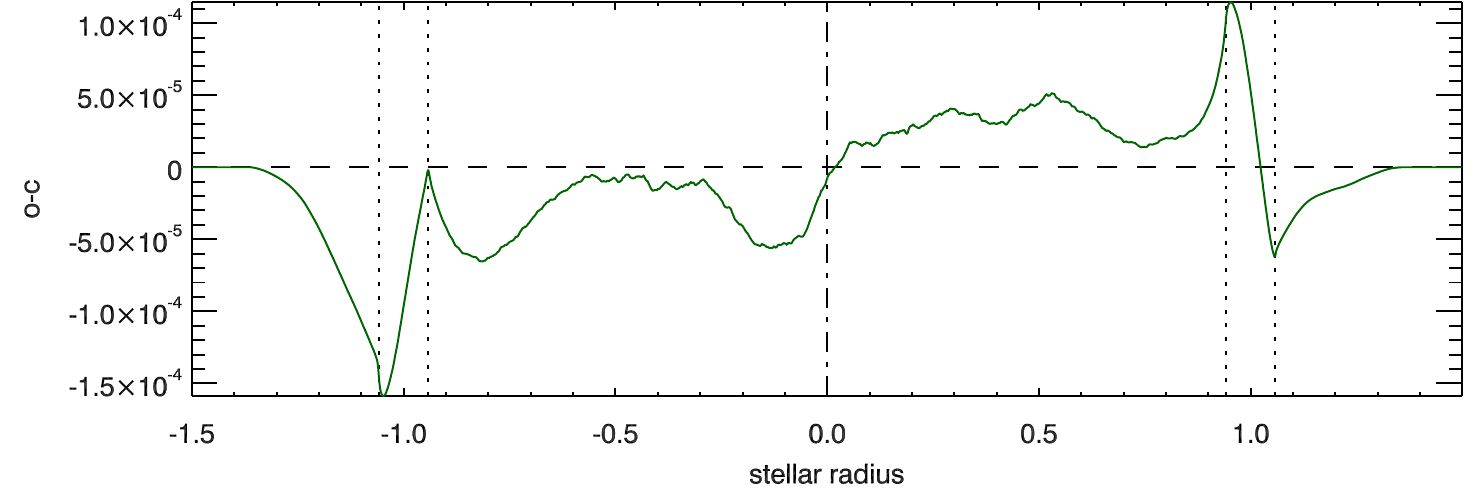}
\caption{Enlarged residuals from Figure~\ref{fig:earth2}. The diameter of the planet is represented of the vertical dashed lines. The center of the star is indicated by the vertical dashed dotted line.\label{fig:earth2oc}}
\end{figure}

If we constrain the limb-darkening parameters to be within $[0\dots1]$  and also the impact parameter to be of positive value, we get $b \sim 0\pm0.015$, $p = 0.0580248\pm0.00001666$, $\gamma_1 = 0.79096\pm0.00147$ and $\gamma_2 = 0$. The limb darkening is far from the theoretical value of $\gamma_1=0.6$ given in \citet{Stix2002}. The initial value for $p$ was estimated from $\sqrt{\delta F}$ which is slightly overestimating, due to the contribution of the limb darkening. A detailed view of the residuals is shown in Figure~\ref{fig:earth2oc}.

Interestingly we see out-of-transit variations, especially before the transit at $R_\ast\sim-1.1$. This feature can be explained by instrumental effects, e.\,g. insufficient baffling, or we see traces of the K~corona, which is also reasonable. The major error results from the imperfect limb darkening function that is unable to describe the limb correctly. The second largest error results from the quadratic limb darkening model which is also insufficient. A fifth degree polynomial is necessary to describe the limb darkening correctly. What remains are traces of active regions between $R_\ast\sim-0.6 \dots -0.3$ and $R_\ast\sim0.05 \dots 0.5$. The asymmetry is caused by instrumental effects, possibly a gradient in the CCD sensitivity or the filter bandwidth. A non-linear response of the CCD might also cause a wrong determination of the limb darkening coefficients.

\subsection{Stellar Activity on CoRoT-6}
The transiting extrasolar planet CoRoT-6b \citep{Fridlund2010} is an interesting case. Especially the discrepancy between the photometrically active lightcurve and the missing activity signal in the spectroscopically measured Ca II H\&K lines, proves the necessity to investigate transit lightcurves. The `warm Jupiter' planet orbits the active star in a $8.9$~day orbit in a distance of $0.08$~AU. The host star is of solar type (F9V) and has an estimated age of 3.0~Gyr.

\begin{figure}
\centering
\includegraphics[width=\textwidth]{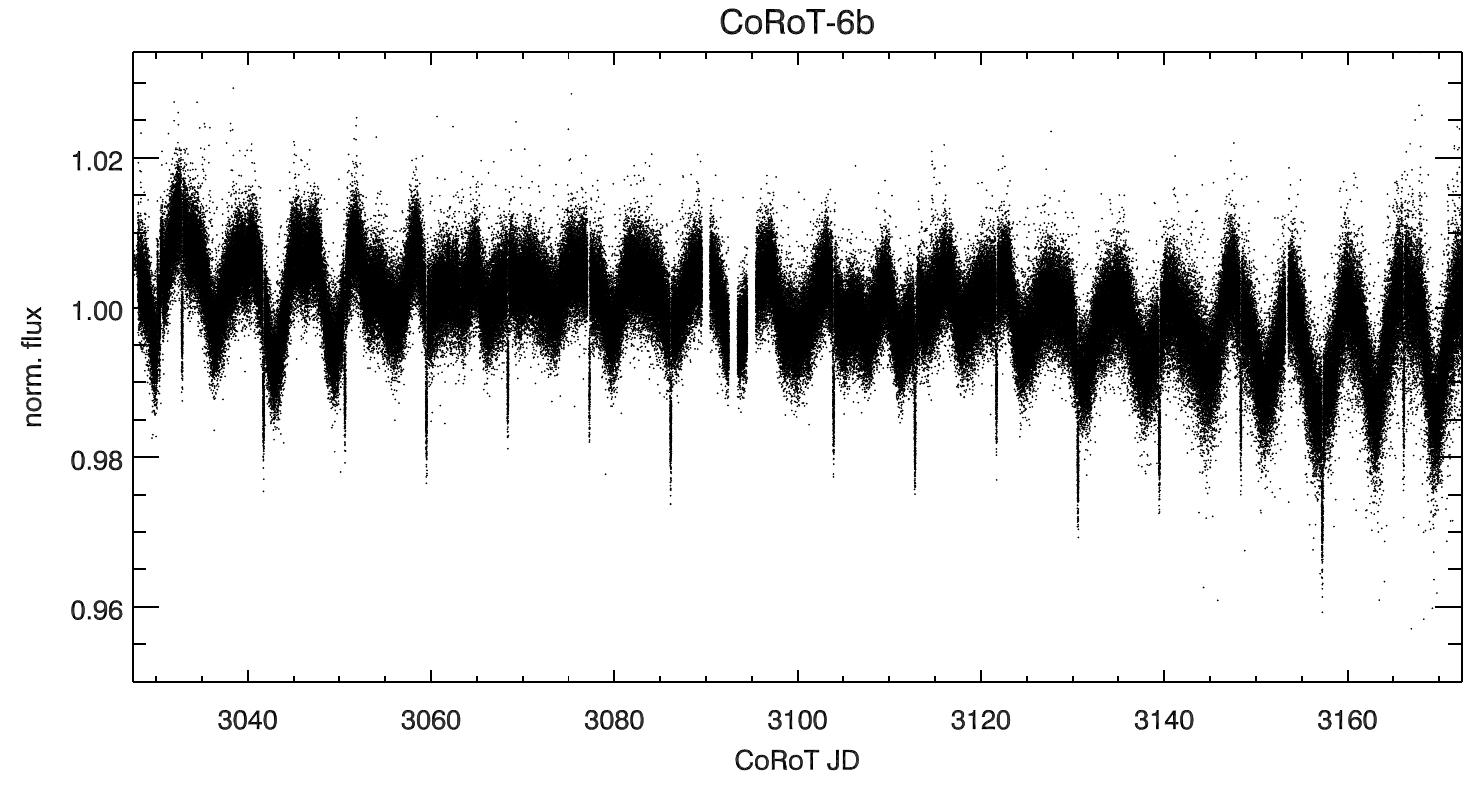}\label{fig:corot6b}
\caption{Lightcurve of CoRoT-6 with the transits of CoRoT-6b. In the last quarter the stellar activity signal is in the order of the transit depth.}
\end{figure}

The level of noise is rather high in the lightcurve of CoRoT-6 (Figure~\ref{fig:corot6b}). Three data gaps are evident around CoRoT JD~3092, wiping out one transit completely.
 
\begin{figure}
\centering
\includegraphics[width=\textwidth]{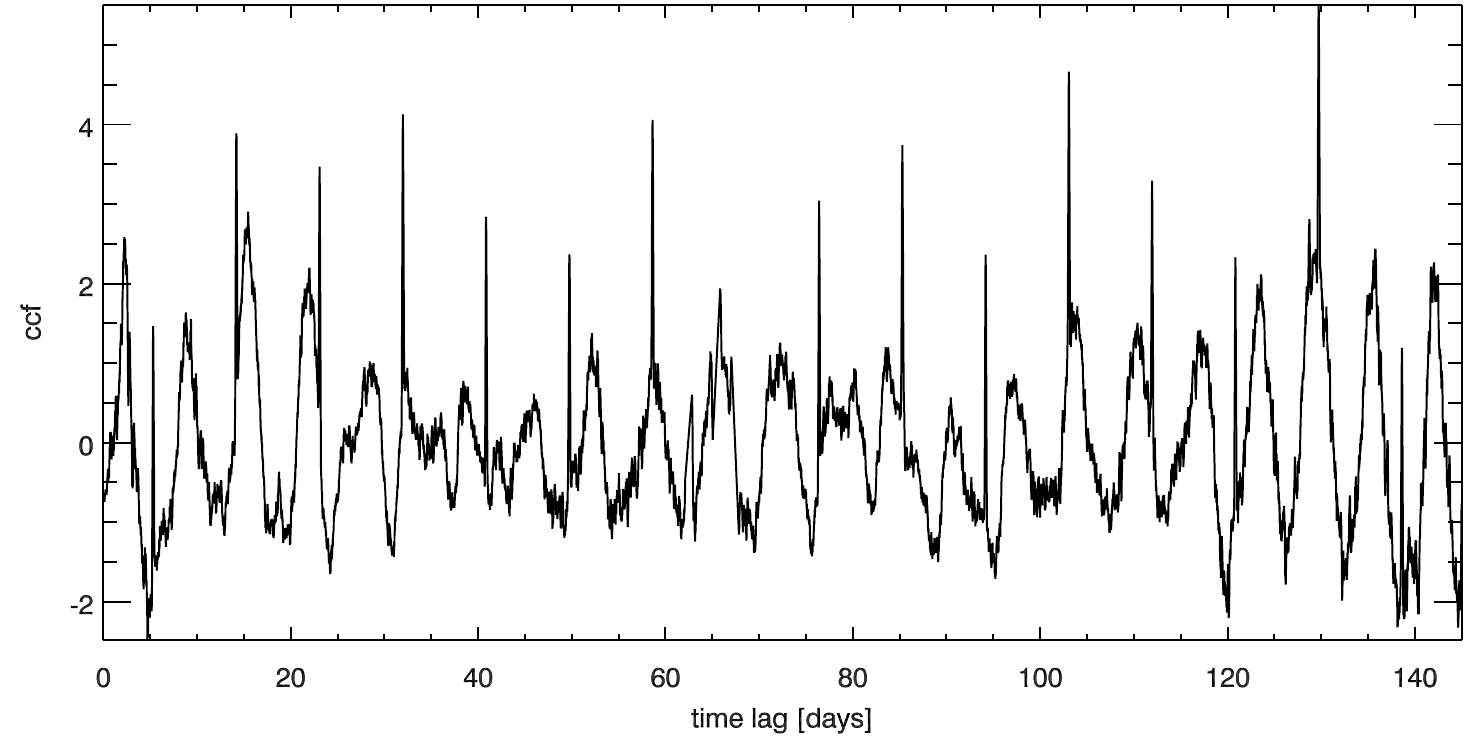}\label{fig:corot6bccf}
\caption{Cross-correlation function of CoRoT-6 with a trapezoidal function.}
\end{figure}

The cross-correlation function with a trapezoidal transit function is shown in Figure~\ref{fig:corot6bccf}. The high stellar activity signal modulates the transit signals.

\begin{figure}
\centering
\includegraphics[width=\textwidth]{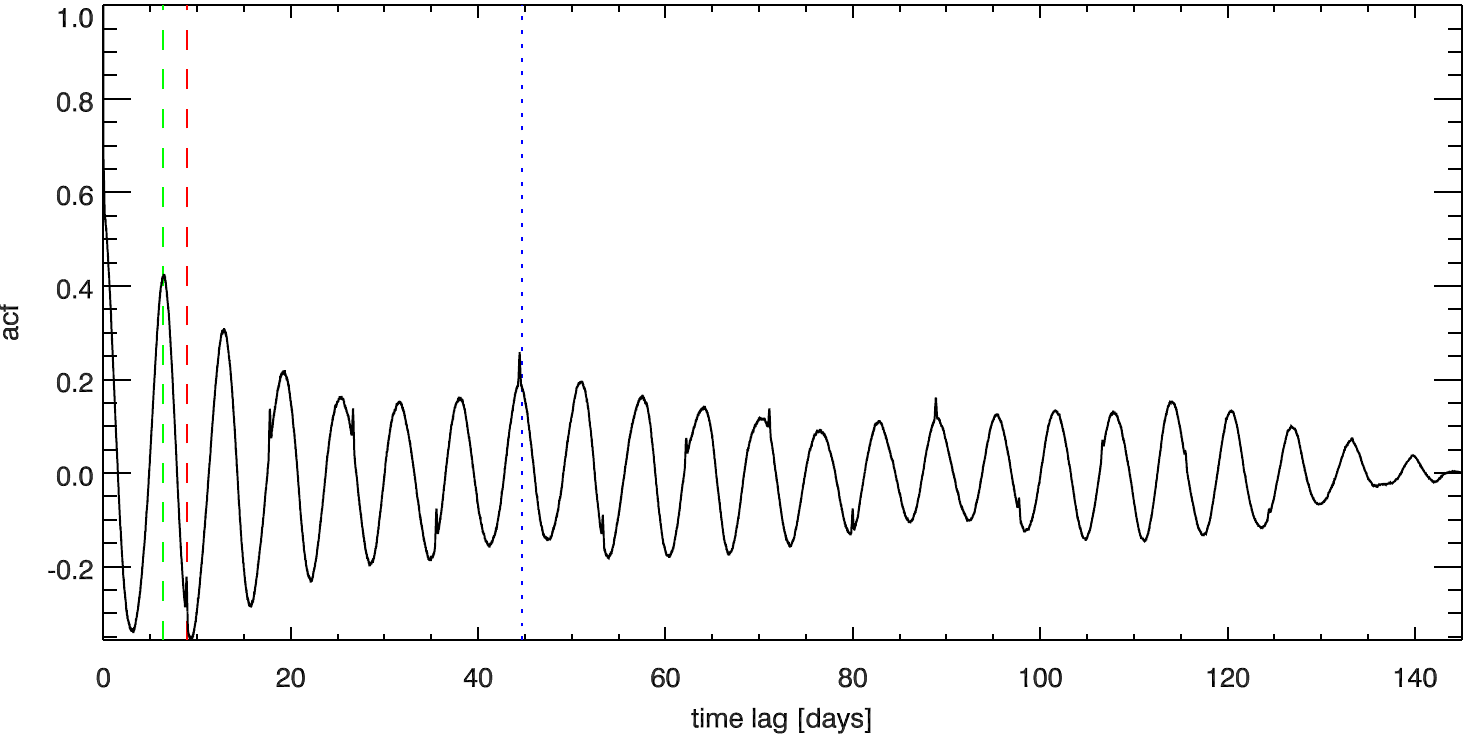}\label{fig:corot6bac}
\caption{Autocorrelation function of CoRoT-6. The orbital period of the planet is indicated by a dashed red line, the orbital period of the star by a dashed green line. The $7:5$ resonance is marked with a blue dotted line.}
\end{figure}

The orbital period of the planet is $8.886593\pm0.000004$\,days according to \citet{Fridlund2010}. The rotational period was estimated with $6.4\pm0.5$\,days differing slightly from the spectral analysis ($6.9\pm0.9$\,days). Looking closely on th autocorrelation function (Figure~\ref{fig:corot6bac}) we clearly see that the transit-peak and the stellar rotation coincide after $\approx44.8$\,days. This is close to a $7:5$-resonance between the stellar rotation and the orbital period. We might see here the evidence of an interaction between the planet and its host star.

\subsection{Stellar Activity on CoRoT-2}
\begin{figure}
\centering
\includegraphics[width=\textwidth]{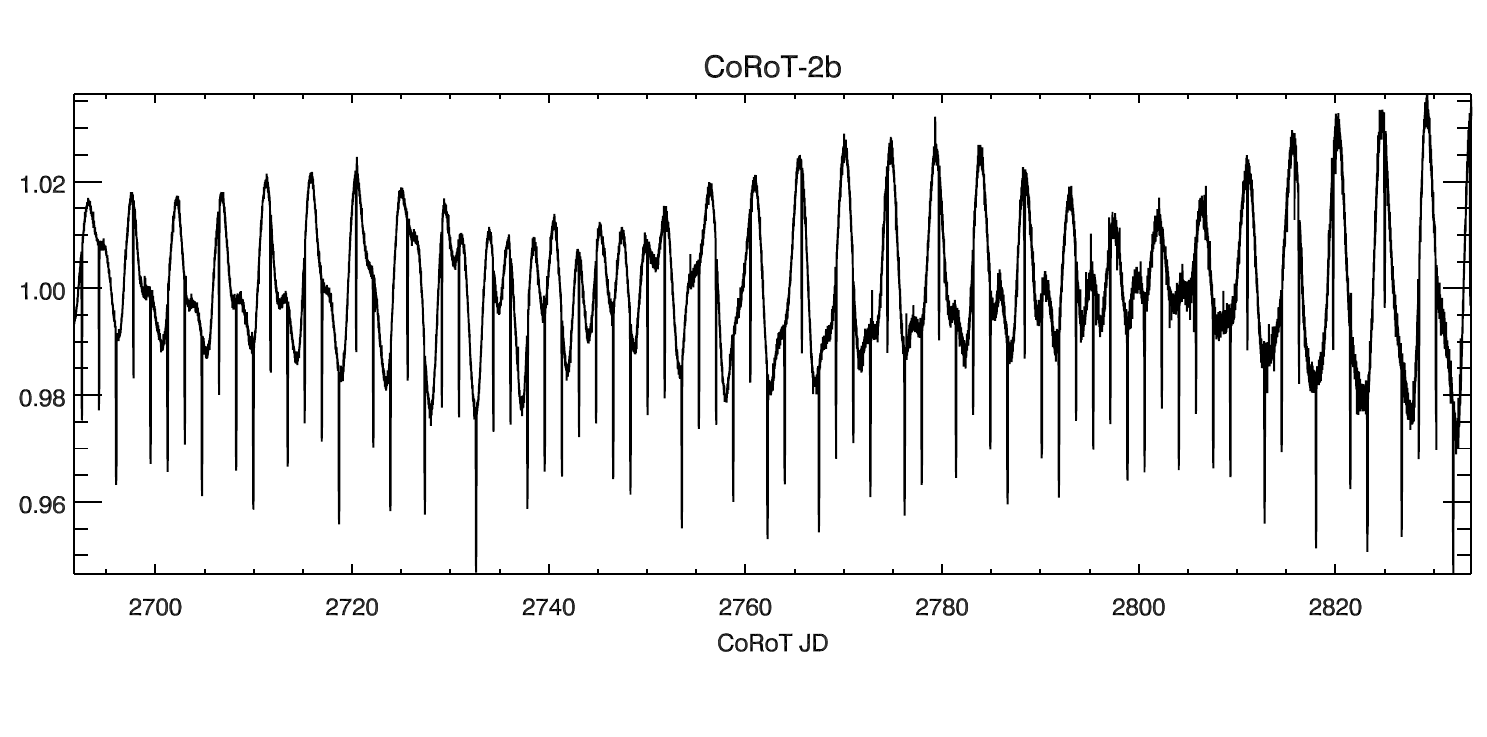}\label{fig:corot2b}
\caption{Lightcurve of CoRoT-2 with the transits of CoRoT-2b. The stellar activity signal is clearly visible and exceeds the depth of the transit at several times.}
\end{figure}

CoRoT-2b is a heavy and inflated planet orbiting an active G7V-star in a $1.7429$-day orbit \citep{Alonso2008}. Since its discovery, it has been an interesting object because of its unique features. The stellar activity signal is heavily modulated and changing over a period of $\approx30$\,days, which is a clear indication of spot evolution on this star (Figure~\ref{fig:corot2b}). The magnetic activity of the star has been thoroughly investigated by \cite{Lanza2009}.

\subsection{CoRoT-4}
\begin{figure}
  \centering
  \includegraphics[width=\textwidth]{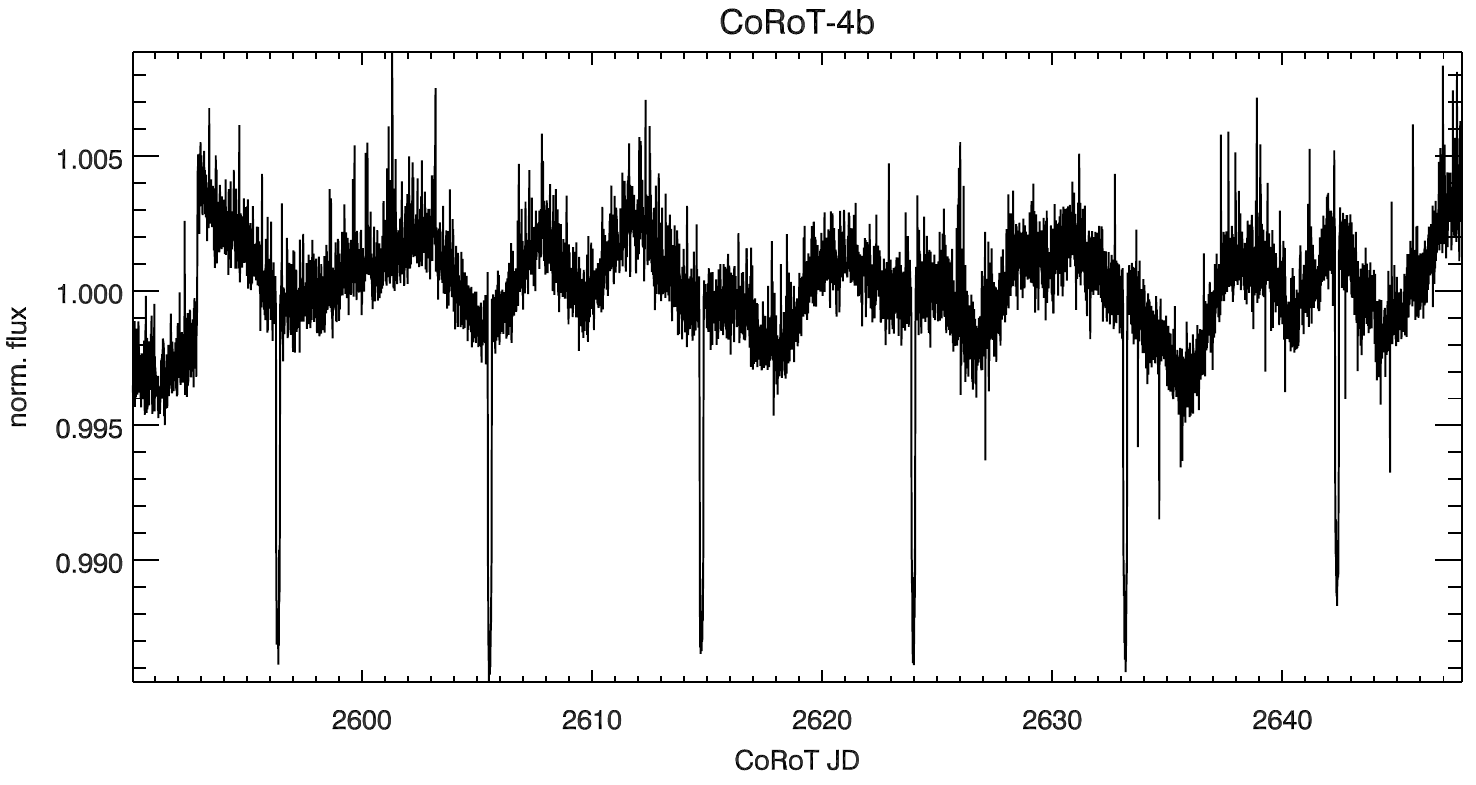}
  \caption{Lightcurve of CoRoT-4 with the transits of CoRoT-4b. If we compare to CoRoT-2, we see that the stellar noise is much higher.}\label{fig:corot4b}
\end{figure}

We want to investigate CoRoT-4b where the planet's orbit of $9.20205 \pm 0.00037$~days and the stellar rotation $8.87 \pm 1.12$~days are closely related \citep{Aigrain2008}. The spectral type of the host star is a G2V dwarf star which was determined from spectroscopic observations as well as the stellar radius $R_\ast=1.17^{+0.01}_{-0.03}$ \citep{Moutou2008}. The lightcurve can be seen in Figure~\ref{fig:corot4b}.

The stellar rotation period was determined by calculating the autocorrelation-function \citep[see][for details]{Aigrain2008}. In order to analyze the stellar activity, the transits signatures have been taken out of the lightcure with the missing data being interpolated linearly. The autocorrelation function of the lightcurve is shown in Figure~ \ref{fig:corot4bac}.

\begin{figure}
  \centering
  \includegraphics[width=\textwidth]{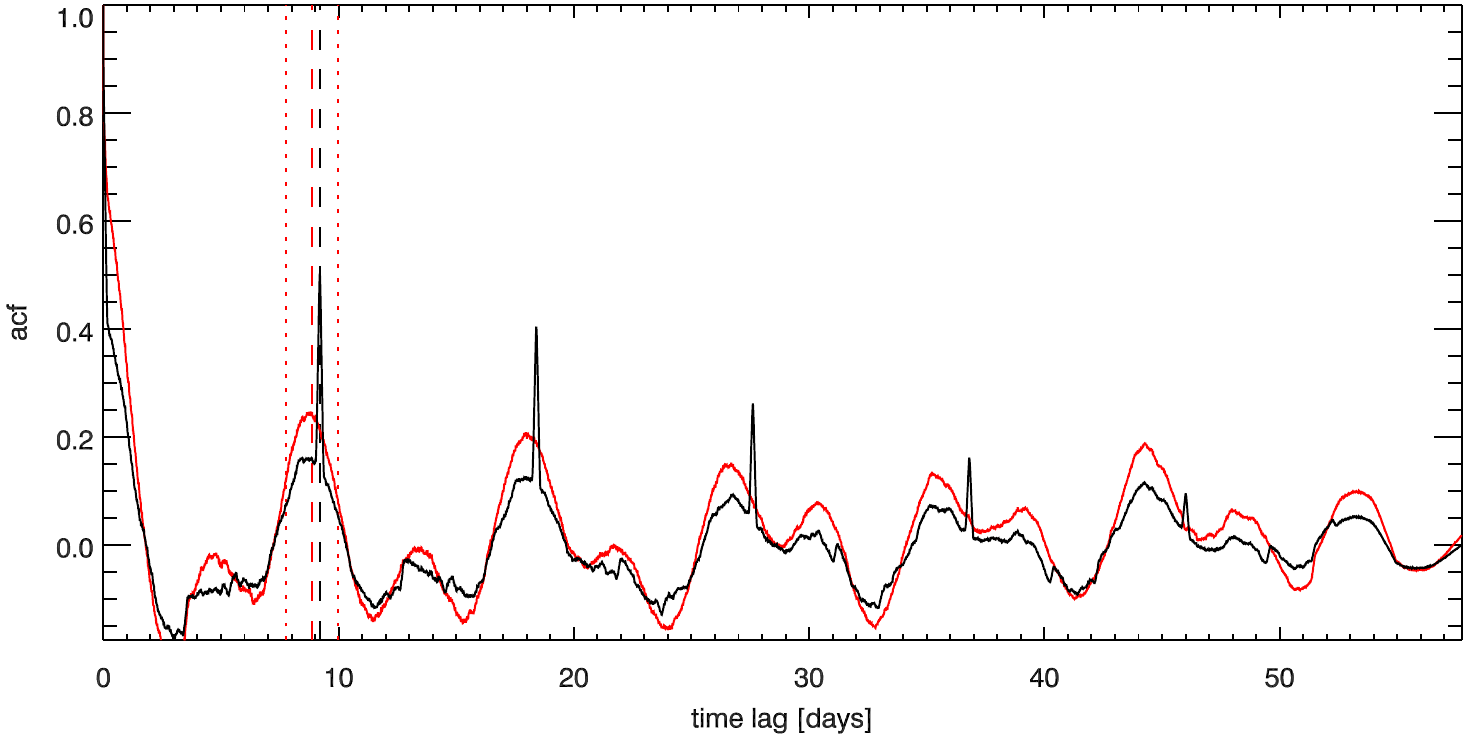}
  \caption{Autocorrelation function (acf) of CoRoT-4b with transits (black) and without transits (red). The stellar rotation period is indicated by a vertical dashed line and its uncertainty with red dotted lines. The orbital period of the planet is marked with a black vertical dashed line.}\label{fig:corot4bac}
\end{figure}
We can see the maxima of the stellar rotation signal at the determined period given by \citet{Aigrain2008}, but we also see a signal, though less significant at half the period at $\sim4.5$~days. This signal reaches its maximum at $\sim39$~days, which can be explained with spot evolution on the rotating host star.

By applying the transit detection algorithm described in Section~\ref{sec:detectionalgorithm} by using a trapezoid as a template for the cross-correlation function, we get a $6.51\sigma$-detection for the transit signal.

\begin{figure}
  \centering
  \includegraphics[width=\textwidth]{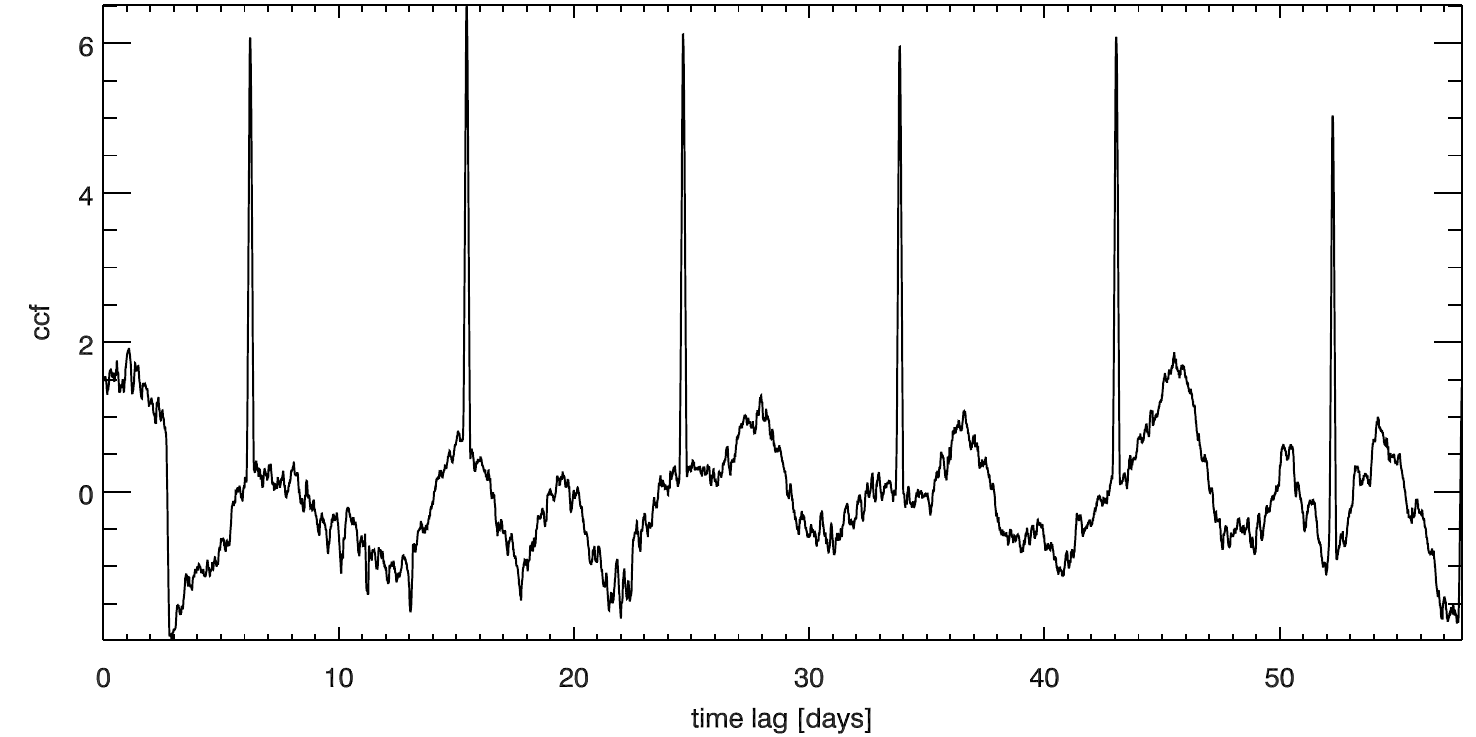}
  \caption{Crosscorrelation function (ccf) of CoRoT-4b. The ccf has been scaled with the standard-deviation of the ccf.}\label{fig:corot4bac}
\end{figure}  

By applying a MCMC-method for the parameters duration and impact-factor of the trapezoid, the signal of the ccf was maximized resulting in an optimal duration of the template transit of $6.15$~hours.

\begin{figure}
  \centering
  \includegraphics[width=\textwidth]{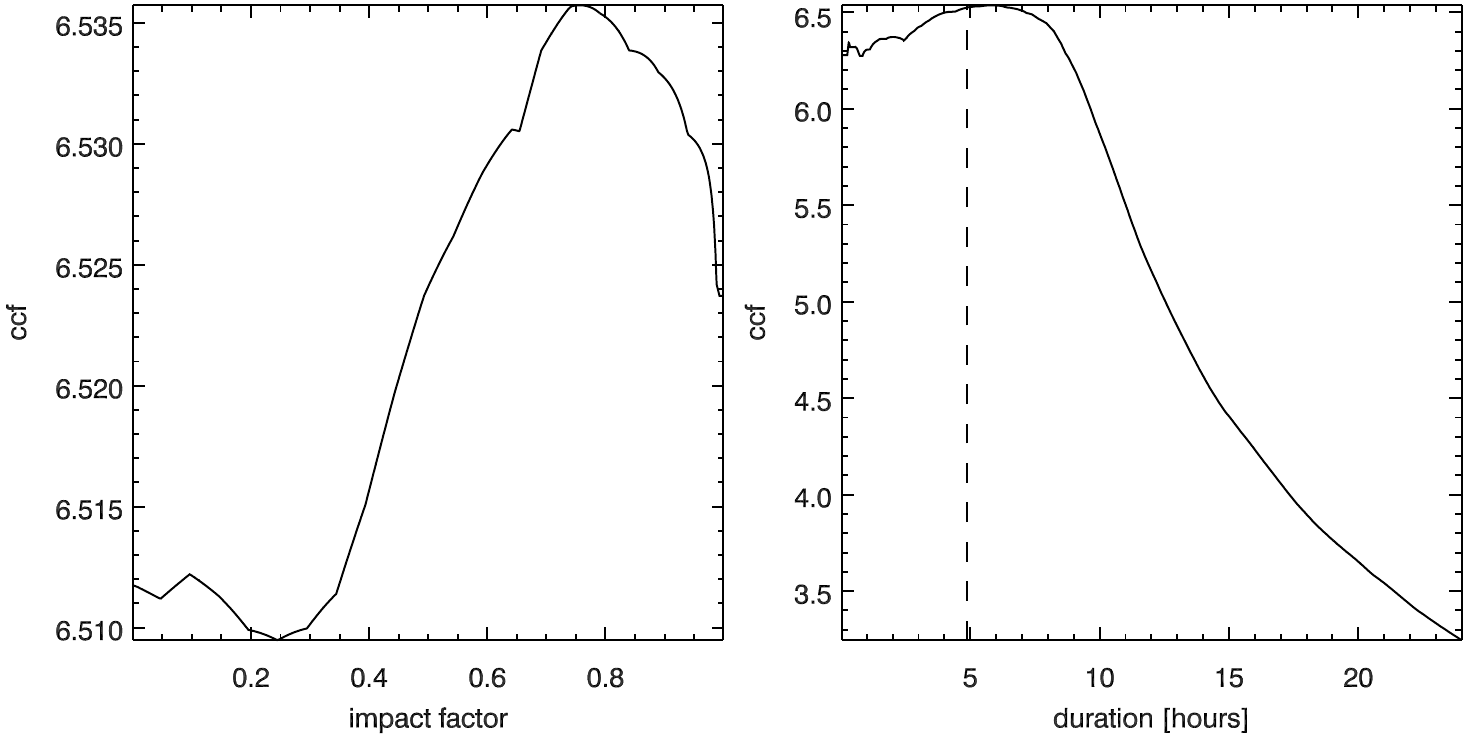}
  \caption{Dependency of the crosscorrelation function (ccf) in the impact factor (left) and the duration (right). The expected maximum duration of $t_T=4.864$~hours is marked by a vertical dashed line.}\label{fig:corot4bimp}
\end{figure}  

From Equation~\ref{eqn:kepler} and a total observation of $57.736296$~days we can observe a maximum period $P=19.245432$~days by assuming three fully observed transits. For a solar like star we would expect $t_T=4.864$~hours by applying Equation~\ref{eqn:transitduration}. An explanation for the difference between the maximum transit duration and the maximum of the ccf lies in the assumption for the solar like star, whereas CoRoT-4 was reported to have $R_\ast \approx 1.17 R_\sun$. We also have to keep in mind that the used template is a trapezoid and not a synthetic transit-shape, which would have introduced more free parameters (the limb darkening coefficients). As seen in Figure~\ref{fig:corot4bimp}, the detection of the transit is insensitive to the impact factor, with a minor difference of $0.025\sigma$ between the optimum $b=0.757$ and the minimum. So the shape (triangular, trapezoidal or box-like) is less critical.

\chapter{Conclusions}
Stellar activity increases the difficulty to observe transiting planets and to measure their masses using radial velocity methods. Stellar spots change the size of the measured transit and increase the uncertainty by 3\% \citep{Czesla2009}. Besides that, the color temperature of the star is decreased by large spot groups which can lead to false characterization of the host stars properties and its mean density.

The mean activity level of all spectral types is $377\pm100$~ppm, which is more than the average sun at $244.13$~ppm. Even if the background noise of 144.3~ppm is of instrumental origin, the majority of stars is more active than the sun. Late type stars seem to be more active than earlier types.

The radial velocity (RV) method is a proper tool to detect planets around main sequence stars \citep{Frink2002}. High stellar activity, however, limits the detection capabilities due to stellar granulation noise. This is a well known fact for radial velocity, but not for transiting systems. The problem of disentanglement has been discussed e.\,g. by \citet{Boisse2011}. CoRoT-2b \citep{Alonso2008} is one of the most famous examples for an extrasolar planet in orbit of an active star. CoRoT-6b is a good candidate for a planet-star interaction, showing a $7:5$-resonance between the orbital cycle of the planet and the rotation period of the star.

Planetary transits can be used to identify the stellar rotation period \citep[e.g.][]{Silva-Valio2008}. Using the information from the transits, especially the inclination can be constrained, and the position of the spot on the star can be assumed. There are still drawbacks, as described by \citet{Silva-Valio2008}, since a spot can not be clearly identified and there is an ambiguity in the observation of HD209458, where the period could either be 9.9 or 11.4 days. Nevertheless this method is a valuable alternative to spectroscopically measured line broadening. 

As seen in Figure~\ref{fig:corot4bimp}, the detection of the transit is independent of the impact factor, regardless of the stellar activity. So the shape (triangular, trapezoidal or box-like) does not favour the detection. This fact is used the the transit detection algorithm by \citet{Kovacs2002} (see~Section~\ref{sec:detectionalgorithms}). 

A robust filtering algorithm is used to clean the lightcurve from instrumental effects and to suppress the stellar activity signal. Wiener filtering is not suitable for filtering stellar activity since it fails to estimate the noise background. Usually the subtraction of a median average or application of a Butterworth filter is sufficient. In rare occasions where pulsations or oscillations pollute the lightcurve, pre-whitening is most effective to enhance the transit detection.

The application of the filter by \citet{Savitzky1964} is essential to the quality of the data, but has a severe impact on the transit-shape transforming it into a Gaussian. Alternatives have to be considered to ensure a robust outlier-detection and removal. Generally simple algorithms like e.g. a three-point median filter is not sufficient.

Nevertheless, a well designed filter system can enhance the transit detection and suppress the stellar activity signal \citep{Weingrill2010}.

Stellar variability can be easily identified by investigating the lightcurve using autocorrelation. Using radial velocity measurements additionally helps to constrain the rotational period of the star.

Transit signals have to be expected in the domain  of stellar rotation periods. CoRoT-2b and CoRoT-6b are strikingly examples for these problems. Nevertheless, the transit signal usually has sufficient power at higher harmonics to be identified.

Except for multi-band observations like the three color channels of CoRoT stellar flares are almost impossible to be distinguished from instrumental effects like cosmic ray hits. Amplitudes and decay-rates and post-event-levels might give indications to detect a flare. Nevertheless such rare events do not harm the transit detection.

A magnetic or gravitational interaction between a planet and its host star is most likely as demonstrated on CoRoT-6b. Tidally locked planets or planets near the Roché-Limit are candidates for direct or resonant interaction.

Stellar pulsations like p-modes do not impair the detection of a transiting planet. This might be the case for radial pulsators like e.\,g. Cepheids. We can overcome this problem either by modelling pulsations or by using additional constraints like the color information. The same holds for binary stars, which were not discussed within the frame of this work, where planetary transit signals can be detected after removing the binary transit signals.
 
We are now able to extend the values given in Table~\ref{tab:cox2000} with semi-major axis and their respective periods for the habitable zone taken from \citet{Kasting1993} with the transit depth estimated for a Super-Earth (Table~\ref{tab:conclusion}).

\begin{table}[ht]
\centering
\captionabove{Spectral types of dwarf stars taken from \citet{Cox2000} with their respective masses $M_\ast$ in solar masses and solar radii $R_\ast$. The table has been extended with the expected semi-major axis $a$ and the Keplerian period $P$ as well as the expected transit depth $\Delta F$ in ppm for a transiting Super-Earth with two Earth-radii and the approximate transit duration $t_T$ in hours.\label{tab:conclusion}}
\begin{tabular}{lrrrrrrr}
\hline \hline
Spectral Type & $M_\ast$ & $R_\ast$ & $a$ & $P$ & $\Delta F$ & $t_T $\\ \hline
A5V &  2.0  & 1.7 & 2.81 & 3.34 & 116 & 52.34\\
F0V &  1.6  & 1.5 & 2.20 & 2.58 & 149 & 45.68\\
F5V &  1.4  & 1.3 & 1.82 & 2.08 & 199 & 38.48\\
G0V &  1.05 & 1.1 & 1.13 & 1.17 & 278 & 29.60\\
G2V &  1.0  & 1.0 & 1.03 & 1.05 & 336 & 26.39\\
G5V &  0.92 & 0.92 & 0.89 & 0.87 & 397 & 23.43\\
K0V &  0.79 & 0.85 & 0.66 & 0.61 & 465 & 20.20\\
K5V &  0.67 & 0.72 & 0.48 & 0.41 & 648 & 15.80\\
M0V &  0.51 & 0.60 & 0.28 & 0.21 & 933 & 11.51\\
M2V &  0.40 & 0.50 & 0.18 & 0.12 & 1344 & 8.62\\
M5V &  0.21 & 0.27 & 0.08 & 0.05 & 4608 & 4.37\\
\end{tabular}
\end{table}

Given the level of stellar activity determined from CoRoT the stellar activity of early types might prevent the successful detection of a Super-Earth in the habitable zone. The sun as a G2V star within an average activity level is the limiting case.

In rare occasions flares are observed during transits \citep{Bentley2009}. But photometric methods that measure the integral flux of the star can not discern between a temporal phenomenon like a flare on the stellar disk or a spatial phenomenon like a stellar spot.

Future missions will utilise new methods like transit spectroscopy or observation in different wavelength regimes to identify stellar activity and its interaction with the hosted planet.

\singlespacing

\bibliographystyle{apalike} 
\bibliography{thesis} 

\begin{thebibliography}{}

\bibitem[{Aigrain} et~al., 2008]{Aigrain2008}
{Aigrain}, S., {Collier Cameron}, A., {Ollivier}, M., {Pont}, F., {Jorda}, L.,
  {Almenara}, J.~M., {Alonso}, R., {Barge}, P., {Bord{\'e}}, P., {Bouchy}, F.,
  {Deeg}, H., {de La Reza}, R., {Deleuil}, M., {Dvorak}, R., {Erikson}, A.,
  {Fridlund}, M., {Gondoin}, P., {Gillon}, M., {Guillot}, T., {Hatzes}, A.,
  {Lammer}, H., {Lanza}, A.~F., {L{\'e}ger}, A., {Llebaria}, A., {Magain}, P.,
  {Mazeh}, T., {Moutou}, C., {Paetzold}, M., {Pinte}, C., {Queloz}, D.,
  {Rauer}, H., {Rouan}, D., {Schneider}, J., {Wuchter}, G., and {Zucker}, S.
  (2008).
\newblock {Transiting exoplanets from the CoRoT space mission. IV.
  CoRoT-Exo-4b: a transiting planet in a 9.2 day synchronous orbit}.
\newblock {\em \aap}, 488:L43--L46.

\bibitem[{Aigrain} et~al., 2004]{Aigrain2004}
{Aigrain}, S., {Favata}, F., and {Gilmore}, G. (2004).
\newblock {Characterising stellar micro-variability for planetary transit
  searches}.
\newblock {\em \aap}, 414:1139--1152.

\bibitem[{Aigrain} et~al., 2003]{Aigrain2003}
{Aigrain}, S., {Gilmore}, G., {Favata}, F., and {Carpano}, S. (2003).
\newblock {The Frequency Content of the VIRGO/SoHO Light Curves: Implications
  for Planetary Transit Detection from Space}.
\newblock In {D.~Deming \& S.~Seager}, editor, {\em Scientific Frontiers in
  Research on Extrasolar Planets}, volume 294 of {\em Astronomical Society of
  the Pacific Conference Series}, pages 441--444.

\bibitem[{Aigrain} et~al., 2009]{Aigrain2009}
{Aigrain}, S., {Pont}, F., {Fressin}, F., {Alapini}, A., {Alonso}, R.,
  {Auvergne}, M., {Barbieri}, M., {Barge}, P., {Bord{\'e}}, P., {Bouchy}, F.,
  {Deeg}, H., {de La Reza}, R., {Deleuil}, M., {Dvorak}, R., {Erikson}, A.,
  {Fridlund}, M., {Gondoin}, P., {Guterman}, P., {Jorda}, L., {Lammer}, H.,
  {L{\'e}ger}, A., {Llebaria}, A., {Magain}, P., {Mazeh}, T., {Moutou}, C.,
  {Ollivier}, M., {P{\"a}tzold}, M., {Queloz}, D., {Rauer}, H., {Rouan}, D.,
  {Schneider}, J., {Wuchter}, G., and {Zucker}, S. (2009).
\newblock {Noise properties of the CoRoT data. A planet-finding perspective}.
\newblock {\em \aap}, 506:425--429.

\bibitem[{Alapini} and {Aigrain}, 2009]{Alapini2009}
{Alapini}, A. and {Aigrain}, S. (2009).
\newblock {An iterative filter to reconstruct planetary transit signals in the
  presence of stellar variability}.
\newblock {\em \mnras}, 397:1591--1598.

\bibitem[{Alonso} et~al., 2009]{Alonso2009}
{Alonso}, R., {Alapini}, A., {Aigrain}, S., {Auvergne}, M., {Baglin}, A.,
  {Barbieri}, M., {Barge}, P., {Bonomo}, A.~S., {Bord{\'e}}, P., {Bouchy}, F.,
  {Chaintreuil}, S., {de La Reza}, R., {Deeg}, H.~J., {Deleuil}, M., {Dvorak},
  R., {Erikson}, A., {Fridlund}, M., {de Oliveira Fialho}, F., {Gondoin}, P.,
  {Guillot}, T., {Hatzes}, A., {Jorda}, L., {Lammer}, H., {L{\'e}ger}, A.,
  {Llebaria}, A., {Magain}, P., {Mazeh}, T., {Moutou}, C., {Ollivier}, M.,
  {P{\"a}tzold}, M., {Pont}, F., {Queloz}, D., {Rauer}, H., {Rouan}, D.,
  {Schneider}, J., and {Wuchterl}, G. (2009).
\newblock {The secondary eclipse of CoRoT-1b}.
\newblock {\em \aap}, 506:353--358.

\bibitem[{Alonso} et~al., 2008]{Alonso2008}
{Alonso}, R., {Auvergne}, M., {Baglin}, A., {Ollivier}, M., {Moutou}, C.,
  {Rouan}, D., {Deeg}, H.~J., {Aigrain}, S., {Almenara}, J.~M., {Barbieri}, M.,
  {Barge}, P., {Benz}, W., {Bord{\'e}}, P., {Bouchy}, F., {de La Reza}, R.,
  {Deleuil}, M., {Dvorak}, R., {Erikson}, A., {Fridlund}, M., {Gillon}, M.,
  {Gondoin}, P., {Guillot}, T., {Hatzes}, A., {H{\'e}brard}, G., {Kabath}, P.,
  {Jorda}, L., {Lammer}, H., {L{\'e}ger}, A., {Llebaria}, A., {Loeillet}, B.,
  {Magain}, P., {Mayor}, M., {Mazeh}, T., {P{\"a}tzold}, M., {Pepe}, F.,
  {Pont}, F., {Queloz}, D., {Rauer}, H., {Shporer}, A., {Schneider}, J.,
  {Stecklum}, B., {Udry}, S., and {Wuchterl}, G. (2008).
\newblock {Transiting exoplanets from the CoRoT space mission. II.
  CoRoT-Exo-2b: a transiting planet around an active G star}.
\newblock {\em \aap}, 482:L21--L24.

\bibitem[{Appourchaux} et~al., 2008]{Appourchaux2008}
{Appourchaux}, T., {Michel}, E., {Auvergne}, M., {Baglin}, A., {Toutain}, T.,
  {Baudin}, F., {Benomar}, O., {Chaplin}, W.~J., {Deheuvels}, S., {Samadi}, R.,
  {Verner}, G.~A., {Boumier}, P., {Garc{\'{\i}}a}, R.~A., {Mosser}, B.,
  {Hulot}, J.-C., {Ballot}, J., {Barban}, C., {Elsworth}, Y.,
  {Jim{\'e}nez-Reyes}, S.~J., {Kjeldsen}, H., {R{\'e}gulo}, C., and {Roxburgh},
  I.~W. (2008).
\newblock {CoRoT sounds the stars: p-mode parameters of Sun-like oscillations
  on HD 49933}.
\newblock {\em \aap}, 488:705--714.

\bibitem[{Auvergne} et~al., 2009]{Auvergne2009}
{Auvergne}, M., {Bodin}, P., {Boisnard}, L., {Buey}, J.-T., {Chaintreuil}, S.,
  {Epstein}, G., {Jouret}, M., {Lam-Trong}, T., {Levacher}, P., {Magnan}, A.,
  {Perez}, R., {Plasson}, P., {Plesseria}, J., {Peter}, G., {Steller}, M.,
  {Tiph{\`e}ne}, D., {Baglin}, A., {Agogu{\'e}}, P., {Appourchaux}, T.,
  {Barbet}, D., {Beaufort}, T., {Bellenger}, R., {Berlin}, R., {Bernardi}, P.,
  {Blouin}, D., {Boumier}, P., {Bonneau}, F., {Briet}, R., {Butler}, B.,
  {Cautain}, R., {Chiavassa}, F., {Costes}, V., {Cuvilho}, J., {Cunha-Parro},
  V., {de Oliveira Fialho}, F., {Decaudin}, M., {Defise}, J.-M., {Djalal}, S.,
  {Docclo}, A., {Drummond}, R., {Dupuis}, O., {Exil}, G., {Faur{\'e}}, C.,
  {Gaboriaud}, A., {Gamet}, P., {Gavalda}, P., {Grolleau}, E., {Gueguen}, L.,
  {Guivarc'h}, V., {Guterman}, P., {Hasiba}, J., {Huntzinger}, G., {Hustaix},
  H., {Imbert}, C., {Jeanville}, G., {Johlander}, B., {Jorda}, L., {Journoud},
  P., {Karioty}, F., {Kerjean}, L., {Lafond}, L., {Lapeyrere}, V., {Landiech},
  P., {Larqu{\'e}}, T., {Laudet}, P., {Le Merrer}, J., {Leporati}, L.,
  {Leruyet}, B., {Levieuge}, B., {Llebaria}, A., {Martin}, L., {Mazy}, E.,
  {Mesnager}, J.-M., {Michel}, J.-P., {Moalic}, J.-P., {Monjoin}, W., {Naudet},
  D., {Neukirchner}, S., {Nguyen-Kim}, K., {Ollivier}, M., {Orcesi}, J.-L.,
  {Ottacher}, H., {Oulali}, A., {Parisot}, J., {Perruchot}, S., {Piacentino},
  A., {Pinheiro da Silva}, L., {Platzer}, J., {Pontet}, B., {Pradines}, A.,
  {Quentin}, C., {Rohbeck}, U., {Rolland}, G., {Rollenhagen}, F., {Romagnan},
  R., {Russ}, N., {Samadi}, R., {Schmidt}, R., {Schwartz}, N., {Sebbag}, I.,
  {Smit}, H., {Sunter}, W., {Tello}, M., {Toulouse}, P., {Ulmer}, B.,
  {Vandermarcq}, O., {Vergnault}, E., {Wallner}, R., {Waultier}, G., and
  {Zanatta}, P. (2009).
\newblock {The CoRoT satellite in flight: description and performance}.
\newblock {\em \aap}, 506:411--424.

\bibitem[{Backer} et~al., 1993]{Backer1993}
{Backer}, D.~C., {Foster}, R.~S., and {Sallmen}, S. (1993).
\newblock {A second companion of the millisecond pulsar 1620 - 26}.
\newblock {\em \nat}, 365:817--819.

\bibitem[{Baglin} et~al., 2009]{Baglin2009}
{Baglin}, A., {Auvergne}, M., {Barge}, P., {Deleuil}, M., {Michel}, E., and
  {The CoRoT Exoplanet Science Team} (2009).
\newblock {CoRoT: Description of the Mission and Early Results}.
\newblock In {\em IAU Symposium}, volume 253 of {\em IAU Symposium}, pages
  71--81.

\bibitem[{Baglin} et~al., 2006]{Baglin2006}
{Baglin}, A., {Auvergne}, M., {Boisnard}, L., {Lam-Trong}, T., {Barge}, P.,
  {Catala}, C., {Deleuil}, M., {Michel}, E., and {Weiss}, W. (2006).
\newblock {CoRoT: a high precision photometer for stellar ecolution and
  exoplanet finding}.
\newblock In {\em 36th COSPAR Scientific Assembly}, volume~36, pages 3749--+.

\bibitem[Bakos et~al., 2002]{Bakos2002}
Bakos, G.~A., Lazar, J., Papp, I., Sari, P., and Green, E.~M. (2002).
\newblock System description and first light-curves of {HAT,} an autonomous
  observatory for variability search.
\newblock {\em astro-ph/0206001}.
\newblock {Publ.Astron.Soc.Pac.114:974-987,2002}.

\bibitem[{Ballot} et~al., 2011]{Ballot2011}
{Ballot}, J., {Gizon}, L., {Samadi}, R., {Vauclair}, G., {Benomar}, O.,
  {Bruntt}, H., {Mosser}, B., {Stahn}, T., {Verner}, G.~A., {Campante}, T.~L.,
  {Garc{\'{\i}}a}, R.~A., {Mathur}, S., {Salabert}, D., {Gaulme}, P.,
  {R{\'e}gulo}, C., {Roxburgh}, I.~W., {Appourchaux}, T., {Baudin}, F.,
  {Catala}, C., {Chaplin}, W.~J., {Deheuvels}, S., {Michel}, E., {Bazot}, M.,
  {Creevey}, O., {Dolez}, N., {Elsworth}, Y., {Sato}, K.~H., {Vauclair}, S.,
  {Auvergne}, M., and {Baglin}, A. (2011).
\newblock {Accurate p-mode measurements of the G0V metal-rich CoRoT target HD
  52265}.
\newblock {\em \aap}, 530:A97+.

\bibitem[{Baudin} et~al., 2011]{Baudin2011}
{Baudin}, F., {Barban}, C., {Belkacem}, K., {Hekker}, S., {Morel}, T.,
  {Samadi}, R., {Benomar}, O., {Goupil}, M.-J., {Carrier}, F., {Ballot}, J.,
  {Deheuvels}, S., {De Ridder}, J., {Hatzes}, A.~P., {Kallinger}, T., and
  {Weiss}, W.~W. (2011).
\newblock {Amplitudes and lifetimes of solar-like oscillations observed by
  CoRoT. Red-giant versus main-sequence stars}.
\newblock {\em \aap}, 529:A84+.

\bibitem[{Bentley} et~al., 2009]{Bentley2009}
{Bentley}, S.~J., {Hellier}, C., {Maxted}, P.~F.~L., {Dhillon}, V.~S., {Marsh},
  T.~R., {Copperwheat}, C.~M., and {Littlefair}, S.~P. (2009).
\newblock {A stellar flare during the transit of the extrasolar planet
  OGLE-TR-10b}.
\newblock {\em \aap}, 505:901--902.

\bibitem[{Bertin} and {Arnouts}, 1996]{Bertin1996}
{Bertin}, E. and {Arnouts}, S. (1996).
\newblock {SExtractor: Software for source extraction.}
\newblock {\em \aaps}, 117:393--404.

\bibitem[{Boisse} et~al., 2011]{Boisse2011}
{Boisse}, I., {Bouchy}, F., {H{\'e}brard}, G., {Bonfils}, X., {Santos}, N., and
  {Vauclair}, S. (2011).
\newblock {Disentangling between stellar activity and planetary signals}.
\newblock {\em \aap}, 528:A4+.

\bibitem[{Bolker}, 1998]{Bolker1998}
{Bolker}, J. (1998).
\newblock {\em Writing Your Dissertation in Fifteen Minutes a Day}.
\newblock Macmillan Us.

\bibitem[{Bord{\'e}} et~al., 2003]{borde2003}
{Bord{\'e}}, P., {Rouan}, D., and {L{\'e}ger}, A. (2003).
\newblock {Exoplanet detection capability of the COROT space mission}.
\newblock {\em \aap}, 405:1137--1144.

\bibitem[{Borucki}, 2011]{Borucki2011}
{Borucki}, W.~J. (2011).
\newblock {Kepler Mission: An Overview of Science Results}.
\newblock In {\em American Astronomical Society Meeting Abstracts \#217},
  volume~43 of {\em Bulletin of the American Astronomical Society}, pages
  \#428.02--+.

\bibitem[{Boss} et~al., 2007]{Boss2007}
{Boss}, A.~P., {Butler}, R.~P., {Hubbard}, W.~B., {Ianna}, P.~A.,
  {K{\"u}rster}, M., {Lissauer}, J.~J., {Mayor}, M., {Meech}, K.~J., {Mignard},
  F., {Penny}, A.~J., {Quirrenbach}, A., {Tarter}, J.~C., and {Vidal-Madjar},
  A. (2007).
\newblock {Working Group on Extrasolar Planets}.
\newblock {\em Transactions of the International Astronomical Union, Series A},
  26:183--186.

\bibitem[{Bouchy} et~al., 2009]{Bouchy2009}
{Bouchy}, F., {Hebrard}, G., {Udry}, S., {Delfosse}, X., and {Boisse}, I.
  (2009).
\newblock {The SOPHIE northern extrasolar planets. I. A companion close to the
  planet/brown-dwarf transition around HD16760}.
\newblock {\em ArXiv e-prints}.

\bibitem[{Brown}, 2008]{Brown2008}
{Brown}, T.~M. (2008).
\newblock {Characterizing extrasolar planets}.
\newblock In {H.~Deeg, J.~A.~Belmonte, \& A.~Aparicio}, editor, {\em Extrasolar
  Planets}, pages 65--+.

\bibitem[{Burke} et~al., 2007]{Burke2007}
{Burke}, C.~J., {McCullough}, P.~R., {Valenti}, J.~A., {Johns-Krull}, C.~M.,
  {Janes}, K.~A., {Heasley}, J.~N., {Summers}, F.~J., {Stys}, J.~E.,
  {Bissinger}, R., {Fleenor}, M.~L., {Foote}, C.~N., {Garc{\'{\i}}a-Melendo},
  E., {Gary}, B.~L., {Howell}, P.~J., {Mallia}, F., {Masi}, G., {Taylor}, B.,
  and {Vanmunster}, T. (2007).
\newblock {XO-2b: Transiting Hot Jupiter in a Metal-rich Common Proper Motion
  Binary}.
\newblock {\em \apj}, 671:2115--2128.

\bibitem[{Carpano} and {Fridlund}, 2008]{carpano2008}
{Carpano}, S. and {Fridlund}, M. (2008).
\newblock {Detecting transits from Earth-sized planets around Sun-like stars}.
\newblock {\em \aap}, 485:607--613.

\bibitem[{Carrington}, 1863]{Carrington1863}
{Carrington}, R.~C. (1863).
\newblock {On the Motion of the Solar System in Space}.
\newblock {\em \mnras}, 23:203--+.

\bibitem[{Catala}, 2009]{Catala2009}
{Catala}, C. (2009).
\newblock {PLATO: PLAnetary Transits and Oscillations of stars}.
\newblock {\em Communications in Asteroseismology}, 158:330--+.

\bibitem[{Charbonneau} et~al., 2005]{Charbonneau2005}
{Charbonneau}, D., {Allen}, L.~E., {Megeath}, S.~T., {Torres}, G., {Alonso},
  R., {Brown}, T.~M., {Gilliland}, R.~L., {Latham}, D.~W., {Mandushev}, G.,
  {O'Donovan}, F.~T., and {Sozzetti}, A. (2005).
\newblock {Detection of Thermal Emission from an Extrasolar Planet}.
\newblock {\em \apj}, 626:523--529.

\bibitem[{Charbonneau} et~al., 2000]{Charbonneau2000}
{Charbonneau}, D., {Brown}, T.~M., {Latham}, D.~W., and {Mayor}, M. (2000).
\newblock {Detection of Planetary Transits Across a Sun-like Star}.
\newblock {\em \apjl}, 529:L45--L48.

\bibitem[{Claret}, 2000]{Claret2000}
{Claret}, A. (2000).
\newblock {A new non-linear limb-darkening law for LTE stellar atmosphere
  models. Calculations for $-5.0 \le \log[M/H] \le +1, 2000 K \le T_{eff} \le
  50000 K$ at several surface gravities}.
\newblock {\em \aap}, 363:1081--1190.

\bibitem[{Cox}, 2000]{Cox2000}
{Cox}, A.~N. (2000).
\newblock {\em {Allen's astrophysical quantities}}.

\bibitem[{Czesla} et~al., 2009]{Czesla2009}
{Czesla}, S., {Huber}, K.~F., {Wolter}, U., {Schr{\"o}ter}, S., and {Schmitt},
  J.~H.~M.~M. (2009).
\newblock {How stellar activity affects the size estimates of extrasolar
  planets}.
\newblock {\em \aap}, 505:1277--1282.

\bibitem[{Debosscher} et~al., 2009]{Debosscher2009}
{Debosscher}, J., {Sarro}, L.~M., {L{\'o}pez}, M., {Deleuil}, M., {Aerts}, C.,
  {Auvergne}, M., {Baglin}, A., {Baudin}, F., {Chadid}, M., {Charpinet}, S.,
  {Cuypers}, J., {De Ridder}, J., {Garrido}, R., {Hubert}, A.~M.,
  {Janot-Pacheco}, E., {Jorda}, L., {Kaiser}, A., {Kallinger}, T., {Kollath},
  Z., {Maceroni}, C., {Mathias}, P., {Michel}, E., {Moutou}, C., {Neiner}, C.,
  {Ollivier}, M., {Samadi}, R., {Solano}, E., {Surace}, C., {Vandenbussche},
  B., and {Weiss}, W.~W. (2009).
\newblock {Automated supervised classification of variable stars in the CoRoT
  programme. Method and application to the first four exoplanet fields}.
\newblock {\em \aap}, 506:519--534.

\bibitem[{Deeg}, 2009]{Deeg2009}
{Deeg}, H. (2009).
\newblock {UTM, a universal simulator for lightcurves of transiting systems}.
\newblock In {\em IAU Symposium}, volume 253 of {\em IAU Symposium}, pages
  388--391.

\bibitem[{Deleuil} et~al., 2009]{Deleuil2009}
{Deleuil}, M., {Meunier}, J.~C., {Moutou}, C., {Surace}, C., {Deeg}, H.~J.,
  {Barbieri}, M., {Debosscher}, J., {Almenara}, J.~M., {Agneray}, F., {Granet},
  Y., {Guterman}, P., and {Hodgkin}, S. (2009).
\newblock {Exo-Dat: An Information System in Support of the CoRoT/Exoplanet
  Science}.
\newblock {\em \aj}, 138:649--663.

\bibitem[{Demory} et~al., 2011]{Demory2011}
{Demory}, B., {Seager}, S., {Kjeldsen}, H., and {Kepler Science Team} (2011).
\newblock {The High Albedo of the Hot-Jupiter Kepler-7b}.
\newblock In {\em American Astronomical Society Meeting Abstracts \#217},
  volume~43 of {\em Bulletin of the American Astronomical Society}, pages
  \#103.08--+.

\bibitem[{Doyle}, 2008]{Doyle2008}
{Doyle}, L.~R. (2008).
\newblock {Overview of extrasolar planet detection methods}.
\newblock In {H.~Deeg, J.~A.~Belmonte, \& A.~Aparicio}, editor, {\em Extrasolar
  Planets}, pages 1--+.

\bibitem[{Dumusque} et~al., 2011]{Dumusque2011}
{Dumusque}, X., {Udry}, S., {Lovis}, C., {Santos}, N.~C., and {Monteiro},
  M.~J.~P.~F.~G. (2011).
\newblock {Planetary detection limits taking into account stellar noise. I.
  Observational strategies to reduce stellar oscillation and granulation
  effects}.
\newblock {\em \aap}, 525:A140+.

\bibitem[{Eggenberger} et~al., 2004]{Eggenberger2004}
{Eggenberger}, P., {Carrier}, F., {Bouchy}, F., and {Blecha}, A. (2004).
\newblock {Solar-like oscillations in Procyon A}.
\newblock {\em \aap}, 422:247--252.

\bibitem[{Finsterle} and {Fr{\"o}hlich}, 2001]{Finsterle2001}
{Finsterle}, W. and {Fr{\"o}hlich}, C. (2001).
\newblock {Low-Order p Modes From Virgo Irradiance Data}.
\newblock {\em Solar Physics}, 200:393--406.

\bibitem[{Fridlund} et~al., 2010]{Fridlund2010}
{Fridlund}, M., {H{\'e}brard}, G., {Alonso}, R., {Deleuil}, M., {Gandolfi}, D.,
  {Gillon}, M., {Bruntt}, H., {Alapini}, A., {Csizmadia}, S., {Guillot}, T.,
  {Lammer}, H., {Aigrain}, S., {Almenara}, J.~M., {Auvergne}, M., {Baglin}, A.,
  {Barge}, P., {Bord{\'e}}, P., {Bouchy}, F., {Cabrera}, J., {Carone}, L.,
  {Carpano}, S., {Deeg}, H.~J., {de La Reza}, R., {Dvorak}, R., {Erikson}, A.,
  {Ferraz-Mello}, S., {Guenther}, E., {Gondoin}, P., {den Hartog}, R.,
  {Hatzes}, A., {Jorda}, L., {L{\'e}ger}, A., {Llebaria}, A., {Magain}, P.,
  {Mazeh}, T., {Moutou}, C., {Ollivier}, M., {P{\"a}tzold}, M., {Queloz}, D.,
  {Rauer}, H., {Rouan}, D., {Samuel}, B., {Schneider}, J., {Shporer}, A.,
  {Stecklum}, B., {Tingley}, B., {Weingrill}, J., and {Wuchterl}, G. (2010).
\newblock {Transiting exoplanets from the CoRoT space mission. IX. CoRoT-6b: a
  transiting ``hot Jupiter'' planet in an 8.9d orbit around a low-metallicity
  star}.
\newblock {\em \aap}, 512:A14+.

\bibitem[{Frink} et~al., 2002]{Frink2002}
{Frink}, S., {Mitchell}, D.~S., {Quirrenbach}, A., {Fischer}, D.~A., {Marcy},
  G.~W., and {Butler}, R.~P. (2002).
\newblock {Discovery of a Substellar Companion to the K2 III Giant {$\iota$}
  Draconis}.
\newblock {\em \apj}, 576:478--484.

\bibitem[{Garc{\'{\i}}a} et~al., 2010]{Garcia2010}
{Garc{\'{\i}}a}, R.~A., {Mathur}, S., {Salabert}, D., {Ballot}, J.,
  {R{\'e}gulo}, C., {Metcalfe}, T.~S., and {Baglin}, A. (2010).
\newblock {CoRoT Reveals a Magnetic Activity Cycle in a Sun-Like Star}.
\newblock {\em Science}, 329:1032--.

\bibitem[{Gazak} et~al., 2011]{Gazak2011}
{Gazak}, J.~Z., {Johnson}, J.~A., {Tonry}, J., {Eastman}, J., {Mann}, A.~W.,
  and {Agol}, E. (2011).
\newblock {Transit Analysis Package (TAP and autoKep): IDL Graphical User
  Interfaces for Extrasolar Planet Transit Photometry}.
\newblock {\em ArXiv e-prints}.

\bibitem[{Gilliland} et~al., 2010]{Gilliland2010}
{Gilliland}, R.~L., {McCullough}, P.~R., {Nelan}, E.~P., {Brown}, T.~M.,
  {Charbonneau}, D., {Nutzman}, P., {Christensen-Dalsgaard}, J., and
  {Kjeldsen}, H. (2010).
\newblock {Asteroseismology of the Transiting Exoplanet Host HD 17156 with HST
  FGS}.
\newblock {\em ArXiv e-prints}.

\bibitem[{Gorlova} et~al., 2006]{Gorlova2006}
{Gorlova}, N., {Rieke}, G.~H., {Muzerolle}, J., {Stauffer}, J.~R., {Siegler},
  N., {Young}, E.~T., and {Stansberry}, J.~H. (2006).
\newblock {Spitzer 24 {$\mu$}m Survey of Debris Disks in the Pleiades}.
\newblock {\em \apj}, 649:1028--1042.

\bibitem[{Guinan} et~al., 2003]{Guinan2003}
{Guinan}, E.~F., {Ribas}, I., and {Harper}, G.~M. (2003).
\newblock {Far-Ultraviolet Emissions of the Sun in Time: Probing Solar Magnetic
  Activity and Effects on Evolution of Paleoplanetary Atmospheres}.
\newblock {\em \apj}, 594:561--572.

\bibitem[{Han}, 2006]{Han2006}
{Han}, C. (2006).
\newblock {Secure Identification of Free-floating Planets}.
\newblock {\em \apj}, 644:1232--1236.

\bibitem[{Harvey} et~al., 1993]{Harvey1993}
{Harvey}, J.~W., {Duvall}, Jr., T.~L., {Jefferies}, S.~M., and {Pomerantz},
  M.~A. (1993).
\newblock {Chromospheric Oscillations and the Background Spectrum}.
\newblock In {T.~M.~Brown}, editor, {\em GONG 1992. Seismic Investigation of
  the Sun and Stars}, volume~42 of {\em Astronomical Society of the Pacific
  Conference Series}, pages 111--+.

\bibitem[{Hatzes} et~al., 2011]{Hatzes2011}
{Hatzes}, A.~P., {Fridlund}, M., {Nachmani}, G., {Mazeh}, T., {Valencia}, D.,
  {Hebrard}, G., {Carone}, L., {Paetzold}, M., {Udry}, S., {Bouchy}, F.,
  {Borde}, P., {Deeg}, H., {Tingley}, B., {Dvorak}, R., {Gandolfi}, D.,
  {Ferraz-Mello}, S., {Wuchterl}, G., {Guenther}, E., {Rauer}, H., {Erikson},
  A., {Cabrera}, J., {Csizmadia}, S., {Leger}, A., {Lammer}, H., {Weingrill},
  J., {Queloz}, D., {Alonso}, R., and {Schneider}, J. (2011).
\newblock {On the Mass of CoRoT-7b}.
\newblock {\em ArXiv e-prints}.

\bibitem[{Hulot} et~al., 2011]{Hulot2011}
{Hulot}, J.~C., {Baudin}, F., {Samadi}, R., and {Goupil}, M.~J. (2011).
\newblock {A quantitative analysis of stellar activity based on CoRoT
  photometric data}.
\newblock {\em ArXiv e-prints}.

\bibitem[{Kalas} et~al., 2009]{Kalas2009}
{Kalas}, P., {Fitzgerald}, M.~P., {Clampin}, M., {Graham}, J.~R., {Chiang}, E.,
  {Kite}, E.~S., {Stapelfeldt}, K., and {Krist}, J. (2009).
\newblock {Fomalhaut b: Direct Detection of a {$\epsilon$} Jupiter-mass Object
  Orbiting Fomalhaut}.
\newblock In {\em American Astronomical Society Meeting Abstracts \#213},
  volume~41 of {\em Bulletin of the American Astronomical Society}, pages
  \#351.02--+.

\bibitem[{Kasting} et~al., 1993]{Kasting1993}
{Kasting}, J.~F., {Whitmire}, D.~P., and {Reynolds}, R.~T. (1993).
\newblock {Habitable Zones around Main Sequence Stars}.
\newblock {\em \icarus}, 101:108--128.

\bibitem[{Kopp} et~al., 2004]{Kopp2004}
{Kopp}, G., {Lawrence}, G.~M., {Rottman}, G., and {Woods}, T. (2004).
\newblock {Total Solar Irradiance Observations of the Oct./Nov. 2003 Solar
  Flares}.
\newblock In {\em American Astronomical Society Meeting Abstracts \#204},
  volume~36 of {\em Bulletin of the American Astronomical Society}, pages
  669--+.

\bibitem[{Kov{\'a}cs} et~al., 2002]{Kovacs2002}
{Kov{\'a}cs}, G., {Zucker}, S., and {Mazeh}, T. (2002).
\newblock {A box-fitting algorithm in the search for periodic transits}.
\newblock {\em \aap}, 391:369--377.

\bibitem[Kretzschmar, 2011]{Kretzschmar2011}
Kretzschmar, M. (2011).
\newblock Sun-as-a-star observation of {White-Light} flares.
\newblock {\em 1103.3125}.
\newblock Astronomy \& Astrophysics, Volume 530, {id.A84}, 2011.

\bibitem[{Lammer} et~al., 2011]{Lammer2011}
{Lammer}, H., {Eybl}, V., {Kislyakova}, K.~G., {Weingrill}, J.,
  {Holmstr{\"o}m}, M., {Khodachenko}, M.~L., {Kulikov}, Y.~N., {Reiners}, A.,
  {Leitzinger}, M., {Odert}, P., {Xiang Gr{\"u}{\ss}}, M., {Dorner}, B.,
  {G{\"u}del}, M., and {Hanslmeier}, A. (2011).
\newblock {UV transit observations of EUV-heated expanded thermospheres of
  Earth-like exoplanets around M-stars: testing atmosphere evolution
  scenarios}.
\newblock {\em \apss}, pages 69--+.

\bibitem[{Landsman}, 1995]{Landsman1995}
{Landsman}, W.~B. (1995).
\newblock {The IDL Astronomy User's Library}.
\newblock In {R.~A.~Shaw, H.~E.~Payne, \& J.~J.~E.~Hayes}, editor, {\em
  Astronomical Data Analysis Software and Systems IV}, volume~77 of {\em
  Astronomical Society of the Pacific Conference Series}, pages 437--+.

\bibitem[Lanza et~al., 2009]{Lanza2009}
Lanza, A.~F., Pagano, I., Leto, G., Messina, S., Aigrain, S., Alonso, R.,
  Auvergne, M., Baglin, A., Barge, P., Bonomo, A.~S., Boumier, P.,
  Collier~Cameron, A., Comparato, M., Cutispoto, G., de~Medeiros, J.~R., Foing,
  B., Kaiser, A., Moutou, C., Parihar, P.~S., {Silva-Valio}, A., and Weiss,
  W.~W. (2009).
\newblock Magnetic activity in the photosphere of {CoRoT-Exo-2a.} active
  longitudes and short-term spot cycle in a young sun-like star.
\newblock {\em Astronomy and Astrophysics}, 493:193--200.

\bibitem[{Lazorenko} et~al., 2011]{Lazorenko2011}
{Lazorenko}, P.~F., {Sahlmann}, J., {S{\'e}gransan}, D., {Figueira}, P.,
  {Lovis}, C., {Martin}, E., {Mayor}, M., {Pepe}, F., {Queloz}, D., {Rodler},
  F., {Santos}, N., and {Udry}, S. (2011).
\newblock {Astrometric search for a planet around VB 10}.
\newblock {\em \aap}, 527:A25+.

\bibitem[{Mallama} et~al., 2002]{Mallama2002}
{Mallama}, A., {Wang}, D., and {Howard}, R.~A. (2002).
\newblock {Photometry of Mercury from SOHO/LASCO and Earth. The Phase Function
  from 2 to 170 deg.}
\newblock {\em \icarus}, 155:253--264.

\bibitem[{Mandel} and {Agol}, 2002]{Mandel2002}
{Mandel}, K. and {Agol}, E. (2002).
\newblock {Analytic Light Curves for Planetary Transit Searches}.
\newblock {\em \apjl}, 580:L171--L175.

\bibitem[{Markwardt}, 2009]{Markwardt2009}
{Markwardt}, C.~B. (2009).
\newblock {Non-linear Least-squares Fitting in IDL with MPFIT}.
\newblock In {D.~A.~Bohlender, D.~Durand, \& P.~Dowler}, editor, {\em
  Astronomical Data Analysis Software and Systems XVIII}, volume 411 of {\em
  Astronomical Society of the Pacific Conference Series}, pages 251--+.

\bibitem[{Mathur} et~al., 2011]{Mathur2011}
{Mathur}, S., {Handberg}, R., {Campante}, T.~L., {Garc{\'{\i}}a}, R.~A.,
  {Appourchaux}, T., {Bedding}, T.~R., {Mosser}, B., {Chaplin}, W.~J.,
  {Ballot}, J., {Benomar}, O., {Bonanno}, A., {Corsaro}, E., {Gaulme}, P.,
  {Hekker}, S., {R{\'e}gulo}, C., {Salabert}, D., {Verner}, G., {White}, T.~R.,
  {Brand{\~a}o}, I.~M., {Creevey}, O.~L., {Do{\u g}an}, G., {Elsworth}, Y.,
  {Huber}, D., {Hale}, S.~J., {Houdek}, G., {Karoff}, C., {Metcalfe}, T.~S.,
  {Molenda-{\.Z}akowicz}, J., {Monteiro}, M.~J.~P.~F.~G., {Thompson}, M.~J.,
  {Christensen-Dalsgaard}, J., {Gilliland}, R.~L., {Kawaler}, S.~D.,
  {Kjeldsen}, H., {Quintana}, E.~V., {Sanderfer}, D.~T., and {Seader}, S.~E.
  (2011).
\newblock {Solar-like Oscillations in KIC 11395018 and KIC 11234888 from 8
  Months of Kepler Data}.
\newblock {\em \apj}, 733:95--+.

\bibitem[{Matthews} et~al., 2004]{Matthews2004}
{Matthews}, J.~M., {Kuschnig}, R., {Guenther}, D.~B., {Walker}, G.~A.~H.,
  {Moffat}, A.~F.~J., {Rucinski}, S.~M., {Sasselov}, D., and {Weiss}, W.~W.
  (2004).
\newblock {No stellar p-mode oscillations in space-based photometry of
  Procyon}.
\newblock {\em \nat}, 430:51--53.

\bibitem[{Mayor} et~al., 1995]{Mayor1995}
{Mayor}, M., {Queloz}, D., {Marcy}, G., {Butler}, P., {Noyes}, R., {Korzennik},
  S., {Krockenberger}, M., {Nisenson}, P., {Brown}, T., {Kennelly}, T.,
  {Rowland}, C., {Horner}, S., {Burki}, G., {Burnet}, M., and {Kunzli}, M.
  (1995).
\newblock {51 Pegasi}.
\newblock {\em \iaucirc}, 6251:1--+.

\bibitem[{Mazeh} et~al., 2009]{Mazeh2009}
{Mazeh}, T., {Guterman}, P., {Aigrain}, S., {Zucker}, S., {Grinberg}, N.,
  {Alapini}, A., {Alonso}, R., {Auvergne}, M., {Barbieri}, M., {Barge}, P.,
  {Bord{\'e}}, P., {Bouchy}, F., {Deeg}, H., {de La Reza}, R., {Deleuil}, M.,
  {Dvorak}, R., {Erikson}, A., {Fridlund}, M., {Gondoin}, P., {Jorda}, L.,
  {Lammer}, H., {L{\'e}ger}, A., {Llebaria}, A., {Magain}, P., {Moutou}, C.,
  {Ollivier}, M., {P{\"a}tzold}, M., {Pont}, F., {Queloz}, D., {Rauer}, H.,
  {Rouan}, D., {Sabo}, R., {Schneider}, J., and {Wuchterl}, G. (2009).
\newblock {Removing systematics from the CoRoT light curves. I.
  Magnitude-dependent zero point}.
\newblock {\em \aap}, 506:431--434.

\bibitem[{Michel} et~al., 2008]{Michel2008}
{Michel}, E., {Baglin}, A., {Auvergne}, M., {Catala}, C., {Samadi}, R.,
  {Baudin}, F., {Appourchaux}, T., {Barban}, C., {Weiss}, W.~W., {Berthomieu},
  G., {Boumier}, P., {Dupret}, M.-A., {Garcia}, R.~A., {Fridlund}, M.,
  {Garrido}, R., {Goupil}, M.-J., {Kjeldsen}, H., {Lebreton}, Y., {Mosser}, B.,
  {Grotsch-Noels}, A., {Janot-Pacheco}, E., {Provost}, J., {Roxburgh}, I.~W.,
  {Thoul}, A., {Toutain}, T., {Tiph{\`e}ne}, D., {Turck-Chieze}, S.,
  {Vauclair}, S.~D., {Vauclair}, G.~P., {Aerts}, C., {Alecian}, G., {Ballot},
  J., {Charpinet}, S., {Hubert}, A.-M., {Ligni{\`e}res}, F., {Mathias}, P.,
  {Monteiro}, M.~J.~P.~F.~G., {Neiner}, C., {Poretti}, E., {Renan de Medeiros},
  J., {Ribas}, I., {Rieutord}, M.~L., {Cort{\'e}s}, T.~R., and {Zwintz}, K.
  (2008).
\newblock {CoRoT Measures Solar-Like Oscillations and Granulation in Stars
  Hotter Than the Sun}.
\newblock {\em Science}, 322:558--.

\bibitem[{Moutou} et~al., 2008]{Moutou2008}
{Moutou}, C., {Bruntt}, H., {Guillot}, T., {Shporer}, A., {Guenther}, E.,
  {Aigrain}, S., {Almenara}, J.~M., {Alonso}, R., {Auvergne}, M., {Baglin}, A.,
  {Barbieri}, M., {Barge}, P., {Benz}, W., {Bord{\'e}}, P., {Bouchy}, F.,
  {Deeg}, H.~J., {de La Reza}, R., {Deleuil}, M., {Dvorak}, R., {Erikson}, A.,
  {Fridlund}, M., {Gillon}, M., {Gondoin}, P., {Hatzes}, A., {H{\'e}brard}, G.,
  {Jorda}, L., {Kabath}, P., {Lammer}, H., {L{\'e}ger}, A., {Llebaria}, A.,
  {Loeillet}, B., {Magain}, P., {Mayor}, M., {Mazeh}, T., {Ollivier}, M.,
  {P{\"a}tzold}, M., {Pepe}, F., {Pont}, F., {Queloz}, D., {Rabus}, M.,
  {Rauer}, H., {Rouan}, D., {Schneider}, J., {Udry}, S., and {Wuchterl}, G.
  (2008).
\newblock {Transiting exoplanets from the CoRoT space mission. V. CoRoT-Exo-4b:
  stellar and planetary parameters}.
\newblock {\em \aap}, 488:L47--L50.

\bibitem[Moutou et~al., 2005]{Moutou2005}
Moutou, C., Pont, F., Barge, P., Aigrain, S., Auvergne, M., Blouin, D.,
  Cautain, R., Erikson, A.~R., Guis, V., Guterman, P., Irwin, M., Lanza, A.~F.,
  Queloz, D., Rauer, H., Voss, H., and Zucker, S. (2005).
\newblock Comparative blind test of five planetary transit detection algorithms
  on realistic synthetic light curves.
\newblock {\em Astronomy and Astrophysics}, 437:355--368.

\bibitem[{Moutou} et~al., 2006]{Moutou2006}
{Moutou}, C., {Pont}, F., and {Halbwachs}, J.-L. (2006).
\newblock {Detection and characterization of extrasolar planets: the transit
  method}.
\newblock {\em Formation plan{\'e}taire et exoplan{\`e}tes, Ecole
  th{\'e}matique du CNRS}, 28:55--79.

\bibitem[Ofir et~al., 2010]{Ofir2010}
Ofir, A., Alonso, R., Bonomo, A.~S., Carone, L., Carpano, S., Samuel, B.,
  Weingrill, J., Aigrain, S., Auvergne, M., Baglin, A., et~al. (2010).
\newblock The {SARS} algorithm: detrending {CoRoT} light curves with sysrem
  using simultaneous external parameters.
\newblock {\em Arxiv preprint {arXiv:1003.0427}}.

\bibitem[{Pepe} et~al., 2004]{Pepe2004}
{Pepe}, F., {Mayor}, M., {Queloz}, D., {Benz}, W., {Bonfils}, X., {Bouchy}, F.,
  {Lo Curto}, G., {Lovis}, C., {M{\'e}gevand}, D., {Moutou}, C., {Naef}, D.,
  {Rupprecht}, G., {Santos}, N.~C., {Sivan}, J., {Sosnowska}, D., and {Udry},
  S. (2004).
\newblock {The HARPS search for southern extra-solar planets. I. HD 330075 b: A
  new ``hot Jupiter''}.
\newblock {\em \aap}, 423:385--389.

\bibitem[{Perryman}, 2011]{Perryman2011}
{Perryman}, M. (2011).
\newblock {\em The Exoplanet Handbook}.
\newblock Cambridge University Press.

\bibitem[{Poddany}, 2008]{Poddany2008}
{Poddany}, S. (2008).
\newblock {Transiting Explanet Light Curve Solution by Phoebe Code}.
\newblock {\em Open European Journal on Variable Stars}, 95:81--+.

\bibitem[Pollacco et~al., 2006]{Pollacco2006}
Pollacco, D.~L., Skillen, I., Cameron, A.~C., Christian, D.~J., Hellier, C.,
  Irwin, J., Lister, T.~A., Street, R.~A., West, R.~G., Anderson, D., Clarkson,
  W.~I., Deeg, H., Enoch, B., Evans, A., Fitzsimmons, A., Haswell, C.~A.,
  Hodgkin, S., Horne, K., Kane, S.~R., Keenan, F.~P., Maxted, P. F.~L., Norton,
  A.~J., Osborne, J., Parley, N.~R., Ryans, R. S.~I., Smalley, B., Wheatley,
  P.~J., and Wilson, D.~M. (2006).
\newblock The {WASP} project and the {SuperWASP} cameras.
\newblock {\em astro-ph/0608454}.
\newblock {Publ.Astron.Soc.Pac.118:1407-1418,2006}.

\bibitem[Press et~al., 1992]{Press1992}
Press, W.~H., Teukolsky, S.~A., Vetterling, W.~T., and Flannery, B.~P. (1992).
\newblock {\em Numerical recipes in {FORTRAN.} The art of scientific
  computing}.

\bibitem[{Pr{\v s}a} and {Zwitter}, 2007]{Prsa2007}
{Pr{\v s}a}, A. and {Zwitter}, T. (2007).
\newblock {Introducing Powell's Direction Set Method to a Fully Automated
  Analysis of Eclipsing Binary Stars}.
\newblock In {O.~Demircan, S.~O.~Selam, \& B.~Albayrak}, editor, {\em Solar and
  Stellar Physics Through Eclipses}, volume 370 of {\em Astronomical Society of
  the Pacific Conference Series}, pages 175--+.

\bibitem[{Queloz} et~al., 2009]{Queloz2009}
{Queloz}, D., {Bouchy}, F., {Moutou}, C., {Hatzes}, A., {H{\'e}brard}, G.,
  {Alonso}, R., {Auvergne}, M., {Baglin}, A., {Barbieri}, M., {Barge}, P.,
  {Benz}, W., {Bord{\'e}}, P., {Deeg}, H.~J., {Deleuil}, M., {Dvorak}, R.,
  {Erikson}, A., {Ferraz Mello}, S., {Fridlund}, M., {Gandolfi}, D., {Gillon},
  M., {Guenther}, E., {Guillot}, T., {Jorda}, L., {Hartmann}, M., {Lammer}, H.,
  {L{\'e}ger}, A., {Llebaria}, A., {Lovis}, C., {Magain}, P., {Mayor}, M.,
  {Mazeh}, T., {Ollivier}, M., {P{\"a}tzold}, M., {Pepe}, F., {Rauer}, H.,
  {Rouan}, D., {Schneider}, J., {Segransan}, D., {Udry}, S., and {Wuchterl}, G.
  (2009).
\newblock {The CoRoT-7 planetary system: two orbiting super-Earths}.
\newblock {\em \aap}, 506:303--319.

\bibitem[Reddy and Chatterji, 1996]{Reddy1996}
Reddy, B.~S. and Chatterji, B.~N. (1996).
\newblock An {FFT-based} technique for translation, rotation, and
  scale-invariant image registration.
\newblock {\em {IEEE} Transactions on Image Processing}, 5(8):1266--1271.

\bibitem[{R{\'e}gulo} et~al., 2007]{Regulo2007}
{R{\'e}gulo}, C., {Almenara}, J.~M., {Alonso}, R., {Deeg}, H., and {Roca
  Cort{\'e}s}, T. (2007).
\newblock {TRUFAS, a wavelet-based algorithm for the rapid detection of
  planetary transits}.
\newblock {\em \aap}, 467:1345--1352.

\bibitem[{R{\'e}gulo} and {Roca Cort{\'e}s}, 2005]{Regulo2005}
{R{\'e}gulo}, C. and {Roca Cort{\'e}s}, T. (2005).
\newblock {Stellar p-mode oscillations signal in Procyon A from MOST data}.
\newblock {\em \aap}, 444:L5--L8.

\bibitem[{Renaud} et~al., 1999]{Renaud1999}
{Renaud}, C., {Grec}, G., {Boumier}, P., {Gabriel}, A.~H., {Robillot}, J.~M.,
  {Cort{\'e}s}, T.~R., {Turck-Chi{\`e}ze}, S., and {Ulrich}, R.~K. (1999).
\newblock {Solar oscillations: time analysis of the GOLF p-mode signal}.
\newblock {\em \aap}, 345:1019--1026.

\bibitem[{Renner} et~al., 2008]{Renner2008}
{Renner}, S., {Rauer}, H., {Erikson}, A., {Hedelt}, P., {Kabath}, P., {Titz},
  R., and {Voss}, H. (2008).
\newblock {The BAST algorithm for transit detection}.
\newblock {\em \aap}, 492:617--620.

\bibitem[{Rowe} et~al., 2006]{Rowe2006}
{Rowe}, J., {Matthews}, J.~M., {Miller-Ricci}, E., {Seager}, S., {Sasselov},
  D., {Kuschnig}, R., {Guenther}, D.~B., {Moffat}, A.~F., {Rucinski}, M.,
  {Walker}, G.~A., and {Weiss}, W. (2006).
\newblock {MOST Spacebased Photometry of Transiting Exoplanet Systems}.
\newblock In {\em American Astronomical Society Meeting Abstracts}, volume~38
  of {\em Bulletin of the American Astronomical Society}, pages \#163.05--+.

\bibitem[{Rowe} et~al., 2008]{Rowe2008}
{Rowe}, J.~F., {Matthews}, J.~M., {Seager}, S., {Miller-Ricci}, E., {Sasselov},
  D., {Kuschnig}, R., {Guenther}, D.~B., {Moffat}, A.~F.~J., {Rucinski}, S.~M.,
  {Walker}, G.~A.~H., and {Weiss}, W.~W. (2008).
\newblock {The Very Low Albedo of an Extrasolar Planet: MOST Space-based
  Photometry of HD 209458}.
\newblock {\em \apj}, 689:1345--1353.

\bibitem[{R\"ukl}, 1979]{Ruekl1979}
{R\"ukl}, A. (1979).
\newblock {\em Welten, Sterne und Planeten}.
\newblock Mosaik Verlag GmbH, München.

\bibitem[{Savitzky} and {Golay}, 1964]{Savitzky1964}
{Savitzky}, A. and {Golay}, M. (1964).
\newblock {Smoothing and Differentiation of Data by Simplified Least Squares
  Procedures}.
\newblock {\em Analytical Chemistry}, 36(8):1627.

\bibitem[{Schneider} et~al., 2011]{Schneider2011}
{Schneider}, J., {Dedieu}, C., {Le Sidaner}, P., {Savalle}, R., and
  {Zolotukhin}, I. (2011).
\newblock {Defining and cataloging exoplanets: the exoplanet.eu database}.
\newblock {\em \aap}, 532:A79+.

\bibitem[{Seager} and {Mall{\'e}n-Ornelas}, 2003]{Seager2003}
{Seager}, S. and {Mall{\'e}n-Ornelas}, G. (2003).
\newblock {A Unique Solution of Planet and Star Parameters from an Extrasolar
  Planet Transit Light Curve}.
\newblock {\em \apj}, 585:1038--1055.

\bibitem[{Silva-Valio}, 2008]{Silva-Valio2008}
{Silva-Valio}, A. (2008).
\newblock {Estimating Stellar Rotation from Starspot Detection during Planetary
  Transits}.
\newblock {\em \apjl}, 683:L179--L182.

\bibitem[{Sing}, 2010]{Sing2010}
{Sing}, D.~K. (2010).
\newblock {Stellar limb-darkening coefficients for CoRot and Kepler}.
\newblock {\em \aap}, 510:A21+.

\bibitem[{Southworth} et~al., 2009]{Southworth2009}
{Southworth}, J., {Hinse}, T.~C., {J{\o}rgensen}, U.~G., {Dominik}, M.,
  {Ricci}, D., {Burgdorf}, M.~J., {Hornstrup}, A., {Wheatley}, P.~J.,
  {Anguita}, T., {Bozza}, V., {Novati}, S.~C., {Harps{\o}e}, K.,
  {Kj{\ae}rgaard}, P., {Liebig}, C., {Mancini}, L., {Masi}, G., {Mathiasen},
  M., {Rahvar}, S., {Scarpetta}, G., {Snodgrass}, C., {Surdej}, J.,
  {Th{\"o}ne}, C.~C., and {Zub}, M. (2009).
\newblock {High-precision photometry by telescope defocusing - I. The
  transiting planetary system WASP-5}.
\newblock {\em \mnras}, 396:1023--1031.

\bibitem[{Spiegel} et~al., 2011]{Spiegel2011}
{Spiegel}, D.~S., {Burrows}, A., and {Milsom}, J.~A. (2011).
\newblock {The Deuterium-burning Mass Limit for Brown Dwarfs and Giant
  Planets}.
\newblock {\em \apj}, 727:57--+.

\bibitem[{Stix}, 2002]{Stix2002}
{Stix}, M. (2002).
\newblock {\em The Sun - An Introduction}.
\newblock Springer-Verlag Berlin, 2nd edition edition.

\bibitem[{Strassmeier} et~al., 2008]{Strassmeier2008}
{Strassmeier}, K.~G., {Briguglio}, R., {Granzer}, T., {Tosti}, G., {Divarano},
  I., {Savanov}, I., {Bagaglia}, M., {Castellini}, S., {Mancini}, A.,
  {Nucciarelli}, G., {Straniero}, O., {Distefano}, E., {Messina}, S., and
  {Cutispoto}, G. (2008).
\newblock {First time-series optical photometry from Antarctica. sIRAIT
  monitoring of the RS CVn binary V841 Centauri and the {$\delta$}-Scuti star
  V1034 Centauri}.
\newblock {\em \aap}, 490:287--295.

\bibitem[{Teixeira} et~al., 2008]{Teixeira2008}
{Teixeira}, T.~C., {Kjeldsen}, H., {Bedding}, T.~R., {Bouchy}, F.,
  {Christensen-Dalsgaard}, J., {Cunha}, M.~S., {Dall}, T., {Frandsen}, S.,
  {Karoff}, C., {Monteiro}, M.~J.~P.~F.~G., and {Pijpers}, F.~P. (2008).
\newblock {Solar-like oscillations in the G8 V star tau Ceti}.
\newblock {\em ArXiv e-prints}.

\bibitem[{Tingley}, 2003]{Tingley2003}
{Tingley}, B. (2003).
\newblock {Improvements to existing transit detection algorithms and their
  comparison}.
\newblock {\em \aap}, 408:L5--L7.

\bibitem[{Weingrill} et~al., 2011]{Weingrill2011}
{Weingrill}, J., {Aigrain}, S., {Bouchy}, F., {Deleuil}, M., and {the CoRoT
  team} (2011).
\newblock {Planetary transit candidates in the CoRoT SRa01 and SRa02 Fields}.
\newblock {\em Astrophysics and Space Science}.
\newblock in preparation.

\bibitem[{Weingrill} et~al., 2010]{Weingrill2010}
{Weingrill}, J., {Lammer}, H., {Khodachenko}, M.~L., and {Hanslmeier}, A.
  (2010).
\newblock {Detection of Transiting Super-Earths around Active Stars}.
\newblock In {V.~Coud{\'e} Du Foresto, D.~M.~Gelino, \& I.~Ribas}, editor, {\em
  Pathways Towards Habitable Planets}, volume 430 of {\em Astronomical Society
  of the Pacific Conference Series}, pages 556--+.

\bibitem[{Willson} and {Mordvinov}, 1999]{Willson1999}
{Willson}, R.~C. and {Mordvinov}, A.~V. (1999).
\newblock {Time-frequency analysis of total solar irradiance variations}.
\newblock {\em \grl}, 26:3613--3616.

\bibitem[{Wilson} and {Devinney}, 1971]{Wilson1971}
{Wilson}, R.~E. and {Devinney}, E.~J. (1971).
\newblock {Realization of Accurate Close-Binary Light Curves: Application to MR
  Cygni}.
\newblock {\em \apj}, 166:605--+.

\bibitem[{Winn}, 2010]{Winn2010}
{Winn}, J.~N. (2010).
\newblock {Transits and Occultations}.
\newblock {\em ArXiv e-prints}.

\bibitem[{Woodard} and {Hudson}, 1983]{Woodard1983}
{Woodard}, M. and {Hudson}, H.~S. (1983).
\newblock {Frequencies, amplitudes and linewidths of solar oscillations from
  total irradiance observations}.
\newblock {\em \nat}, 305:589--593.

\bibitem[{Woods} and {Kopp}, 2005]{Woods2005}
{Woods}, T.~N. and {Kopp}, G. (2005).
\newblock {Contributions of the Solar Ultraviolet Irradiance to the Total Solar
  Irradiance During Large Flares}.
\newblock {\em AGU Fall Meeting Abstracts}, pages A7+.

\end{thebibliography}

\listoftables
\listoffigures

\appendix
\chapter{Appendix}

\section{The drizzling algorithm}
\label{sec:drizzlealgorithm}
\begin{lstlisting}
function drizzle, t, y, err=err, LOCATIONS=locations, BINSIZE=BINSIZE, $
                                 _EXTRA = extra
  ;last bin is empty!
  if N_ELEMENTS(t) LE 1 then $ 
    message,'insufficient times'
  if N_ELEMENTS(t) NE N_ELEMENTS(y) then $
    message,'times and data not of same length'
  IF BINSIZE EQ 0.0 then begin
    message,'binsize zero!', /continue
    err = REPLICATE(STDDEV(y),N_ELEMENTS(y))
    locations = t
    return,y
  endif
  if not Keyword_set(NBINS) then $
    NBINS = CEIL((MAX(t,/NAN)-MIN(t,/NAN))/BINSIZE)
  h = histogram(t,REVERSE_INDICES = r,/NAN, LOCATIONS = locations, $
      BINSIZE = BINSIZE, _EXTRA = extra)
  result = DBLARR(NBINS, /NOZERO)
  err = DBLARR(NBINS, /NOZERO)
  result[*] = !VALUES.F_NAN
  for i = 0L, nbins-1L do begin
    IF R[i] NE R[i+1L] THEN begin
      ybin = y[R[R[I] : R[i+1L]-1L]]
      result[i] = MEAN(ybin,/NAN)
      ; WARNING: stddev ignores the nan keyword!	  
      err[i] = STDDEV(ybin,/NAN) 
      ; alternatively:
	  ; err[i] = RMS(ybin)
    endif
  endfor
  if N_ELEMENTS(result) NE N_ELEMENTS(locations) then $ 
    message,'Warning: unequal arrays in result', /CONTINUE
  
  return,result
end
\end{lstlisting}

\section{Cosine Fit}
\label{sec:cosinefit}
\begin{lstlisting}
PRO COSFUNC, X, A, F, pder
  ; A[0] ... x1 Phase in days
  ;   1  ... x2 duration in days
  ;   2  ... x3 amplitude
  ;   3  ... x4 offset
  F = cos((X+ A[0])*2.0*!DPI/A[1])*A[2]+A[3]
  IF N_PARAMS() GE 4 THEN $ 
    pder = [ [-2.0*!DPI*A[2]*sin(2.0*!DPI*(X+A[0])/A[1])/A[1]], $
             [2*!DPI*A[2]*(X + A[0])*sin(2.0*!DPI*(X+A[0])/A[1])/(A[1]^2)], $
             [cos(2.0*!DPI*(X+A[0])/A[1])], $
             [replicate(1.0, N_ELEMENTS(X))]]
;  RETURN, RMS(l1 - (cos((t1+ P0[0])*2.0*!DPI/P0[1])*P0[2]+P0[3]))
  ;D[cos((t + x1)*2.0*Pi/x2)*x3+x4,x1]
  ;d/dx1 = -2.0*!DPI*x3*sin(2.0*!DPI*(t+x1)/x2)/x2
  ;d/dx2 = 2*!DPI*x3*(t + x1)*sin(2.0*!DPI*(t+x1)/x2)/(x2^2)
  ;d/dx3 = cos(2.0*!DPI*(t+x1)/x2)
  ;d/dx4 = 1.0
END

FUNCTION COSFIT, X, Y, Weights, A, Sigma
  RETURN, CURVEFIT(X, Y, Weights, A, Sigma, FUNCTION_NAME='COSFUNC')
END
\end{lstlisting}

\section{Fourier Autocorrelation}
\label{sec:acorrpro}
\begin{lstlisting}
FUNCTION ACORR, x, NORMALIZE=NORMALIZE
;+
; NAME:
;         ACORR
; PURPOSE:
;         Compute the fourier autocorrelation function
; CALLING SEQUENCE:
;         ACORR,x
; INPUTS:
;         x         : data to autocorrelate
; OPTIONAL INPUTS:
;         none
; OUTPUTS:
;         result    : autocorrelated data
; KEYWORD PARAMETERS:
;         NORMALIZE : performs normalization in respect of the finite length
;                     of the dataset. 
; MODIFICATION HISTORY:
;         Version 1.0, 04.03.2011, Joerg Weingrill SRI/AAS.
;                                 (joerg.weingrill@oeaw.ac.at)   
;- 
  n = N_ELEMENTS(x)
  k = DINDGEN(n)/n
  x1 = [x,REPLICATE(MEAN(x), N_ELEMENTS(x))]
  fftx = fft(x1,-1,/DOUBLE)
  ac = REAL_PART(fft(fftx*conj(fftx),1))
  if KEYWORD_SET(NORMALIZE) then RETURN, (ac/ac[0])/(1.0 - k) else $
    RETURN, (ac/ac[0])
END


\end{lstlisting}

\section{Fourier Cross-correlation}
\label{sec:ccorrpro}
\begin{lstlisting}
FUNCTION CCORR, x,y
  ; calculate the cross-correlation between x and y
  RETURN,REAL_PART(fft(conj(fft(x,-1))*fft(y,-1),1))
END

\end{lstlisting}

\section{Fourier Interpolation}
\label{sec:finterpolpro}
\begin{lstlisting}
; compute the fourier interpolation of a series
FUNCTION FFTFUNCT, P, X=X, Y=Y
  j = WHERE(~FINITE(Y))
  m = N_ELEMENTS(P)
  Y1 = Y
  Y1[j] = P 
  RETURN, REPLICATE(TOTAL(ABS(FFT(Y1))),m)/m
END


function FINTERPOL, V, X, XOUT
  if N_ELEMENTS(XOUT) EQ 0 then begin
    ; if we do not have XOUT, assume non finite values to be interpolated.
    i = WHERE(~FINITE(V))
    XOUT = X[i]
  endif
  ; use linear interpolation as a first guess
  ; usually zeros would do the same
  P0 = INTERPOL(V,X,XOUT)
  FA = {X:x, Y:v}
  ; MPFIT is far more robust then AMOEBA algorithm
  R = MPFIT('FFTFUNCT', P0, functargs=FA,FTOL=1e-5,MAXITER=10)
  RETURN, R
end
\end{lstlisting}

\section{Trapeozid Function}
\label{sec:trapozoidpro}
\begin{lstlisting}
function trapezoid, x, x0, height, duration, impactfactor
;+
;         --------
;        /        \
; ______/          \______
;       1 2      3 4
; x0 ... time of mid trapez (t3-t2)/2
; height ... hight of trapez
; duration ... t4-t1
; impactfactor ... (t3-t2)/(t4-t1)
; flatpart = (t3-t2) = duration*impactfactor
;-  
  n = N_ELEMENTS(x)
; constraints
  if impactfactor GE 1.0 then impactfactor = 0.999
  if impactfactor LT 0.0 then impactfactor = 0.0
  duration = ABS(duration)
  
  y = DBLARR(n)
  if (height eq 0.0) then return,y
  dt = (x[n-1]-x[0])/(n-1)
  duration2 = duration/(2.0d)
  x1 = DBLARR(6)
  y1 = DBLARR(6)
  
  x0dt = x0;/dt
  t1 = (x0 - duration2)
  t2 = (x0 - (duration2*impactfactor)) 
  t3 = (x0 + (duration2*impactfactor)) 
  t4 = (x0 + duration2)
  x1 = [t1-1.0,  t1,     t2,     t3, t4 , t4+1.0]
  y1 = [0.0   , 0.0, height, height, 0.0, 0.0]
  y = INTERPOL(y1, x1, x)  
  return, y
end
\end{lstlisting}

\section{Trapeozid Fit}
\label{sec:mpfindtrapezpro}
\begin{lstlisting}
FUNCTION TRAPEZFUNC,X , P
  return, trapezoid(x, P[0], P[1], P[2], P[3])
END

pro mpfindtrapez, x, y, err, x0=x0, height=height, duration=duration, $
                impactfactor=impactfactor, BESTNORM=BESTNORM, PERROR=PERROR
  P0 = [x0, height, duration, impactfactor]

  R = mpfitfun('TRAPEZFUNC', x, y, err, p0, $ 
               PERROR=PERROR, BESTNORM=BESTNORM, /NaN, /QUIET) ;
  if n_elements(r) gt 1 then begin
    x0 = R[0]
    height=R[1]
    duration=R[2]
    impactfactor=R[3]
  endif
end
\end{lstlisting}

\section{Root Mean Square}
\label{sec:rmspro}
\begin{lstlisting}
FUNCTION RMS, X
  ; compute the root mean square of the vector X
  RETURN, SQRT(MEAN(X^2,/NAN))
END
\end{lstlisting}

\section{Folding of Data}
\label{sec:foldedpro}
\begin{lstlisting}
pro folded,t,l,period,tf=tf,lf=lf,ferr=ferr
  t0 = t[0]
  tf = (t-t0) mod Period
  i = SORT(tf)
  tp = tf[i]
  lp = l[i]
  k = max(t-t0) / Period
  lf = drizzle(tp, lp, err=ferr,loc=tf, BINSIZE=median(deriv(t)))
end
\end{lstlisting}

\section{Period Testing}
\label{sec:periodtestpro}
\begin{lstlisting}
FUNCTION PERIODFIND,P0
  COMMON PERIODFUNC, ct, cl 
  ; fold mod P0
  tf = ct - ct[0]
  P0[0] = P0[0] < max(tf)/3.0
  P0[0] = P0[0] > 1.0/max(tf)
  tf = tf mod P0[0]
  i = SORT(tf)
  tp = tf[i]
  lp = cl[i]
  n = CEIL(0.25*P0[0]*86400.0d/512.0d)
  if n GT 1 then begin
    ;tp1 = P0[0]*DINDGEN(n)/(n-1)
    ; drizzle with new period
    dl = drizzle(tp, lp, err=s,loc=tp1)
    ; return rms of stddev in drizzle bins
    r = max(s)
  endif else r = RMS(cl)
  RETURN, r
END

pro periodtest
  outpath = 'c:\Users\jwe.SATGEO\Documents\Dissertation\thesis\figures\'
  loadlightcurve,'Q:\SRa02\data\SRa02_E1_0359.fits',t,l,$
                 fluxdev=ld,name=name ; Cepheid
  n = N_ELEMENTS(l)
  t = REFORM(t[1,*])
  lr = REFORM(l[0,*])
  lg = REFORM(l[1,*])
  lb = REFORM(l[2,*])
  lrd = REFORM(ld[0,*])
  n = (max(t)-min(t))*86400./512.
  t1 = FINDGEN(n)*(max(t)-min(t))/n + min(t)
  lr1 = drizzle(t,lr,nbins = n,loc=t1)
  lg1 = drizzle(t,lg,nbins = n,loc=t1)
  lb1 = drizzle(t,lb,nbins = n,loc=t1)
  lrd1 = drizzle(t,lrd,nbins = n,loc=t1)
  i = WHERE(lr1 GE 0.01)
  lr1 = lr1[i]
  lrd1 = lrd1[i]
  i = WHERE(lg1 GE 0.01)
  lg1 = lg1[i]
  i = WHERE(lb1 GE 0.01,n)
  lb1 = lb1[i]
  t1 = t1[i]
  mr = -2.5*ALOG10(lr1)*0.937906 + 25.5486
  mg = -2.5*ALOG10(lg1)*0.703167 + 21.9155
  mb = -2.5*ALOG10(lb1)*0.963139 + 25.6203  

  p = 8 ;t2[i[j]]
  l1 = lr1 + lg1 + lb1
  l1 /= MEAN(l1)
  COMMON PERIODFUNC,ct,cl
  ct = t1
  cl = l1
  n = 500
  v = dblarr(n)
  p = dblarr(n)
  for i=0,n-1 do begin
    ;p[i] = 6.5 + i*(1.0/n)
    p[i] = 0.5 + 10.*i*(1.0/n)
    v[i] = periodfind(p[i])
  endfor
  
  mv = min(v,i)
  print,p[i],mv
  PS_Start, FILENAME=outpath+'periodtest.eps', /European, FONT=1, TT_FONT='Arial', $
    LANDSCAPE=0, INCHES=0, XSIZE=15, YSIZE=10, XOffset=0, YOffset=0, /ENCAPSULATED
  plot,p,v,/xs,/ys,title=name,xtitle='period [days]', $
      ytitle='max(sigma)',/nodata,yrange=[0.0,0.1]
  oplot,p,v,color=FSC_COLOR('red')
 
  PS_END    
  p0 = p[i]
  l1 -= mean(l1)
  ;findperiod,t1,l1,p=p0, chisq=chisq
  n = N_ELEMENTS(l1)
  fdr = FFT([REPLICATE(0.0d,n/2),l1,REPLICATE(0.0d,n/2)])
  n2 = N_ELEMENTS(l1)
  ac = acorr(l1)
  ac2 = A_CORRELATE(l1,INDGEN(n))
  freqs =  86400.0d*FINDGEN(n2/2+1)/(n2*512.0d)
  periods = 1.0/freqs
  plot,periods,ABS(fdr[0:n-1])^2,/xs,/ys,/ylog, $
       xtitle='period [days]',xrange=[0.5,11.5]
  plot,t1[0:n-1]-t1[0],ac
 ; print,p0,chisq
end
\end{lstlisting}

\section{Prewhitening}
\label{sec:prewitpro}
\begin{lstlisting}
PRO gfunct, X, A, F, pder 
  n = n_elements(x)
  F = A[0] * SIN(A[1]*2.*!DPI*X/n + A[2]) 
  da0 =  SIN(A[1]*2.*!DPI*X/n + A[2])
  da1 =  A[0] * COS(A[1]*2.*!DPI*X/n+ A[2])*2*!DPI*X/n
  da2 =  A[0] * COS(A[1]*2.*!DPI*X/n + A[2])
  IF N_PARAMS() GE 4 THEN $
    pder = [[da0], [da1], [da2]] 
END 


FUNCTION prewit,x,y,maxiter,freqs = freqs, amps = amps, phas = phas
;+
;prewitning of function y(x) with maxiter iterations 
;-
  freqs = fltarr(maxiter)
  amps = fltarr(maxiter)
  phas = fltarr(maxiter)
  ;freqs_sig = fltarr(maxiter)
  ;amps_sig = fltarr(maxiter)
  ;phas_sig = fltarr(maxiter)
  n = N_ELEMENTS(y)
  dt = x[1]-x[0]
  
  
  for iter = 0,maxiter-1 do begin
    f = (abs(fft(y)))[0:n/2-1]
    pha = (ATAN(fft(y),/PHASE))[0:n/2-1]
    freq = findgen(n/2)/(n*dt)
  
    weights = replicate(1.0,n)
    
    mf = max(f,i)
    mfi = n/(1./freq[i])
    A = [mf*2.,mfi,pha[i]+((!DPI*0.5) MOD !DPI)]
    status = 0
    sigma = 0
    ;plot,x,y,psym=3
    yfit = CURVEFIT(X, Y, weights, A, SIGMA, FUNCTION_NAME='gfunct',status=status)
    ;gfunct,x,a,yfit 
    ;oplot,x,yfit,color=1
    amps[iter] = a[0]
    freqs[iter] = a[1]
    phas[iter] = a[2]
    if status eq 0 then y -= yfit else message,'no fit!'
    ;print,a,sigma
  endfor
  return,y
END
\end{lstlisting}

\end{document}